\documentclass[12pt,a4paper]{article}
\usepackage{hyperref}
\usepackage{nameref}

\usepackage[top=0.85in,left=1.5in,footskip=0.75in]{geometry}

\usepackage{changepage}

\usepackage[utf8]{inputenc}

\usepackage{textcomp,marvosym}

\usepackage{fixltx2e}

\usepackage{amsmath,amssymb}

\usepackage{cite}

\usepackage[right]{lineno}

\usepackage{microtype}
\DisableLigatures[f]{encoding = *, family = * }

\usepackage{rotating}

\usepackage{caption}
\usepackage{subcaption}
\usepackage{url}

\usepackage{makeidx}
\usepackage{fancybox}
\usepackage{float}

\usepackage{graphicx}
\newcommand*{\Comb}[2]{{}^{#1}C_{#2}}%

\raggedright
\usepackage[aboveskip=1pt,labelfont=bf,labelsep=period,justification=raggedright,singlelinecheck=off]{caption}

\makeatletter
\renewcommand{\@biblabel}[1]{\quad#1.}
\makeatother

\date{}

\begin{document}
\vspace*{0.35in}

\begin{flushleft}
{\Large
\textbf{Information Theory and the Length Distribution of all Discrete Systems}
}
\newline
\\
Les Hatton\textsuperscript{1*}
Gregory Warr\textsuperscript{2**}
\\
\bigskip
\bf{1} Faculty of Science, Engineering and Computing, Kingston University, London, UK
\\
\bf{2} Medical University of South Carolina, Charleston, South Carolina, USA.
\\
\bigskip

* lesh@oakcomp.co.uk
** gregory.warr@cantab.net

\end{flushleft}

\begin{verbatim}
Revision: $Revision: 1.35 $
Date:     $Date: 2017/09/06 07:30:43 $
\end{verbatim}

\section*{Abstract}
We begin with the extraordinary observation that the length distribution of 80 million proteins in UniProt, the Universal Protein Resource, measured in amino acids, is qualitatively identical to the length distribution of large collections of computer functions measured in programming language tokens, \textit{at all scales}.  That two such disparate discrete systems share important structural properties suggests that yet other apparently unrelated discrete systems might share the same properties, and certainly invites an explanation.

We demonstrate that this is inevitable for all discrete systems of components built from tokens or symbols.  Departing from existing work by embedding the Conservation of Hartley-Shannon information (CoHSI) in a classical statistical mechanics framework, we identify two kinds of discrete system, \textit{heterogeneous} and \textit{homogeneous}.  Heterogeneous systems contain components built from a unique alphabet of tokens and yield an implicit CoHSI distribution with a sharp unimodal peak asymptoting to a power-law.  Homogeneous systems contain components each built from just one kind of token unique to that component and yield a CoHSI distribution corresponding to Zipf's law.  

This theory is applied to heterogeneous systems, (proteome, computer software, music); homogeneous systems (language texts, abundance of the elements); and to systems in which both heterogeneous \textit{and} homogeneous behaviour co-exist (word frequencies and word length frequencies in language texts).  In each case, the predictions of the theory are tested and supported to high levels of statistical significance.   We also show that in the same heterogeneous system, different but consistent alphabets must be related by a power-law. We demonstrate this on a large body of music by excluding and including note duration in the definition of the unique alphabet of notes.

\section*{Statement of reproducibility}
This paper adheres to the transparency and reproducibility principles espoused by \cite{Popper1959,Ziolkowski1982,Claerbout1992,HatRob94,Donoho2009,Ince2012} and includes references to all methods and source code necessary to reproduce the results presented.  These are referred to here as the \textit{reproducibility deliverables} and are available in several packages\footnote{\url{http://leshatton.org/index_RE.html}}, one for each published paper \cite{HatTSE14,HattonWarr2015} and another covering the work presented here.  All data used are openly available, thanks to the efforts of scientists everywhere.

Each reproducibility deliverable allows all results, tables and diagrams to be reproduced individually for that paper, as well as performing verification checks on machine environment, availability of essential open source packages, quality of arithmetic and regression testing of the outputs \cite{HattonWarr2016}.  Note that these packages are designed to run on Linux machines for no other reason than to guarantee the absence of any closed source and therefore potentially opaque contributions to these results.

\newpage

\section*{Preface: an interesting observation}
\begin{figure}[H]
    \captionsetup[subfigure]{labelformat=empty}
    \centering
    \begin{subfigure}[t]{0.5\textwidth}
        \centering
        \caption{A}
        \includegraphics[width=6cm]{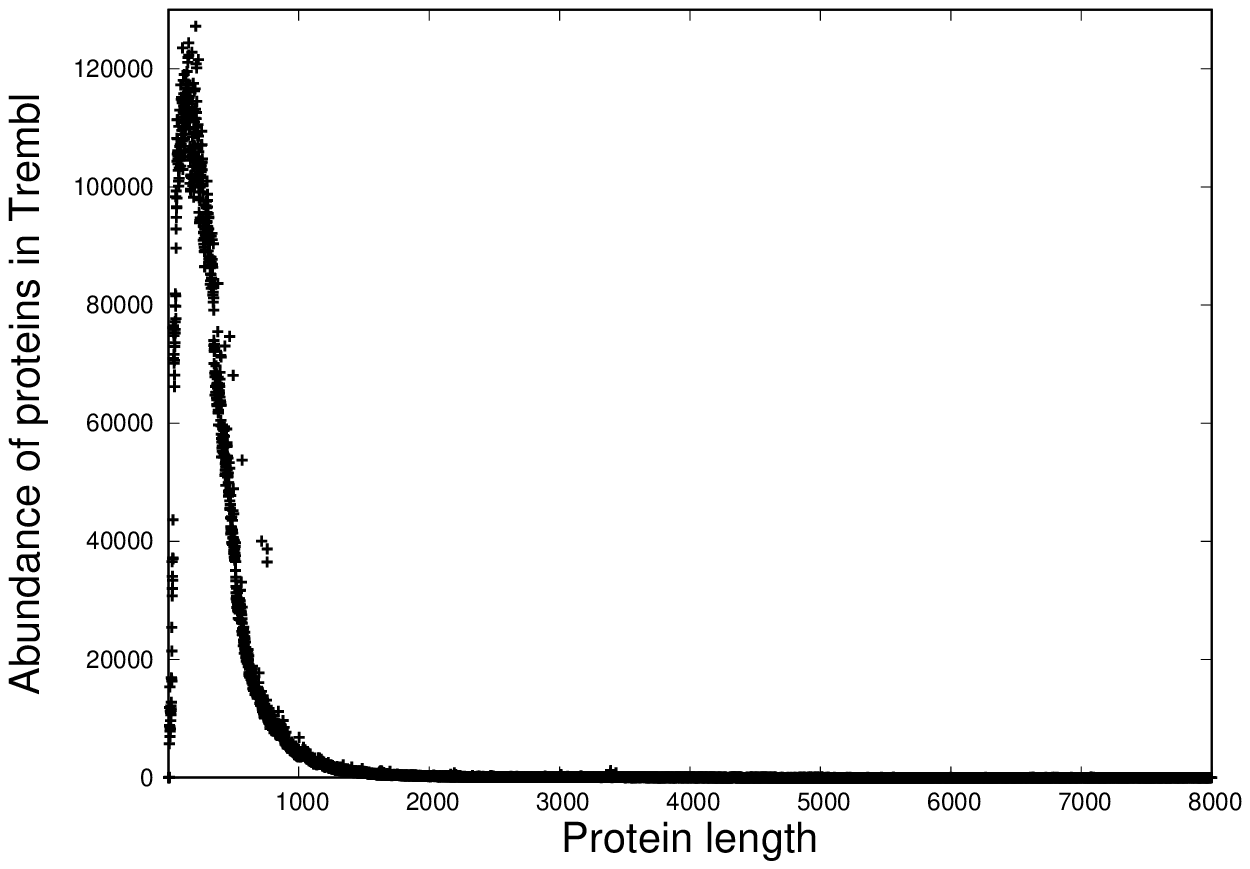}
        \label{fig:trembl_pdf}
    \end{subfigure}%
    ~ 
    \begin{subfigure}[t]{0.5\textwidth}
        \centering
        \caption{B}
        \includegraphics[width=6cm]{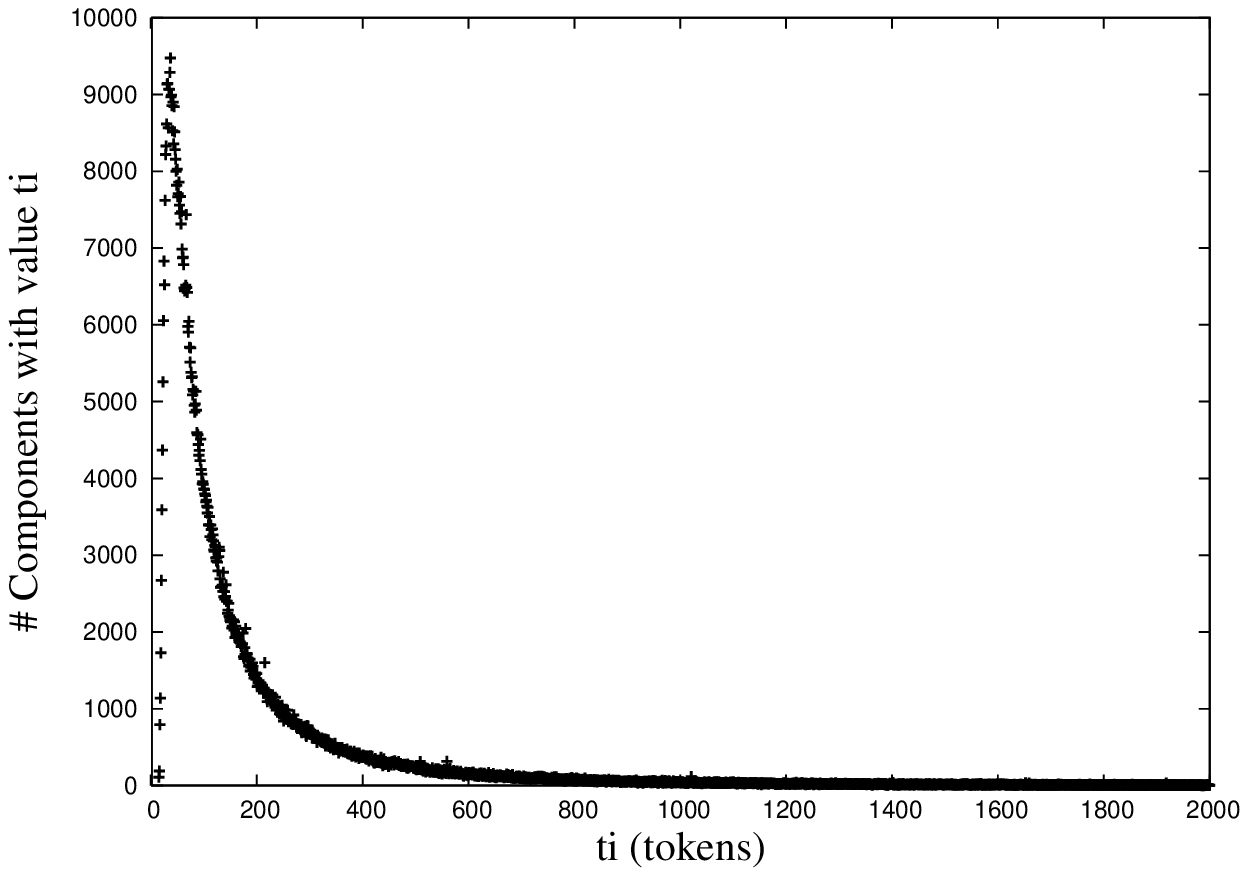}
        \label{fig:c_pdf}
    \end{subfigure}%

    \caption{The frequency distributions of component lengths in two radically different discrete systems - the known proteome as represented by version 17-03 of the TrEMBL distribution (A), and a large collection of computer programs (B) \cite{HatTSE14}.  In each case the y-axis is the frequency of occurrence and the x-axis the length.}
\end{figure}

We start with an interesting observation.  If we \textit{measure} in large populations, how often proteins of different lengths occur (measured in amino acids), and how often computer functions of different lengths occur (measured in programming tokens), we find the following frequency distributions, Figs. \ref{fig:trembl_pdf}, \ref{fig:c_pdf}.

We observe the following:-

\begin{itemize}
 \item Fig. \ref{fig:trembl_pdf} is derived from the European Protein Database TrEMBL version 17-03 (March 2017), \cite{SwissProt2017}.  It is a very large and rapidly growing dataset currently containing around 80 million proteins built from almost 27 billion amino acids.  Fig. \ref{fig:c_pdf} is derived from an analysis of 80 million lines of computer programs written in the programming language C, corresponding to around 500 million programming tokens, around 99\% of which is open source, downloaded from various online archives, \cite{HatTSE14}.
 \item The two systems arose over very different timescales - proteins first appeared perhaps as long ago as 4 billion years, and have been evolving since under natural selection, whilst the C software is definitely less than 40 years old (the C language did not exist much before this).
 \item They are of very different sizes.  The protein data are around 50 times larger than the software data.
 \item They arose through very different processes; proteins are the result of natural selection - nature's ``Blind Watchmaker'' \cite{Dawkins86}, whilst computer programs are the result of deliberate human intellectual endeavour.
 \item Protein data are inherently less accurate than software data \cite{Medigue1999}.  Protein data sequencing\footnote{\url{https://en.wikipedia.org/wiki/Protein_sequencing}, accessed 03-Jun-2017} is subject to experimental error whereas software data sequencing (or tokenization) is precisely defined by programming language standards (ISO/IEC 9899:2011 in the case of C \cite{ISOC11}) and is therefore repeatable.
 \item The frequency distributions of Figs. \ref{fig:trembl_pdf} and \ref{fig:c_pdf} are quite extraordinarily alike.  Both have a sharp unimodal peak with almost linear slopes away from the peaks and from the peak onwards towards longer components, both obey an astonishingly accurate power-law as we shall see.
\end{itemize}

\paragraph{}
\fbox{%
\begin{minipage}{12cm}
\textbf{So, why are they so alike ?  Note that it is not simply a serendipitous choice of datasets, because we will see this pattern appearing again and again at different levels of aggregation and in entirely different systems.  \paragraph{}Instead, as we will show here, this actually arises from the operation of a conservation principle for discrete systems - Conservation of Hartley-Shannon Information (referred to generically as \textit{CoHSI} here to distinguish from other uses of the word ``information'') - acting at a deeper level than either natural selection or human volition.}
\end{minipage}}
\paragraph{}

\newpage
\section*{Introduction}
Such uncanny similarity in very disparate systems strongly suggests the action of an external principle independent of any particular system, so to explore the concepts, let us first consider some examples of discrete systems.  Here a discrete system is considered to be a set of \textit{components}, each of which is built from a \textit{unique alphabet} of discrete choices or \textit{tokens}.  Table \ref{tab:entity} illustrates this nomenclature and its equivalents in various kinds of system.

\begin{table}[h!]
\centering
\begin{tabular}{p{4cm}p{4cm}p{4cm}}
\hline
System & Components & Building blocks (tokens) \\
\hline
Proteome & Proteins & Amino Acids \\
Computer Program & Function & Language tokens \\
Music & Musical composition & Notes \\
Text & Documents & Words, Letters \\
The Universe & Elements & Atomic Numbers \\
\hline
\end{tabular}
\caption{Comparable entities in discrete systems considered in this paper.}
\label{tab:entity}
\end{table}

The first thing to notice is that this seems a very coarse taxonomy.  In the case of proteins, there is no mention of domain of life or species or any other kind of aggregation.  Similarly with computer programs, we do not include the language in which they were written or the application area.  The reason for this as will be seen later is that these considerations are irrelevant.

We will expand on each system later as we apply the theory and discuss its ramifications for each, but the most important concept to grasp now is that there are two measures, \textit{total length} and \textit{unique alphabet}, which will turn out to be fundamental across all such systems.  To illustrate, consider two hypothetical components consisting of the strings of letters shown as rows in Table \ref{tab:letters}.

\begin{table}[H]
\centering
\begin{tabular}{c}
\hline  \\
String \\
\hline \\
AAABA FGACD FFGAB BBBBQ QQQQQ\\
XYZZY TWIST YZZYX TSIWT \\
\hline
\end{tabular} 
\caption{Two simple strings of letters.}
\label{tab:letters}
\end{table}

The \textit{total length} in letters of the first string in Table \ref{tab:letters} is 25.  This is made up of six occurrences of A, 6 of B, 1 of C, 1 of D, 3 of F, 2 of G and 6 of Q.  We therefore define the \textit{unique alphabet} of this component to be of size 7, corresponding to ABCDFGQ.  In other words, the string is built up from one or more occurrences of each and every letter in its unique alphabet.

The second string is made up of 20 characters consisting of 2 of X, 4 of Y, 4 of Z, 4 of T, 2 of W, 2 of I and 2 of S.  Its unique alphabet is therefore XYZTWIS.  In other words, these two strings have different lengths but they have the same size unique alphabet.  As we will see by considering the information content shortly, this property will turn out to be fundamental - the actual letters making up the unique alphabet will turn out to be irrelevant and the two strings inextricably linked in an information theoretic sense.

\newpage
\section*{Methodology}
The methodology we use combines two disparate but long-established methodologies - Statistical Mechanics and Information Theory in a novel way using the simplest possible definition of Information originally defined by Hartley \cite{Hartley1928}.  \textit{We will show that this alone is sufficient to predict all the observed features of Figs. \ref{fig:trembl_pdf} and \ref{fig:c_pdf} and why indeed they are so similar.}

Statistical Mechanics can be used to predict component distributions of general systems made from discrete tokens subject to restrictions known as constraints.  Its classical origins can be found in the Kinetic Theory of Gases \cite{Sommerfeld56} (p.217-) wherein constraints are applied by fixing the total number of particles and the total energy \cite{GlazerWark2001}.  However, the methodology is very general and can equally be used with different constraints on collections of proteins (made from amino acids), software (made from programming language tokens) and, as we shall see, simple boxes containing coloured beads.

Hartley-Shannon Information theory is the result of the pioneering works of Ralph Hartley \cite{Hartley1928} as developed later by Claude Shannon \cite{Shannon1948,Shannon1949}.  It forms the backbone of modern digital communication theory and is also astonishingly versatile.

\paragraph{}
\fbox{%
\begin{minipage}{12cm}
\textbf{The Hartley-Shannon Information Content of a component, in the sense we use here is simply defined to be \textit{the natural logarithm of the total number of distinct ways of arranging the tokens of that component, without any regard for what those tokens actually mean.}}
\end{minipage}}
\paragraph{}

The motivation behind the choice of this form of information is that Figs. \ref{fig:trembl_pdf} and \ref{fig:c_pdf} derive from systems with little if anything in common, but Hartley's definition of information is \textit{token-agnostic}; in other words the meaning of the tokens is irrelevant.  Furthermore, its use favours the ergodic nature of classical Statistical Mechanics with token choice equally likely.

Combining Statistical Mechanics and some form of Information Theory is not new.  For example, building on the maximum entropy framework of \cite{Jaynes2003} rooted in probability theory, Frank demonstrates that by combining Shannon Information \cite{Shannon1948} in a maximum entropy context, the common patterns of nature - Gaussian, exponential, power-law - as predicted by neutral generative processes, naturally emerge \cite{Frank2009}.  Here a neutral generative process assumes that each microscopic process follows random stochastic fluctuations.  Frank's use of \textit{information} can be interpreted as additional \textit{knowledge} about a system constraining the possible patterns which might result.  For example, amongst other things, he demonstrates that simply from an assumption about the measurement scale and knowledge about the geometric mean, a power-law arises. Frank \cite{Frank2009} also stresses the need to distinguish between the generation of patterns by purely random or \textit{neutral} process on the one hand, and the generation of patterns by aggregation of non-neutral processes in which non-neutral fluctuations cancel in the aggregate.

\subsection*{Power-laws}
Power-laws (a.k.a the Pareto distribution) are ubiquitous in nature and are emphatically present in all of the datasets analysed in this paper.  In essence they have a probability distribution which depends on a power $b > 1$ of the independent variable.

\begin{equation}
 p(x) \sim x^{-b}
\end{equation} 

It is important to note that there are numerous known processes which lead to power-laws \cite{Newman2006} and indeed the literature abounds with studies, from the original empirical work of Zipf \cite{Zipf35}, and the earliest generative models such as preferential attachment \cite{Simon1955} onwards.  It is also important to note that other statistical distributions notably lognormal frequently occur in natural phenomena \cite{Mitzenmacher2003}.

In earlier work \cite{HatTSE08,HatTSE14,HattonWarr2015} using the original and arguably the most parsimonious definition of information, \cite{Hartley1928}, embedded as a constraint directly within the classical Statistical Mechanics framework, we demonstrated with compelling support from measurement, that \textit{for large components}, this alone was enough to generate the extraordinarily precise power-laws observed not only in the length distribution of proteins and software, \textit{but also in the distributions of the alphabet of unique tokens.}  

However given the ubiquity of power-laws (and indeed the reasons for this \cite{Frank2009}), perhaps the most compelling reason for accepting the power-law generation inherent in CoHSI as opposed to other generative mechanisms, is to realise that conservation principles cannot be applied selectively.  They are inherently global and either apply everywhere or nowhere and therefore our use of CoHSI \textit{must explain satisfactorily \textbf{all} of the observed properties of the length distributions which appear as Figs. \ref{fig:trembl_pdf}, \ref{fig:c_pdf}, including the sharply unimodal behaviour for smaller values of the independent variable.}

Our novel contributions to the existing body of work are:-

\begin{itemize}
 \item We show that the token-agnostic and scale-independent CoHSI does indeed predict \textit{all} the qualitative features of the distributions of Figs. \ref{fig:trembl_pdf} and \ref{fig:c_pdf} with no other assumptions apart from the conventional use of Stirling's theorem to approximate factorials (Appendix A p. \pageref{app:heterogeneous}).  This is particularly significant because distributions of the nature of Figs. \ref{fig:trembl_pdf} and \ref{fig:c_pdf} are often treated by combining two separate distributions, such as lognormal with a power-law tail \cite{Montroll1982}.  As we show, a single \textit{implicit} distribution which naturally follows from CoHSI is sufficient, thereby emphasizing the parsimony of this approach,
 \item CoHSI naturally leads to an alternative proof of Zipf's law (Appendix A p. \pageref{app:homogeneous}), as we might expect for this Conservation principle,
 \item We enlarge on the results originally derived asymptotically, \cite{HattonWarr2015} that average component lengths (protein, software function ...) are highly conserved across aggregations.  We also point out why very long components naturally must appear quite frequently without any obvious domain-based reason, (Appendix B p. \pageref{app:averagecomplength}),
 \item We show that the asymptotic duality first reported in \cite{HattonWarr2015} between length distribution and alphabet size distribution, naturally implies that different but consistent alphabets for the same system must also be related by power-law, (Appendix C p. \pageref{app:alphabets}).  It also follows that the maximum size of a unique alphabet is intimately related to the total number of components through the slope of the corresponding power-law,
 \item We give experimental confirmation to high levels of significance in multiple disparate datasets at different levels of aggregation for these predictions including systems which contain both heterogeneous and homogeneous behaviour.
\end{itemize}

We stress we are not data-fitting here, and we are not explicitly applying constraints on knowledge of types of mean or variance.  Instead, both the sharp unimodal peak and the very precise power-law tail of Figs. \ref{fig:trembl_pdf} and \ref{fig:c_pdf} naturally emerge from the single Conservation principle, just as the Maxwell-Boltzmann distribution naturally emerges from the Conservation of Energy in Kinetic Theory \cite{Sommerfeld56}.

\subsection*{Why conserve information ?}
To understand this seemingly ad hoc assumption, we must delve into \textit{ergodicity} and consider exactly what happens in classical statistical mechanics when we apply the constraints of total size and total energy to find the most likely distribution of particles amongst energy levels.  Perhaps the most important thing to realise about the statistical mechanics methodology is that it is simply a mathematical technique.  The fact that it is energy (a physical quantity) which is being conserved along with the total number of contributing particles in kinetic theory is irrelevant - conventionally, anything additive can be conserved, however abstract.  The real world only intrudes into the statistical mechanics of Kinetic Theory via Clausius' entropic version of the Second Law of Thermodynamics.  Without this, statistical mechanics simply answers the mathematical question of the most likely distribution of particles or tokens when their total number and their total payload (in this case energy) is conserved.

\textit{In other words, amongst all the possible systems with that number of particles and that total payload (the ergodic ensemble), then presented with one of them, it is most likely to follow the distribution predicted by statistical mechanics for the ensemble.}  It doesn't have to, but it is overwhelmingly likely that it does.  In kinetic theory, the payload happens to be a physical additive quantity, the energy, and the corresponding distribution of particles is then the Maxwell-Boltzmann distribution, but to say something about a system, statistical mechanics does not \textit{need} to be rooted in tangibly defined entities in the physical world.  We simply have to interpret the result appropriately.

Now it so happens that Hartley-Shannon information content, like energy, is also additive for independent sub-systems.  The total energy $E$ of two sub-systems with individual energies $E_{1}$ and $E_{2}$ is $E = E_{1} + E_{2}$.  Similarly, by virtue of its logarithmic definition, the total Hartley-Shannon information content $I$ of two sub-systems with individual information content $I_{1}$ and $I_{2}$ is $I = I_{1} + I_{2}$.  The difference between the two is that energy is a physical quantity, whereas Hartley-Shannon information content is just the $log$ of the total number of ways of arranging something \cite{Cherry1963}.  Mathematically it resembles entropy but we should be hesitant about reading too much into this \cite{Lemons2013}, p. 144.  However, we may still use the formalism of statistical mechanics, which we do here.

\paragraph{}
\fbox{%
\begin{minipage}{12cm}
\textbf{It is in this sense that we assert that Conservation of Hartley-Shannon Information underlies the length distribution of discrete systems whatever their provenance.  It is a natural consequence of statistical mechanics that if we are presented with a system with a total number of tokens and a total information payload, then it is overwhelmingly likely to follow a certain size distribution as described in what follows.  The scale-independence of the results follows from the fact that, given any system, it is the properties of the ergodic system of the same parameters which defines the most likely distributions to occur in any one of its constituents.  This paper contains many examples of real systems at all levels of aggregation where precisely this size distribution is found, just as we expect from the theory.}
\end{minipage}}
\paragraph{}

\subsection*{Statistical Mechanics}
Statistical Mechanics connects the minutiae of systems of large numbers of small particles to macroscopic properties of those systems \cite{GlazerWark2001}.  Like Hartley-Shannon Information, it is quite astonishingly versatile and arose originally in the Kinetic Theory of Gases, leading eventually in the hands of James Clerk Maxwell and later Ludwig Boltzmann in the 19th century, to the statistical distribution of velocities and on to the concept of Entropy\footnote{\url{https://en.wikipedia.org/wiki/Kinetic_theory_of_gases}, accessed 25-May-2017.}.

The methodology of Statistical Mechanics leads naturally to links with probability distributions and energy in the case of gases.  Here we use the same methodology but by embedding CoHSI as a constraint in Statistical Mechanics rather than Conservation of Energy, we demonstrate that this links Hartley-Shannon Information directly to probability distributions of component length and unique alphabet size in discrete systems.

We distinguish between two fundamental types of system which lead naturally to two different definitions of information.  Both contain components made from discrete tokens as described above but with one fundamental difference.

\begin{description}
 \item[Heterogeneous] We define heterogeneous systems here as systems wherein a component has more than one kind of distinct token.  This would include systems as disparate as the proteome, software and digital representations of music.  Appendix A p. \pageref{app:heterogeneous} contains a detailed development of these systems.
 \item[Homogeneous] We define homogeneous systems here as systems wherein a component has only one kind of distinct token and each distinct token is unique to one component.  This would include textual documents and word counts as well as the distribution of elements in the universe.  In such systems, a heterogeneous definition of information would be degenerate and a different definition is necessary.  Appendix A
 p. \pageref{app:homogeneous} contains the detailed development for this kind of system leading directly to an alternative proof of Zipf's law.
\end{description}

However, this is irrelevant as far as statistical mechanics goes because for a given definition of Hartley-Shannon Information, the methodology simply tells us the most likely, or \textit{canonical} distribution for ergodic systems with the same fixed size and fixed information content, howsoever defined.

For heterogeneous systems, we will refer to the resulting distributions as the \textit{heterogeneous CoHSI distribution.}  The corresponding distribution for homogeneous systems is simply Zipf's law.

\subsection*{The heterogeneous CoHSI distribution}
The theory described in Appendix A p. \pageref{app:sm} predicts that the length distribution of a heterogeneous discrete system such as the proteome or software systems, at all scales with total number of tokens $T$ and total Hartley-Shannon Information $I$ is the solution $(t_{i},a_{i})$ of the implicit pdf corresponding to

\begin{equation}
\log t_{i} = -\alpha -\beta ( \frac{d}{dt_{i}} \log N(t_{i}, a_{i}; a_{i} ) ),    \label{eq:cohsi}
\end{equation}

with

\begin{equation}
T = \sum_{i=1}^{M} t_{i}
\end{equation}

and

\begin{equation}
I = \sum_{i=1}^{M} I_{i}
\end{equation}

where $t_{i}, a_{i}, I_{i}$ are the \textit{length in tokens}, the \textit{size of unique alphabet of tokens} and the \textit{Hartley-Shannon Information content}, respectively, of the $i^{th}$ component of a system containing $M$ components in all.  $\alpha$ and $\beta$ are Lagrange undetermined multipliers.  $N(t_{i}, a_{i}; a_{i})$ is the total number of ways of choosing $t_{i}$ tokens at random, choosing from a replaceable unique set of tokens $a_{i}$.  We write $N(t_{i}, a_{i}; a_{i})$ in this special form to remind us of the recursive nature of its construction.  Here, $t_{i}$ is the independent variable and $a_{i}$ plays a dual role acting also as the scaled frequency of occurrence.

In addition, for components which are much longer than their unique alphabet, $t_{i} \gg a_{i}$, the full solution (\ref{eq:cohsi}) tends to the asymptotic pdf (probability distribution function) \cite{HatTSE14} given by

\begin{equation}
p_{i} \equiv \frac{t_{i}}{T} = \frac{a_{i}^{-\beta}}{\sum_{i=1}^{M} a_{i}^{-\beta}},		
\label{eq:pwrlaw}
\end{equation}

which has an algebraic dual pdf \cite{HattonWarr2015} given by

\begin{equation}
q_{i} \equiv \frac{a_{i}}{A} = \frac{t_{i}^{-1/\beta}}{\sum_{i=1}^{M} t_{i}^{-1/\beta}},		\label{eq:pwrlaw2}
\end{equation} 

where

\begin{equation}
A = \sum_{i=1}^{M} a_{i}
\end{equation} 

(\ref{eq:pwrlaw}) and (\ref{eq:pwrlaw2}) show that the tails of both unique alphabet size distributions and length distributions respectively of the M components will be asymptotically power-law as emphatically confirmed in \cite{HattonWarr2015}.

\begin{figure}[H]
\centering
\includegraphics[width=8cm,height=6cm]{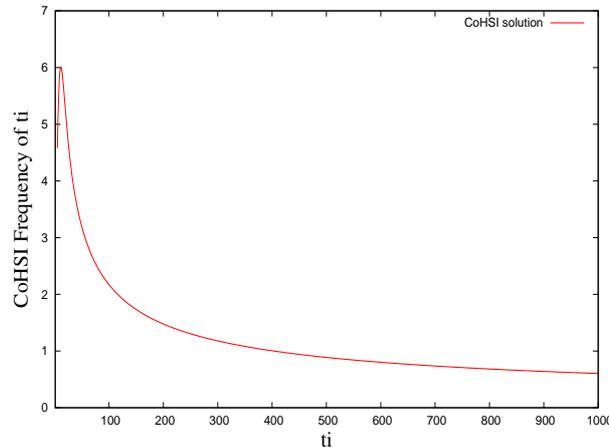}
\caption{A typical solution of (\ref{eq:cohsi}) shown as a pdf.  Both the sharp unimodal peak and power-law tail can be seen clearly.}
\label{fig:cohsi_pdf}
\end{figure}

A typical solution of (\ref{eq:cohsi}) is shown as Fig. \ref{fig:cohsi_pdf}.

Before applying this theory predictively to various systems so that we may test it, we make two comments.

\begin{itemize}
 \item The Lagrange multipliers $\alpha, \beta$ are undetermined by the methodology of statistical mechanics.
 \begin{itemize}
  \item $\alpha$ parameterises the total size of the system and therefore emerges naturally as a normalisation condition so that a pdf results.
  \item $\beta$ is more interesting.  It parameterises the total payload.  The payload in our theory is Hartley-Shannon information which as described in Appendix A p. \pageref{app:heterogeneous}, depends on the size of the alphabet we use to categorise a discrete system.  Small alphabets correspond to large $\beta$ and vice versa.  The implication of this indeterminism is that the range of values of $\beta$ which emerge via information theory can be much wider than those tied to physical systems, (which are mostly in the range 1.5-4).  (In the Maxwell-Boltzmann distribution where the payload is energy, Appendix A p. \pageref{app:sm}, it is fixed by being closely linked by Boltzmann's constant to the temperature, through the Second Law of Thermodynamics.)
 \end{itemize}
 \item Dual regime behaviour is often identified and modelled as lognormal transitioning to power-law \cite{Montroll1982,Mitzenmacher2002}.  We stress here that no such juxtaposition of distinct pdfs is necessary with the theory presented here.  Instead the sharp unimodal peak and the power-law regime of Figs. \ref{fig:trembl_pdf}, \ref{fig:c_pdf} naturally emerge according to the \textit{implicit} solution of (\ref{eq:cohsi}) as $t_{i} \rightarrow 1$ from large values as can be seen in Fig. \ref{fig:cohsi_pdf}.  \textit{This is a direct consequence of CoHSI in an ergodic system as described in Appendix A p. \pageref{app:heterogeneous}.}  The observed shape of the sharply unimodal regime of Fig. \ref{fig:trembl_pdf} simply lends itself to a lognormal fit.  We also note that the unusual implicit behaviour inherent in (\ref{eq:cohsi}) is also a feature of modified entropy definitions such as Tsallis entropy \cite{Tsallis1988,Tsallis1999} constructed to account for non-additive entropy.  Tsallis entropy is a modification of Shannon entropy with an additional parameter.  In our development, the implicit behaviour emerges naturally from the one assumption of CoHSI.
\end{itemize}

We now apply these conclusions predictively to various systems.

\newpage
\section*{Results}
\subsection*{Proteins}
Proteins are constructed as strings of amino acids corresponding to the heterogeneous model we describe here, Appendix A p. \pageref{app:heterogeneous}.  They are represented in exactly the same way as the strings of letters in Table \ref{tab:letters} but the unique alphabet from which the letters are chosen is the 22 unmodified amino acids which are coded directly from DNA \cite{Srinivasan2002,Gladyshev2010}, supplemented with modified versions produced by a process known as Post-Translational Modification, of which there are already thousands known \cite{KhouryBalibanFloudas2011,PrabakarnLippensSteenGunawardena2012,Campbell01012014}.  Table \ref{tab:proteins} shows two small proteins, one from an Archaean and the other from a virus \cite{SwissProt2017} along with their sequences in their single letter abbreviations for compactness \cite{HattonWarr2015}

\begin{table}[H]
\centering
\begin{tabular}{p{5cm}p{7cm}}
\hline  \\
Protein & Sequence \\
\hline \\
FLA1\_METHU & FSGLEAAIVL IAFVVVAAVF \\
(Methanospirillum hungatei) & SYVMLGAGFF AT \\
VE5\_PAPVR & MNHPGLFLFL GLTFAVQLLL \\
(Reindeer papillomavirus)  & LVFLLFFFLV WWDQFGCRCD GFIL \\
\hline
\end{tabular} 
\caption{Sequences of two small proteins.}
\label{tab:proteins}
\end{table}

Biologically, these two proteins differ significantly.  They have different lengths, (32 and 44 amino acids respectively); are built using different amino acids; and they have very distinct structures and functions.  FLA1\_METHU is built from the unique amino acid alphabet FSGLEAIVLYMGT (13) and VE5\_PAPVR is built from the unique amino acid alphabet MNHPGLFTAVQWDCRI (16).  Following our earlier argument about unique alphabets, it does not matter if an amino acid is present once or more often in the sequence.  If it is present at all, then it contributes a count of 1 to the unique amino acid count.  While these two numbers are clearly independent of any physicochemical properties of the amino acids, they are fundamental in determining the length distribution of any aggregation of proteins.

The protein sequences are collected in public databases from which they can be downloaded and analysed \cite{SwissProt2017}.  There are currently more than 80 million proteins in TrEMBL version 17-03 built from almost 27 billion amino acids, (the most recent version analysed before writing this paper).  The proteins vary in length from just four amino acids to over 36,000 amino acids but their average length is only around 300 amino acids.  The reason for the existence of such long proteins is directly predicted by the development of theory which follows later in this paper, Appendix B, p. \pageref{app:averagecomplength}.

The length of a protein is of course one of the factors which determines its folding properties and therefore its functionality \cite{LiKlimov2002,Ivankov2003,LanePande2013}.

\subsubsection*{Power-law tails of alphabet and length}
As originally shown in \cite{HattonWarr2015}, the tails of the alphabet and length distributions are both power-law to a high degree of precision.  The protein alphabets are analysed in Appendix C p. \pageref{app:alphabets}.

\begin{figure}[H]
\centering
\includegraphics[width=8cm,height=6cm]{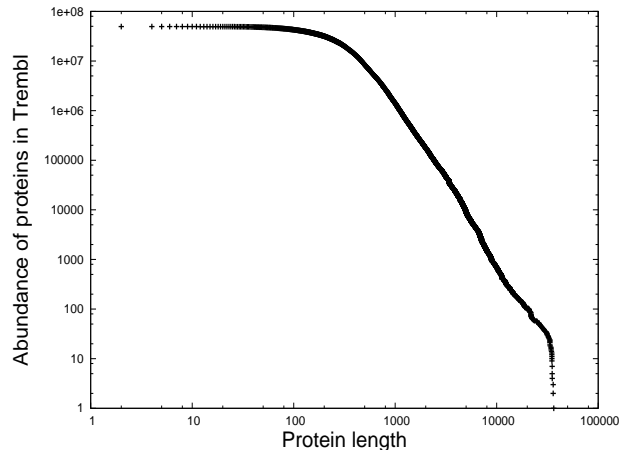}
\caption{The data of Fig. \ref{fig:trembl_pdf} shown as a $\log-\log$ ccdf.}
\label{fig:trembl_cdf}
\end{figure}

The ccdf (complementary cumulative distribution function) of the length distribution corresponding to Fig. \ref{fig:trembl_pdf} is shown as Fig \ref{fig:trembl_cdf}.

\paragraph{}
\fbox{%
\begin{minipage}{12cm}
\textbf{The linearity in the tail of Fig. \ref{fig:trembl_cdf} provides striking confirmation of equation (\ref{eq:pwrlaw2}).  The R lm() function reports that the associated p-value matching the power-law tail linearity is $< 2.2 \times e^{-16}$ over the range $300.0-30000.0$, with an adjusted R-squared value of $0.9942$.  The slope is $-3.13 \pm 0.20$.}
\end{minipage}}
\paragraph{}

\subsubsection*{Aggregations by domain and species}
As stated earlier, the canonical shape of Figs. \ref{fig:trembl_pdf}, \ref{fig:c_pdf} occurs at each level of aggregation in these systems.  This we now show for both the domains of life and also down to individual species.

\begin{figure}[H]
    \captionsetup[subfigure]{labelformat=empty}
    \centering
    \begin{subfigure}[t]{0.5\textwidth}
        \centering
        \caption{A}
        \includegraphics[width=6cm]{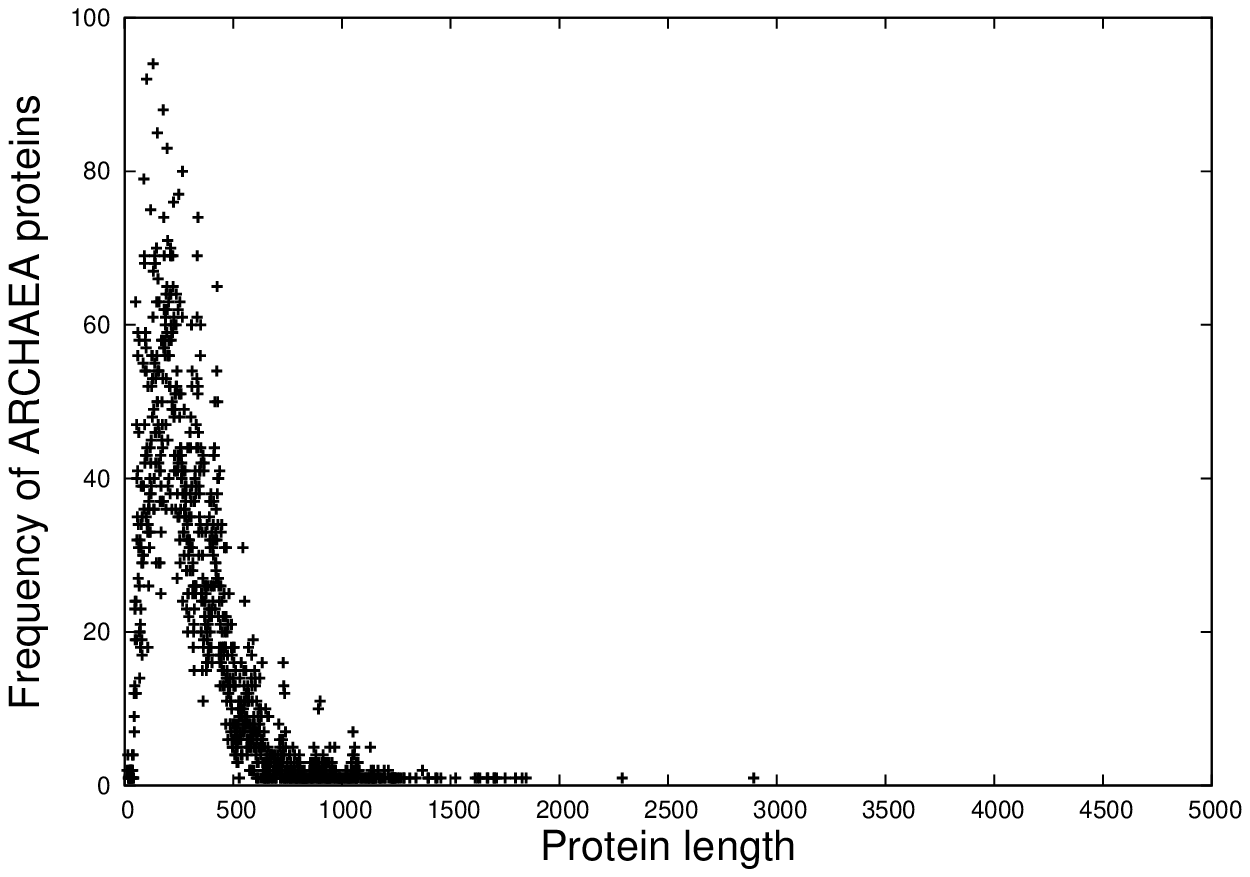}
        \label{fig:archaea_pdf}
    \end{subfigure}%
    ~ 
    \begin{subfigure}[t]{0.5\textwidth}
        \centering
        \caption{B}
        \includegraphics[width=6cm]{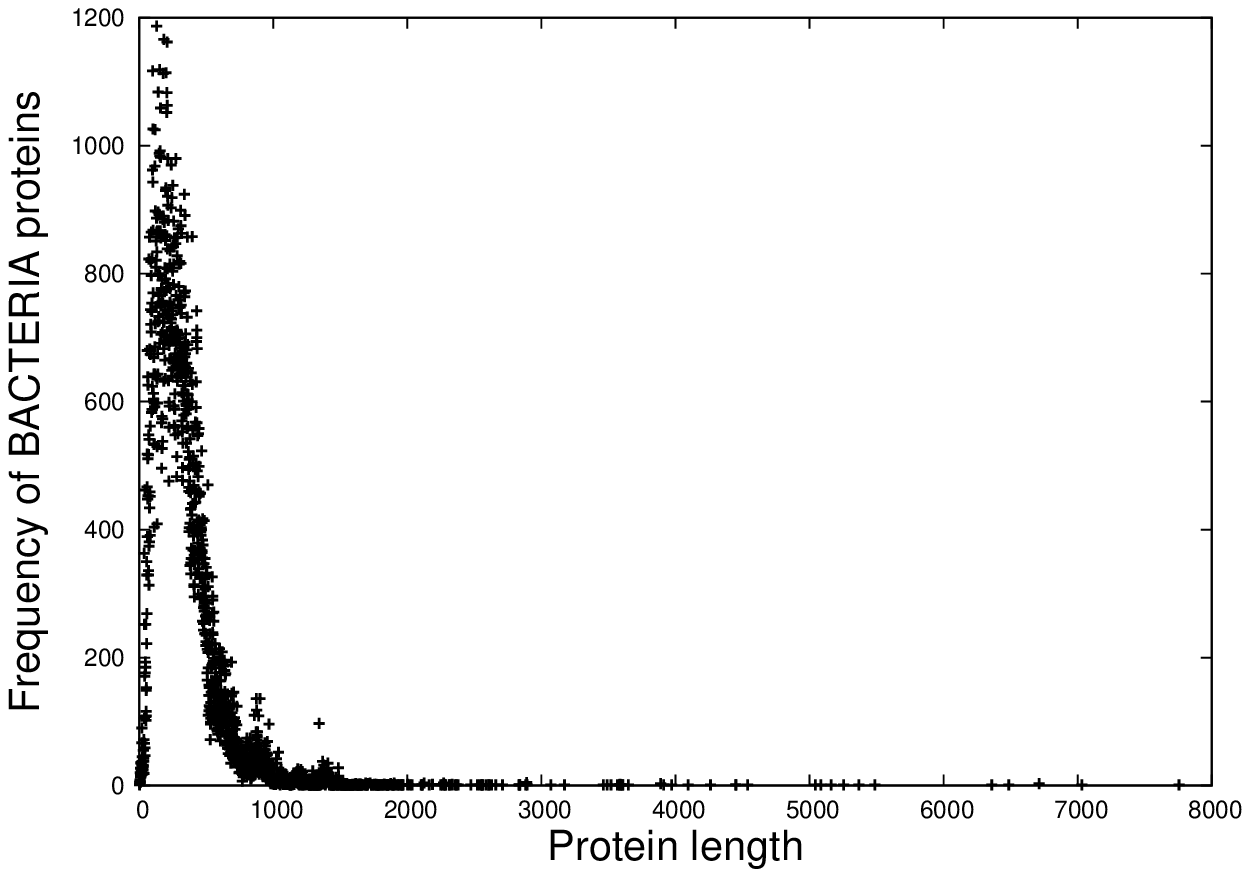}
        \label{fig:bacteria_pdf}
    \end{subfigure}%

    \begin{subfigure}[t]{0.5\textwidth}
        \centering
        \caption{C}
        \includegraphics[width=6cm]{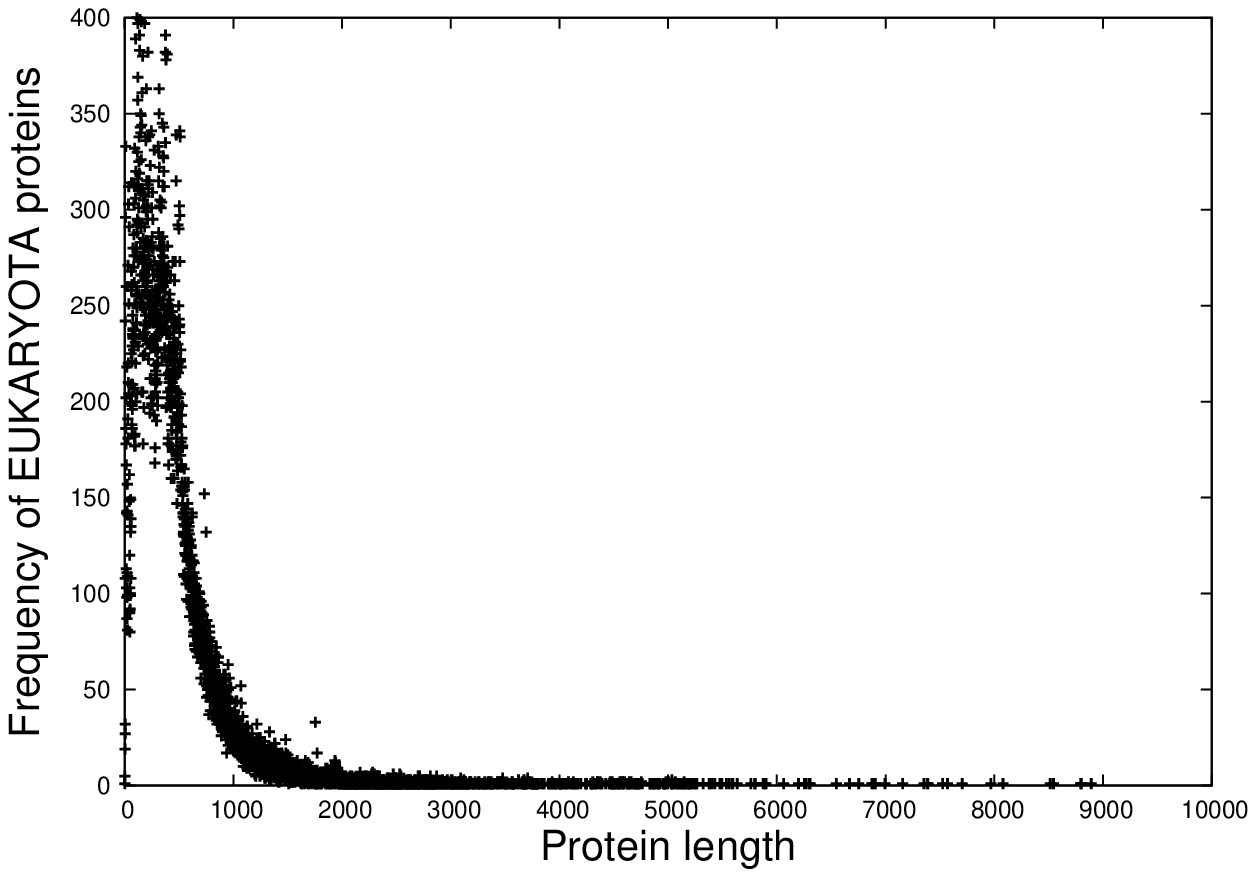}
        \label{fig:eukaryota_pdf}
    \end{subfigure}%
    ~ 
    \begin{subfigure}[t]{0.5\textwidth}
        \centering
        \caption{D}
        \includegraphics[width=6cm]{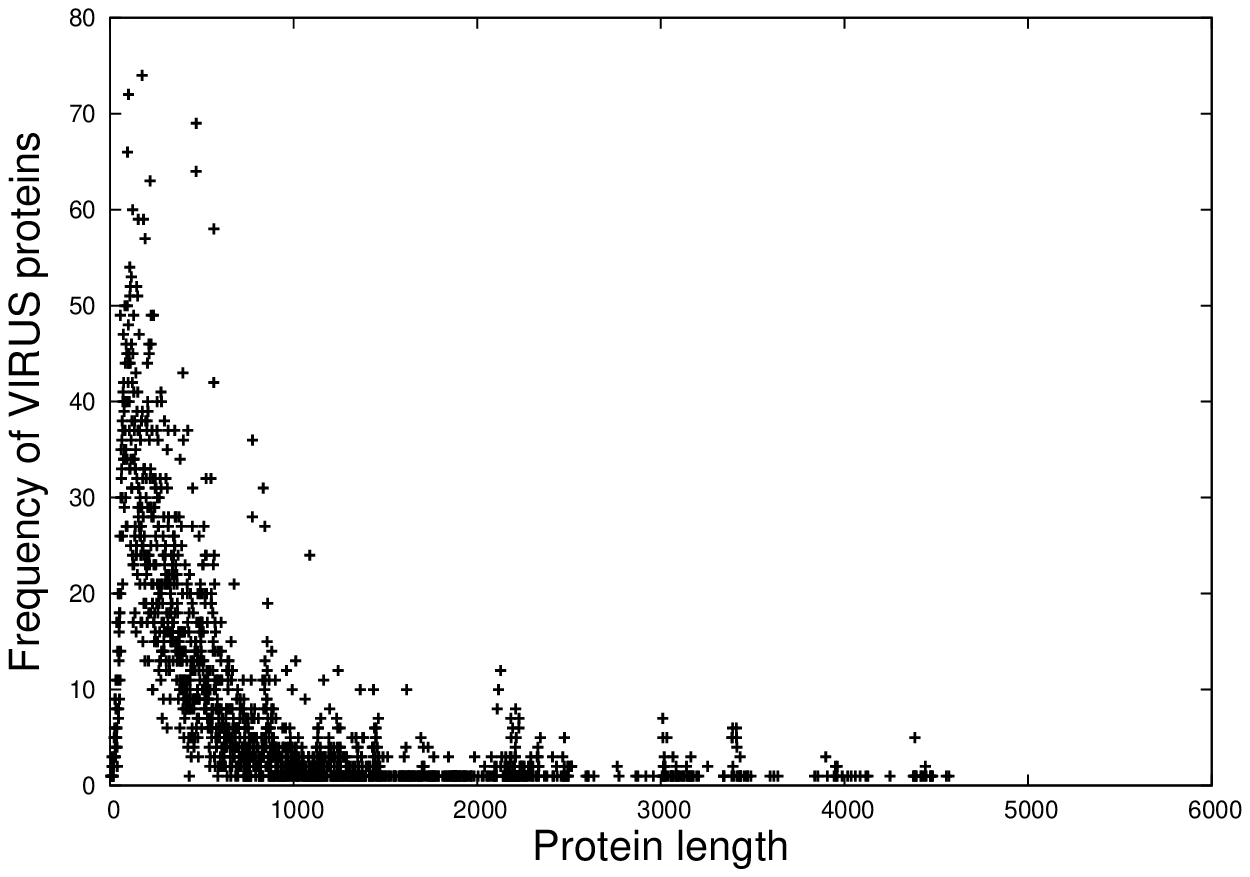}
        \label{fig:viruses_pdf}
    \end{subfigure}%
    \caption{The frequency distributions of protein lengths in the three domains of life, (A): Archaea, (B): Bacteria and (C): Eukarya, along with (D): Viruses.}
\end{figure}

\begin{figure}[H]
\centering
\includegraphics[width=8cm,height=6cm]{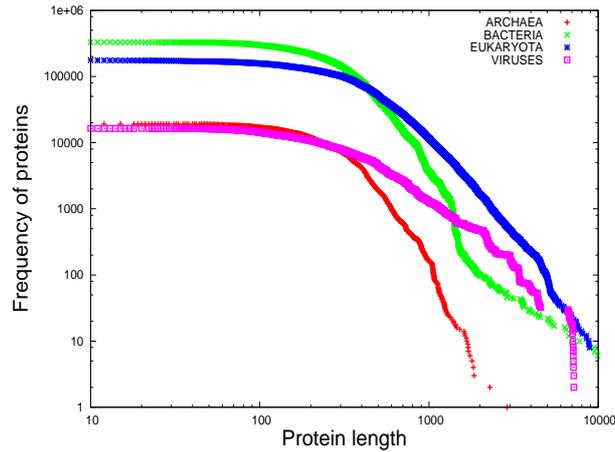}
\caption{The three domains of life and viruses shown as a $\log-\log$ ccdf.  The significantly shallower slope in Viruses is notable.}
\label{fig:cdf_domainsoflife}
\end{figure}

In contrast to TrEMBL, the SwissProt database \cite{SwissProt2015} provides a smaller but well annotated set of data suitable for the extraction and analysis of data from taxa at diverse levels of aggregation, from the highest taxonomic classification shown (the three domains of life) down to individual species.  In Figs. \ref{fig:archaea_pdf}-\ref{fig:viruses_pdf}: are Archaea (Figs \ref{fig:archaea_pdf}: 19,063 species), Bacteria Figs (\ref{fig:bacteria_pdf}: 329,526 species) and Eukarya Figs \ref{fig:eukaryota_pdf}: 177,020 species).  Included for comparison are the viruses Figs (\ref{fig:viruses_pdf}: 16,423 species).  In every case, the characteristic qualitative signature of Fig. \ref{fig:trembl_pdf} is evident in the domain of life (with variations inevitably depending on the sample size), and even (in the case of viruses), a dataset outside the domains of life.  The pdfs are scaled separately to show the qualitative similarity whilst Fig. \ref{fig:cdf_domainsoflife} shows the matching absolutely-scaled ccdfs and the emergence of the power-law tail in each collection.

\begin{figure}[H]
    \captionsetup[subfigure]{labelformat=empty}
    \centering
    \begin{subfigure}[t]{0.5\textwidth}
        \centering
        \caption{A}
        \includegraphics[width=6cm]{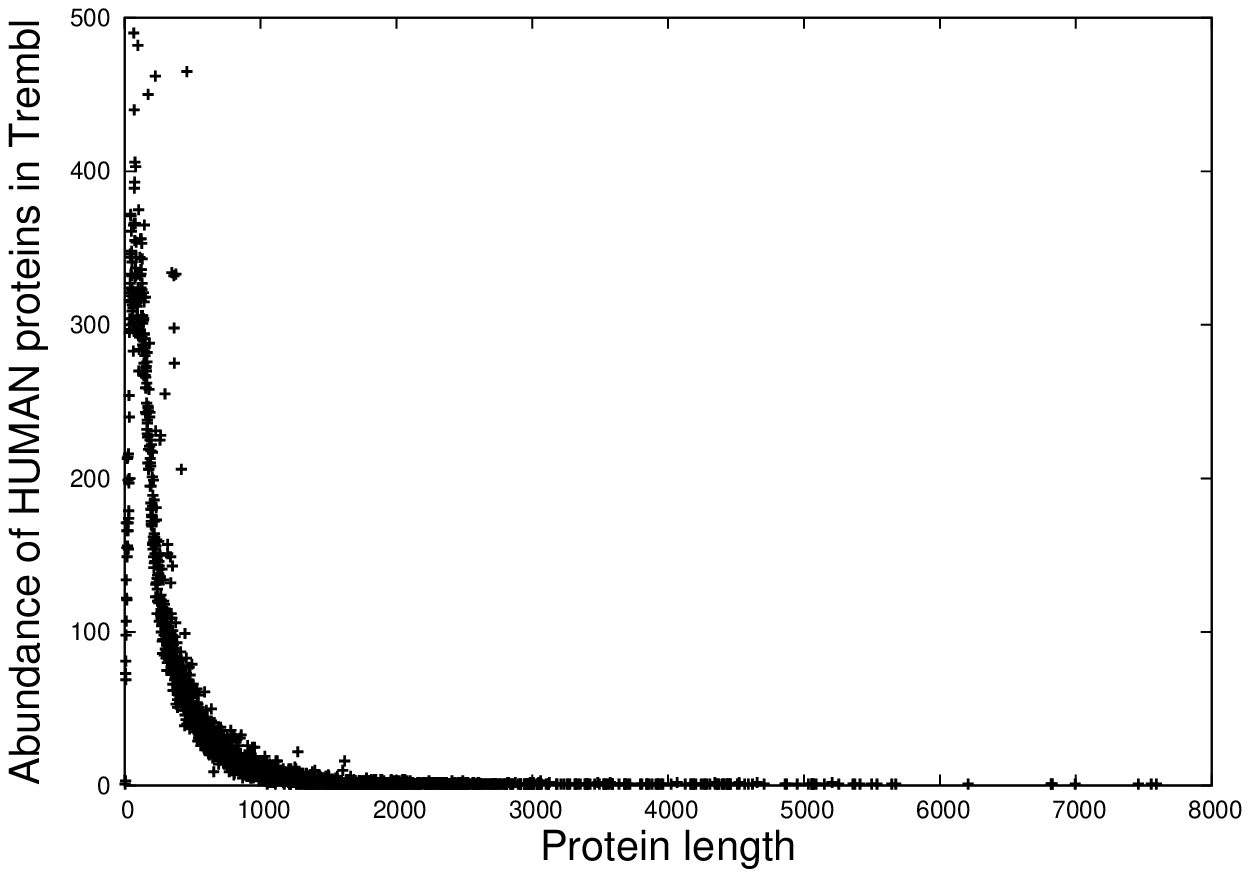}
        \label{fig:human_pdf}
    \end{subfigure}%
    ~ 
    \begin{subfigure}[t]{0.5\textwidth}
        \centering
        \caption{B}
        \includegraphics[width=6cm]{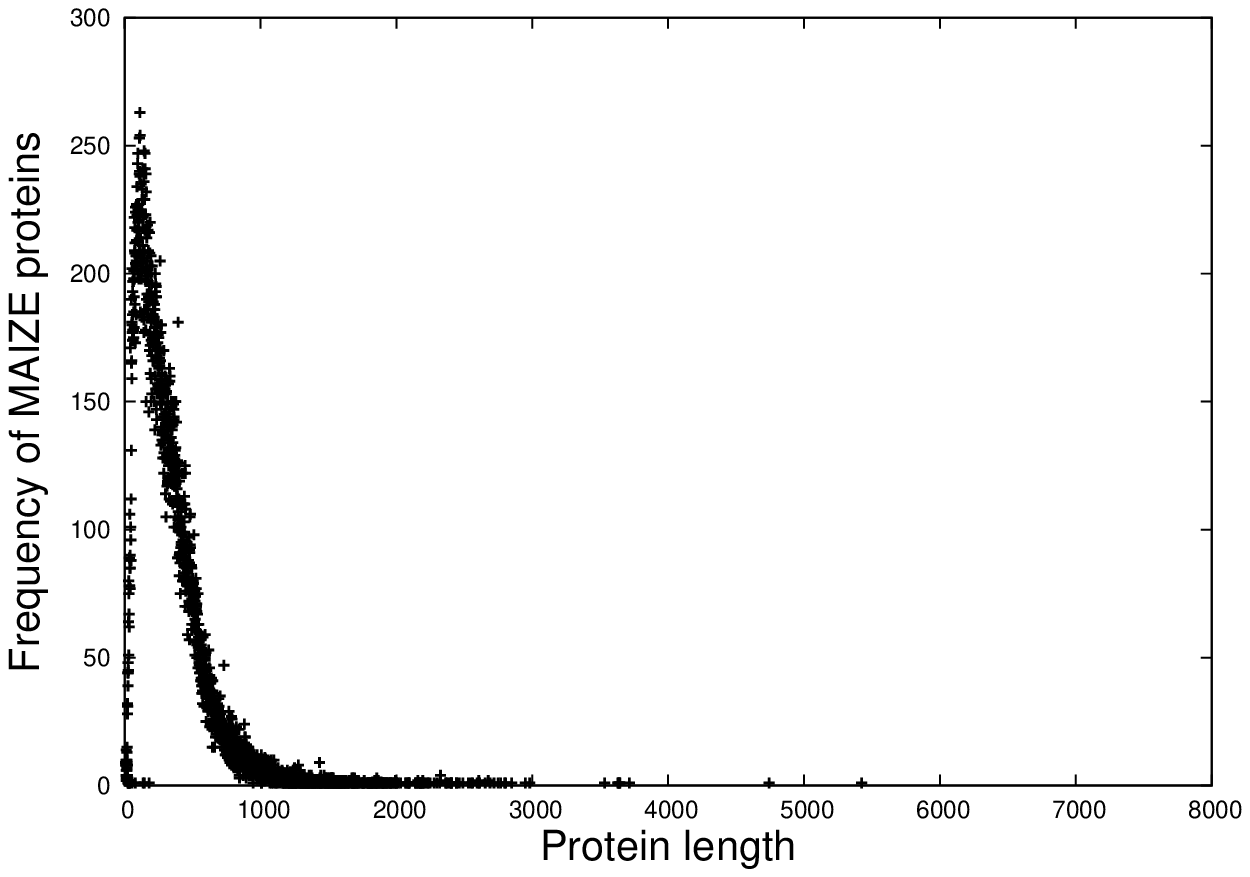}
        \label{fig:maize_pdf}
    \end{subfigure}%

    \begin{subfigure}[t]{0.5\textwidth}
        \centering
        \caption{C}
        \includegraphics[width=6cm]{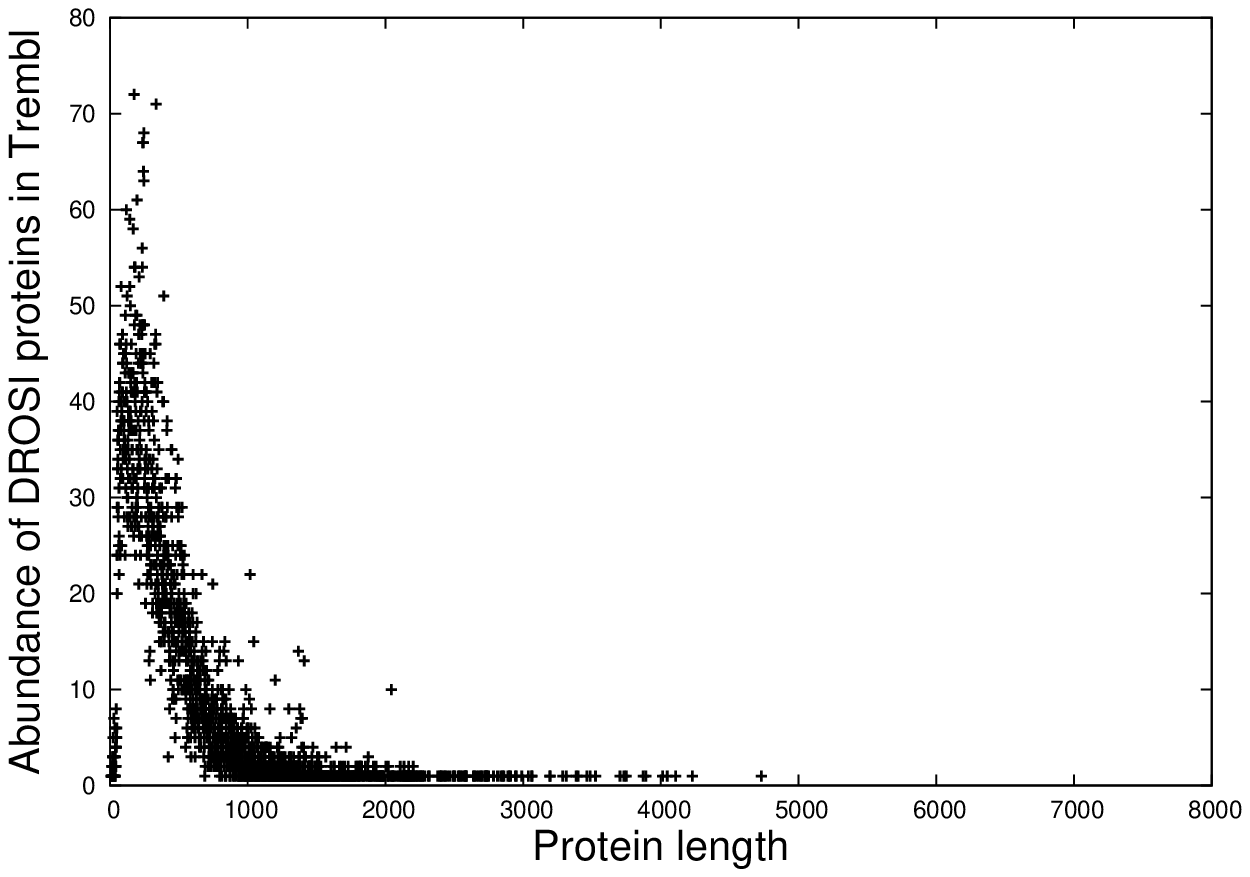}
        \label{fig:drosi_pdf}
    \end{subfigure}%
    ~ 
    \begin{subfigure}[t]{0.5\textwidth}
        \centering
        \caption{D}
        \includegraphics[width=6cm]{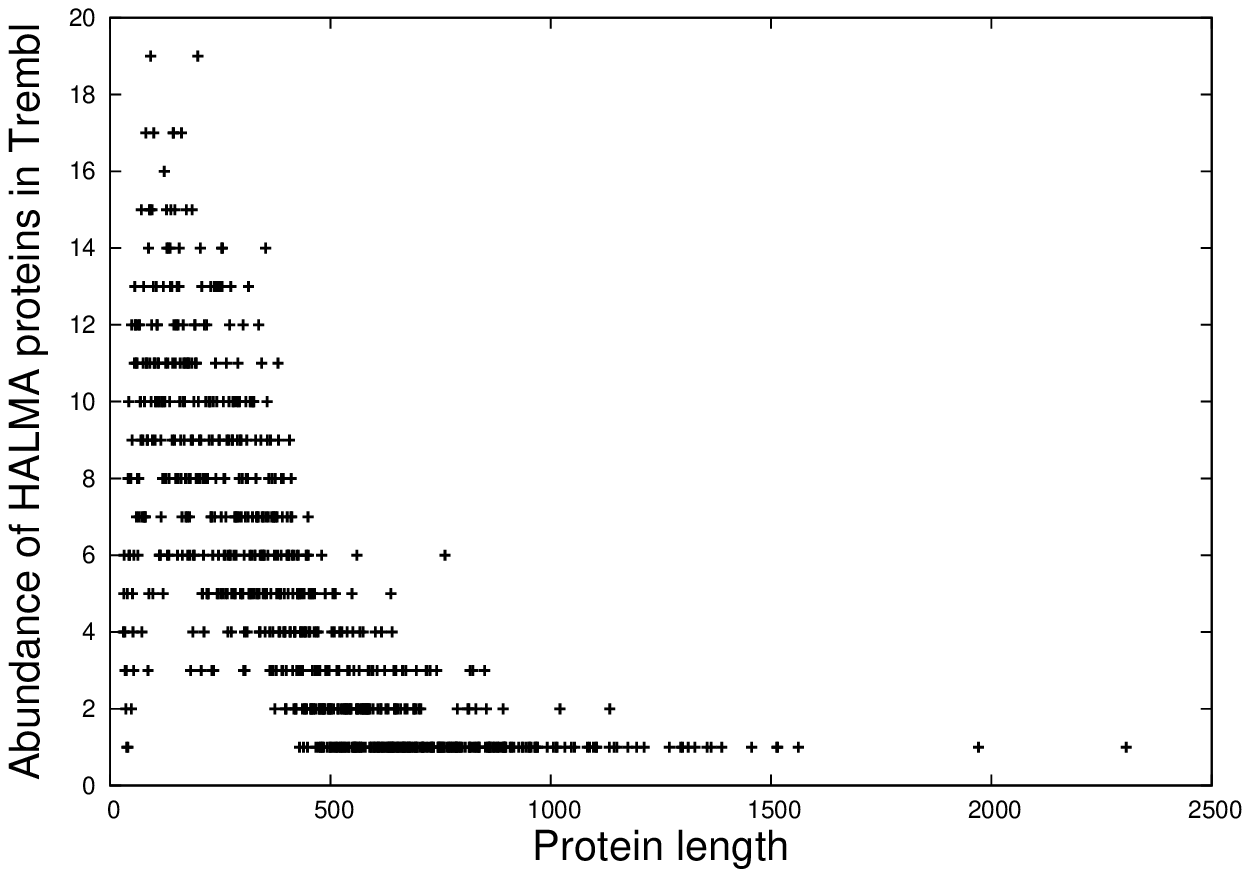}
        \label{fig:halma_pdf}
    \end{subfigure}%
    \caption{The frequency distributions of protein lengths in four species, (A): Human, (B): Maize and (C): Fruit fly, along with (D): Haloarcula marismortui.}
\end{figure}

\begin{figure}[H]
\centering
\includegraphics[width=8cm,height=6cm]{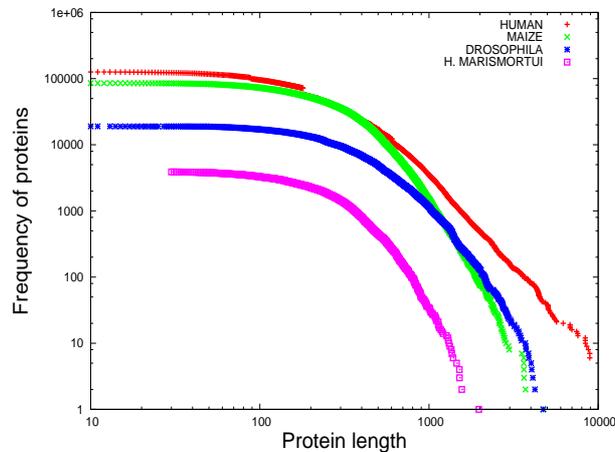}
\caption{Four species shown together as a $\log-\log$ ccdf.}
\label{fig:cdf_species}
\end{figure}

The only remaining approximation of CoHSI is that the number of components (i.e. proteins in this case) be reasonably large, so a critical test case is the analysis of individual species, where the protein databases allow us to analyze small sets of proteins naturally defined by species.  To demonstrate the resilience of CoHSI, we consider species with very different numbers of unique proteins.  Figs. \ref{fig:human_pdf}-\ref{fig:halma_pdf} show the length distributions of proteins in (Fig. \ref{fig:human_pdf}) humans (126,468 proteins); (Fig. \ref{fig:maize_pdf}) maize (85,311 proteins); (Fig. \ref{fig:drosi_pdf}) fruit fly (18,966 proteins); and (Fig. \ref{fig:halma_pdf}) \textit{Haloarcula marismortui} (3,892 proteins) respectively.  Even in the smallest of these datasets, \textit{H. marismortui}, a halophilic red Archaeon found in the extreme environment of the Dead Sea, the canonical shape of Fig. \ref{fig:trembl_pdf} is apparent.  The pdfs are again scaled separately to show the qualitative similarity whilst the corresponding ccdfs are shown absolutely scaled in Fig. \ref{fig:cdf_species}.

\subsection*{Computer programs}
Computer programs are an invention of the human mind following the ground-breaking work of Alan Turing.  In the 50 or so years since they first appeared, many programming languages have arisen, from which computer programs of almost limitless functionality are built.

The individual bases or alphabet of a programming language are called \textit{tokens} and may take two forms; the \textit{fixed} tokens of the language as provided by the language designers, and the \textit{variable} tokens.  Fixed tokens include (in the languages C and C++ for example) keywords such as \textit{if}, \textit{else}, \textit{while}, \textit{\{}, \textit{\}}.  These can not be changed, the programmer can only choose to use them or not. Variable tokens, with some small lexical restrictions (such as the common requirement for identifiers to begin with a letter), can be arbitrarily invented by the programmer whilst constructing their program.  These might be names such as \textit{numberOfCandidateCollisions} or \textit{lengthOfGene} or constants such as 3.14159265.  There are many programming languages but all obey the same principles and every form of software system evolves from such tokens.  They are therefore another example of the heterogeneous model we describe here, Appendix A p. \pageref{app:heterogeneous}.

It should be noted that classifying programs in terms of fixed and variable tokens is not new and appeared at least as early as 1977 in the influential work of Halstead who called them \textit{operators} and \textit{operands}, \cite{Halstead77}. He developed his work to define various dependent concepts such as software \textit{volume} and \textit{effort} and tested them against programs of the time. This was further elaborated by Shooman \cite{Shoo85}. A different approach is used here which borrows from the methods of variational calculus.

Computer programs are often huge.  The software deployed in the search for the recently discovered Higg's boson comprises around 4 million lines of code \cite{Rousseau2012}.  At an average of around six tokens per line of code, this corresponds to some 20 million tokens, although this is still less than 1\% of the human genome in which the tokens are the four bases adenine, cytosine, guanine and thymine.  The largest systems in use today appear to be around 100 million lines of source code \cite{Mossinger2010}, corresponding to perhaps 15\% of the number of tokens of the human genome.  The (largely) open systems used to test the model described here total almost 100 million lines, (specifically 98,476,765 lines), totalling some 600 million tokens.  (If around 6 tokens per line seems a little low, it should be recalled that lines of code include comment lines here in line with common practice, whilst token counts do not.)

As an example of the nomenclature used here, consider the following simple sorting algorithm written in C, for example \cite{Sedgewick1990}.

\begin{verbatim}
void bubble( int a[], int N )
{
  int i, j, t;
  for( i = N; i >= 1; i--)
  {
    for( j = 2; j <= i; j++)
    {
      if ( a[j-1] > a[j] )
      {
        t = a[j-1];a[j-1] = a[j];a[j] = t;
      }
    }
  }
}
\end{verbatim}

This algorithm contains 94 tokens in all based on 18 of the fixed tokens and 8 of the variable tokens of ISO C, so the size of the unique alphabet for this component is $18+8 = 26$.

Note that extracting the tokens of programming languages to assemble these measures requires the development of compiler front-end tools \cite{Aho1977,Schreiner1985}.  These are included in the reproducibility materials, notably associated with \cite{HatTSE14}.

\subsubsection*{Power-law tails of alphabet and length}
\begin{figure}[H]
    \captionsetup[subfigure]{labelformat=empty}
    \centering
    \begin{subfigure}[t]{0.5\textwidth}
        \centering
        \caption{A}
        \includegraphics[width=6cm]{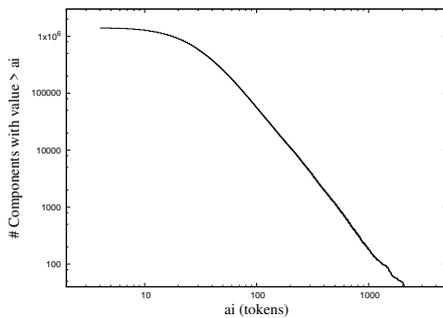}
        \label{fig:software_ai_cdf}
    \end{subfigure}%
    ~ 
    \begin{subfigure}[t]{0.5\textwidth}
        \centering
        \caption{B}
        \includegraphics[width=6cm]{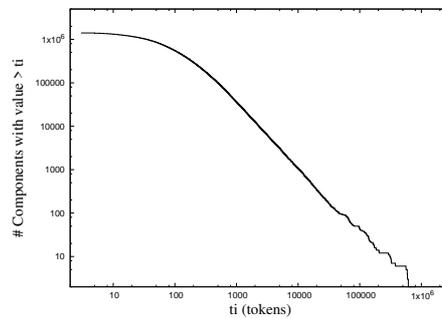}
        \label{fig:software_ti_cdf}
    \end{subfigure}%

    \caption{The unique alphabet $a_{i}$ (A) and length distributions $t_{i}$ (B) of 100 million lines of source code in 7 different programming languages shown as ccdfs.}
\end{figure}

We have from (\ref{eq:pwrlaw}) and (\ref{eq:pwrlaw2}) that power-laws in both the unique alphabet distributions and length distributions are overwhelmingly likely to appear in the tails of the distributions.  Figs. \ref{fig:software_ai_cdf} and \ref{fig:software_ti_cdf} show the $\log-\log$ ccdf plots for the unique alphabet and length distributions respectively of 100 million lines of source code in seven different programming languages \cite{HatTSE14}.

\paragraph{}
\fbox{%
\begin{minipage}{12cm}
\textbf{The linearity in each tail is striking confirmation of (\ref{eq:pwrlaw}) and (\ref{eq:pwrlaw2}).  For Fig. \ref{fig:software_ai_cdf}, R lm() reports that the associated p-value matching the power-law tail linearity is $< 2.2 \times e^{-16}$ over the range $80.0-3500.0$, with an adjusted R-squared value of $0.9975$.  The slope is $-2.15 \pm 0.08$.  For Fig. \ref{fig:software_ti_cdf}, R lm() reports that the associated p-value matching the power-law tail linearity is $< 2.2 \times e^{-16}$ over the range $200.0-68000.0$, with an adjusted R-squared value of $0.9995$.  The slope is $-1.47 \pm 0.03$.}
\end{minipage}}
\paragraph{}

We note in passing that (\ref{eq:pwrlaw}) and (\ref{eq:pwrlaw2}) suggest that the slopes are reciprocals of each other, ($\beta$ and $1/\beta$).  They are clearly not so here but this raises an interesting question concerning the choice of alphabets which is explained in the Appendix C p. \pageref{app:musicalphabets}.

\subsubsection*{Aggregations by language and package}
\begin{figure}[H]
    \captionsetup[subfigure]{labelformat=empty}
    \centering
    \begin{subfigure}[t]{0.5\textwidth}
        \centering
        \caption{A}
        \includegraphics[width=6cm]{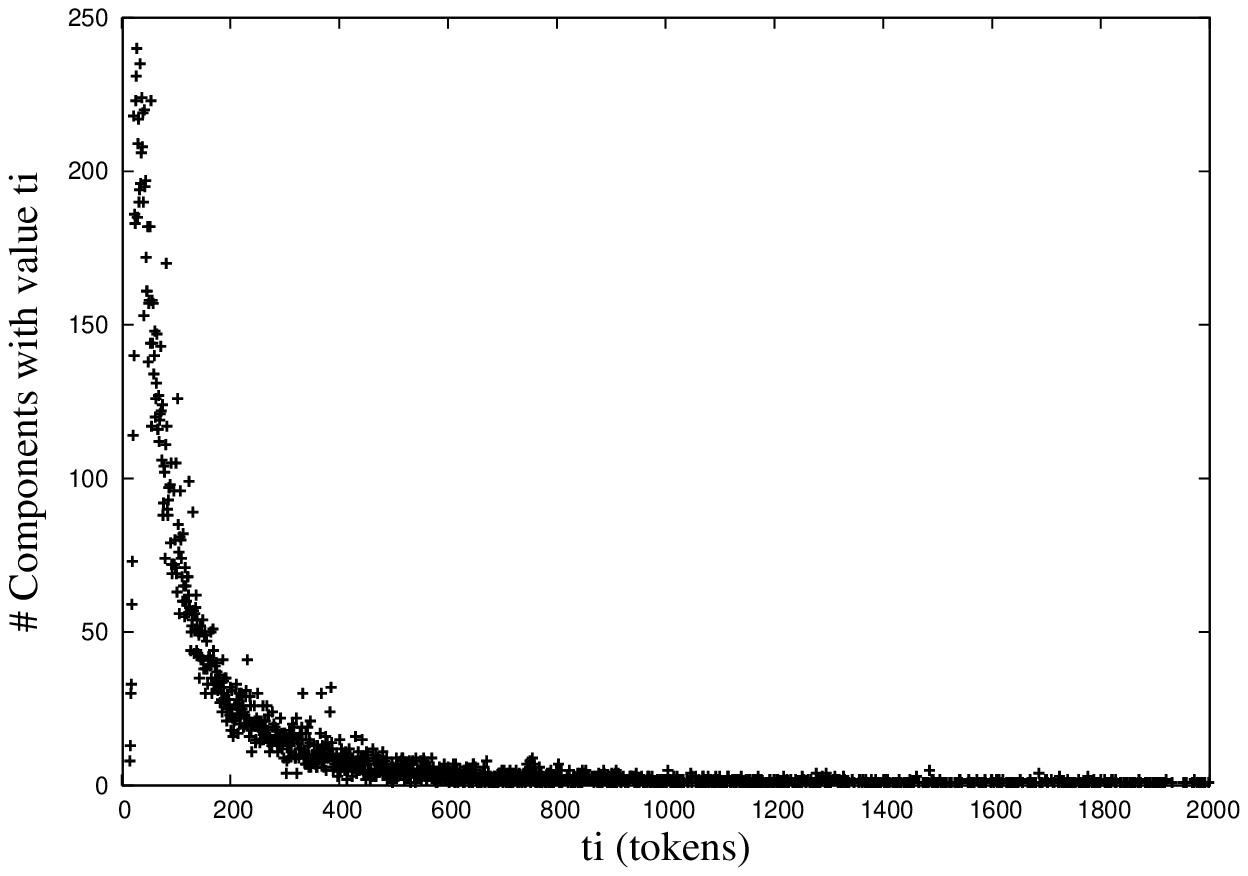}
        \label{fig:c++_pdf}
    \end{subfigure}%
    ~ 
    \begin{subfigure}[t]{0.5\textwidth}
        \centering
        \caption{B}
        \includegraphics[width=6cm]{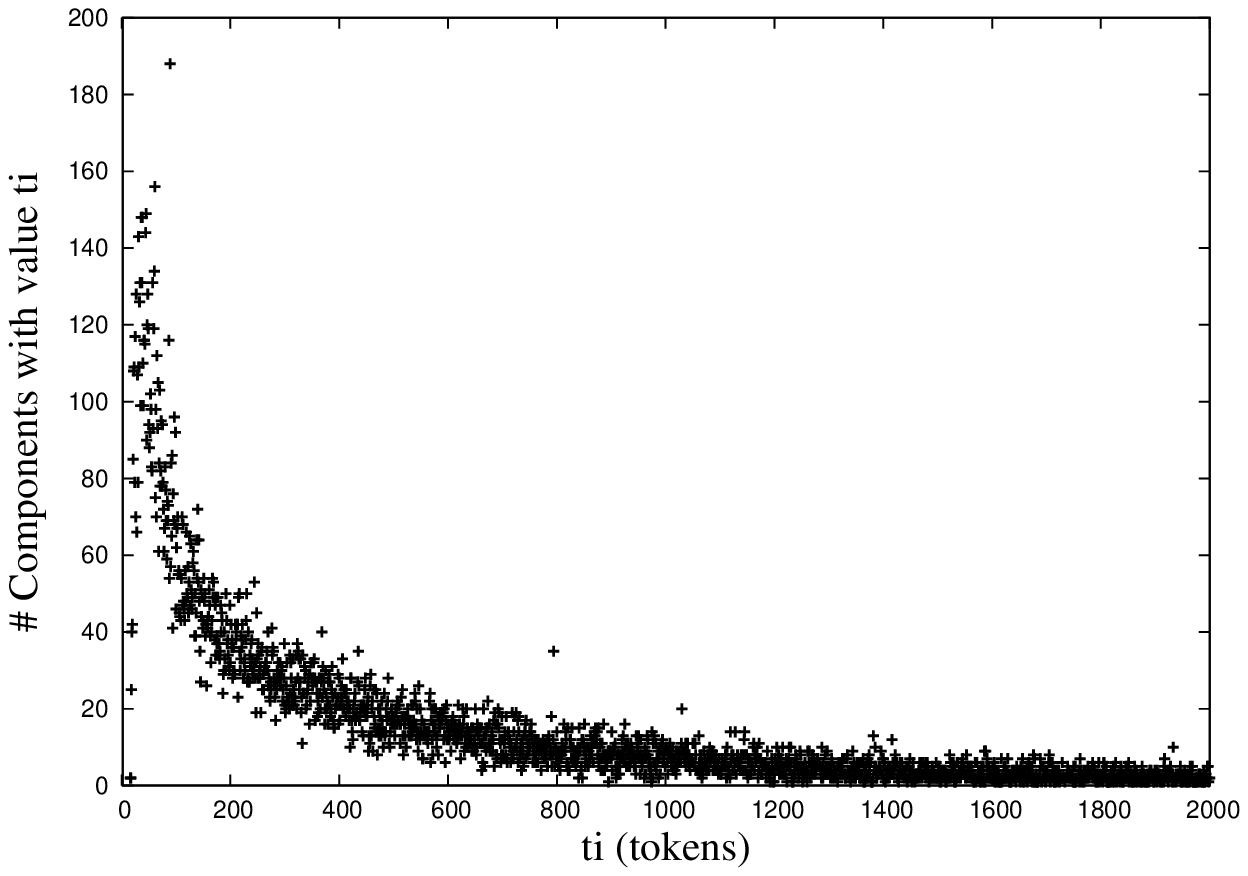}
        \label{fig:java_pdf}
    \end{subfigure}%

    \begin{subfigure}[t]{0.5\textwidth}
        \centering
        \caption{C}
        \includegraphics[width=6cm]{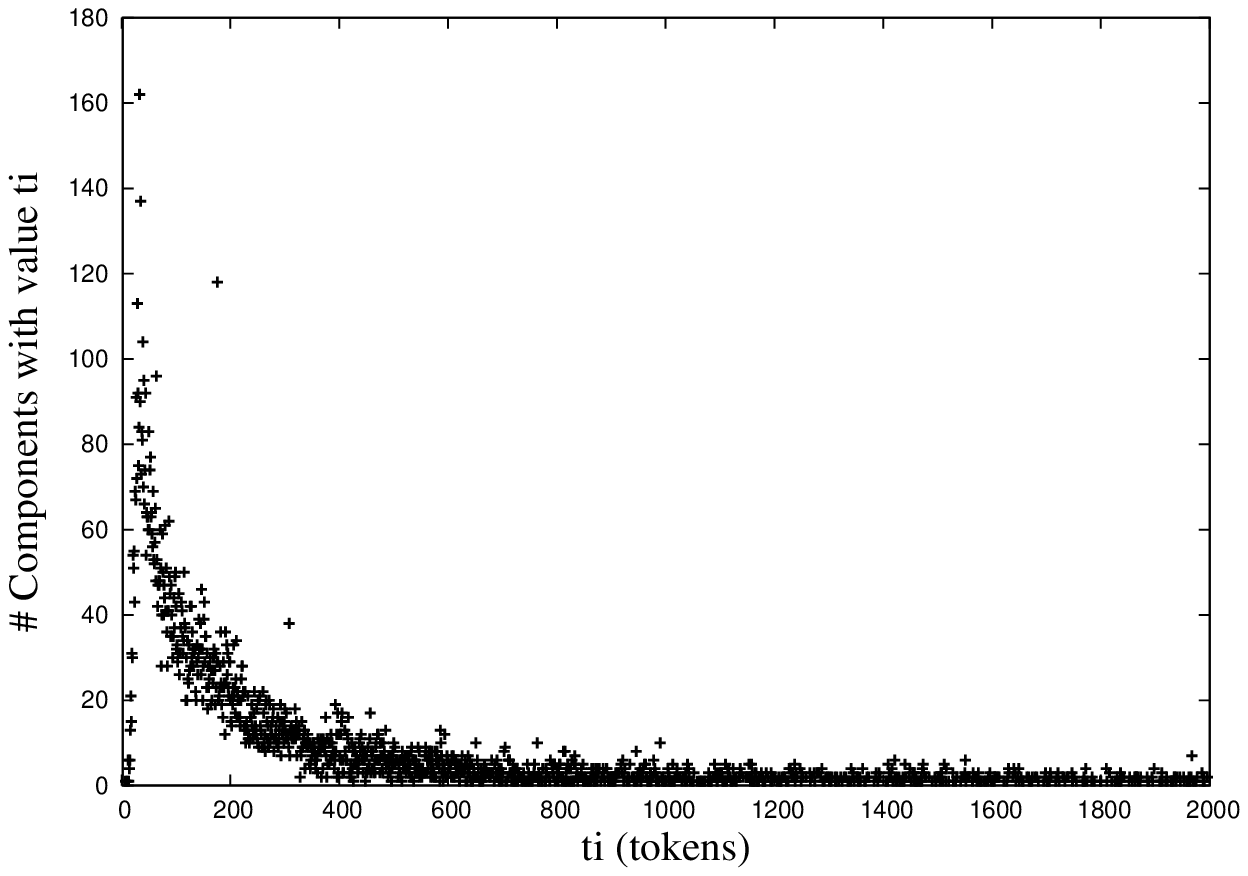}
        \label{fig:f_pdf}
    \end{subfigure}%
    ~ 
    \begin{subfigure}[t]{0.5\textwidth}
        \centering
        \caption{D}
        \includegraphics[width=6cm]{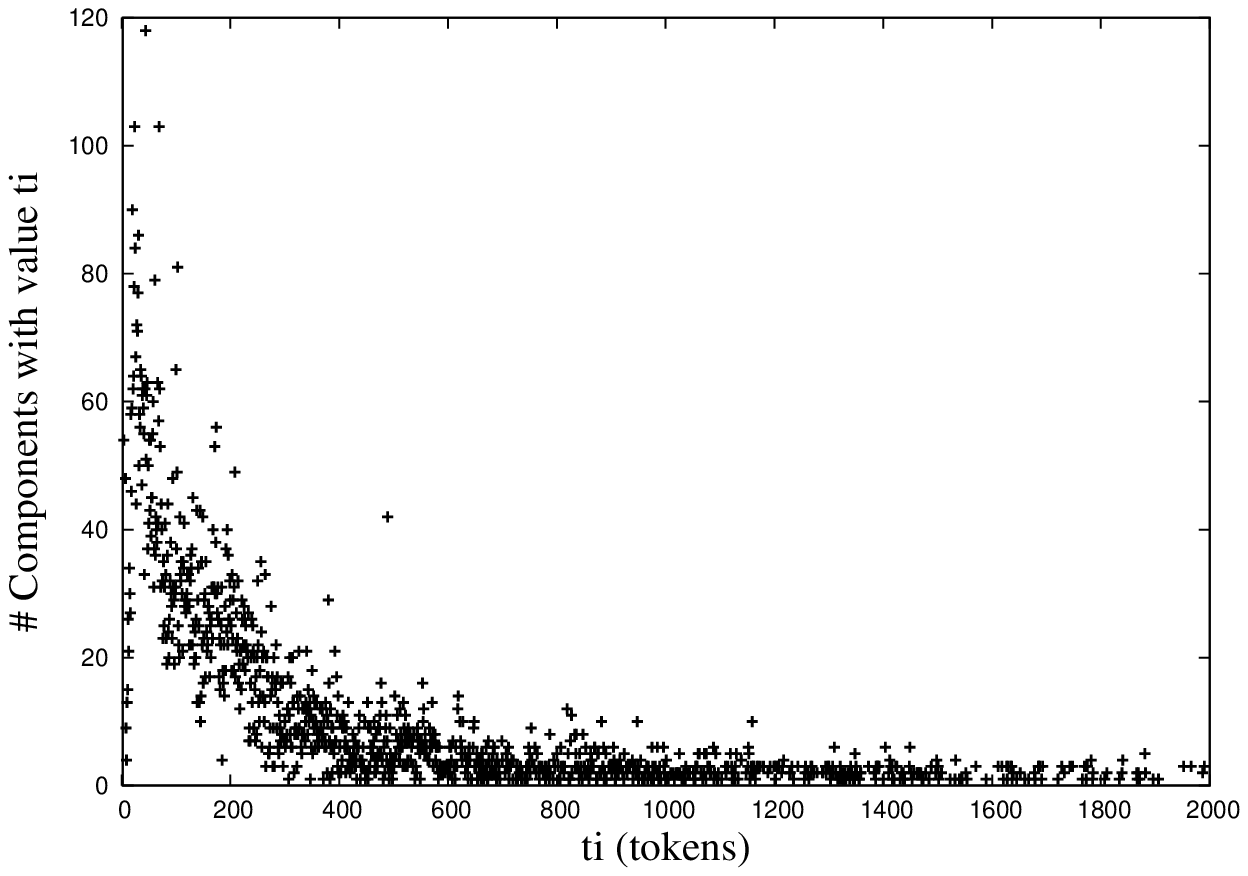}
        \label{fig:ada_pdf}
    \end{subfigure}%
    \caption{The length distributions of functions in large collections of software in four programming languages, (A): C++, (B): Java (C): Fortran and (D): Ada.}
\end{figure}

Although the collections of software available for analysis are many fewer than for proteins, we can still identify, in software, collections equivalent to the domains of life on the basis of software written in different programming languages.  Figs. \ref{fig:ada_pdf}-\ref{fig:java_pdf} illustrate the length distributions of collections of components (software functions) in four programming languages (Fig. \ref{fig:c++_pdf} C++ 22,628 components; Fig. \ref{fig:java_pdf} Java 32,552 components; Fig. \ref{fig:f_pdf} Fortran 14,028 components and Fig. \ref{fig:ada_pdf} Ada 12,680 components).  Despite the disparity in their sizes, each of these collections again shows striking similarity to the canonical form of length distributions of Figs. \ref{fig:trembl_pdf} and \ref{fig:c_pdf}, as predicted by (\ref{eq:cohsi}) and manifest in proteins, as we have already seen.  In each case, the values in the x-axis scale are the same whilst the y-axis is scaled according to the size of the packages to make the peaks of approximately the same vertical extent.

\begin{figure}[H]
    \captionsetup[subfigure]{labelformat=empty}
    \centering
    \begin{subfigure}[t]{0.5\textwidth}
        \centering
        \caption{A}
        \includegraphics[width=6cm]{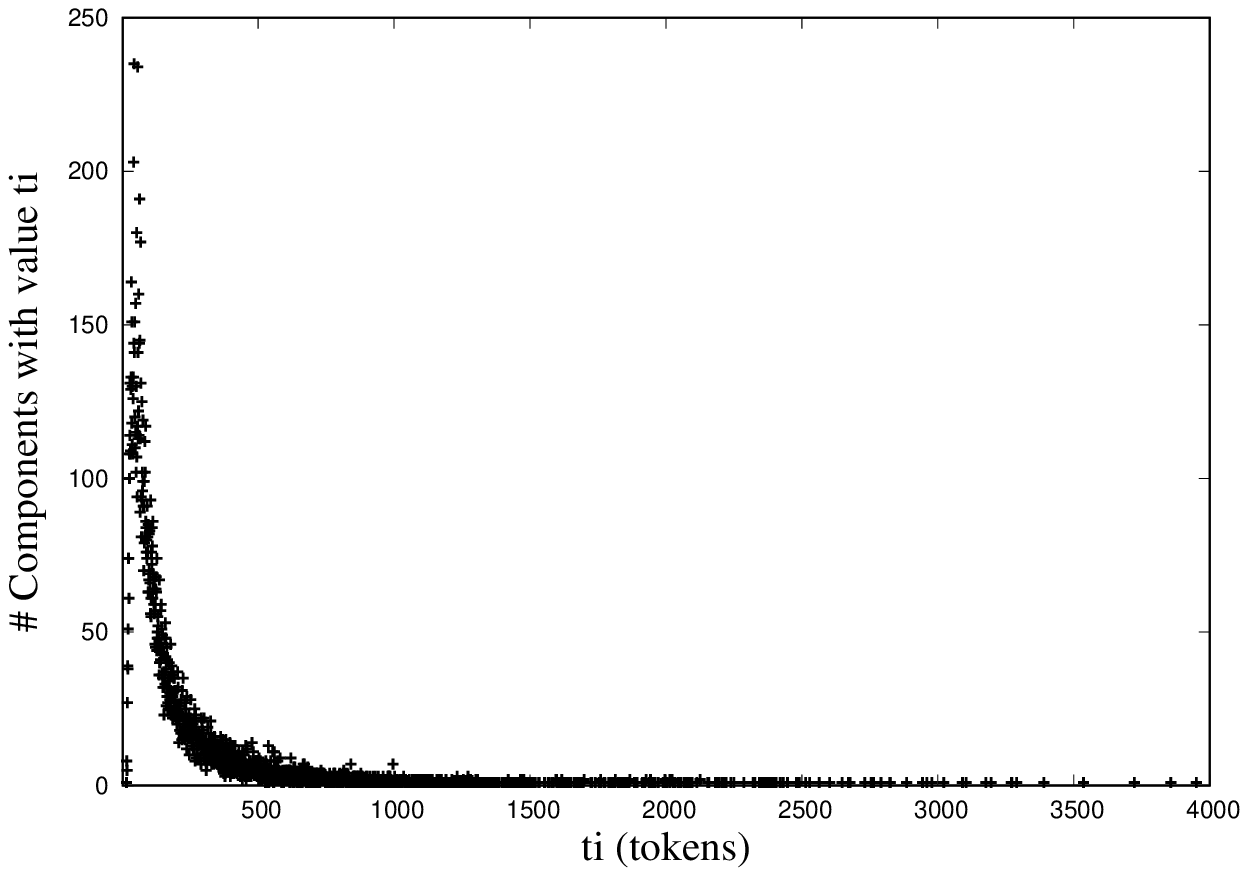}
        \label{fig:gimp_pdf}
    \end{subfigure}%
    ~ 
    \begin{subfigure}[t]{0.5\textwidth}
        \centering
        \caption{B}
        \includegraphics[width=6cm]{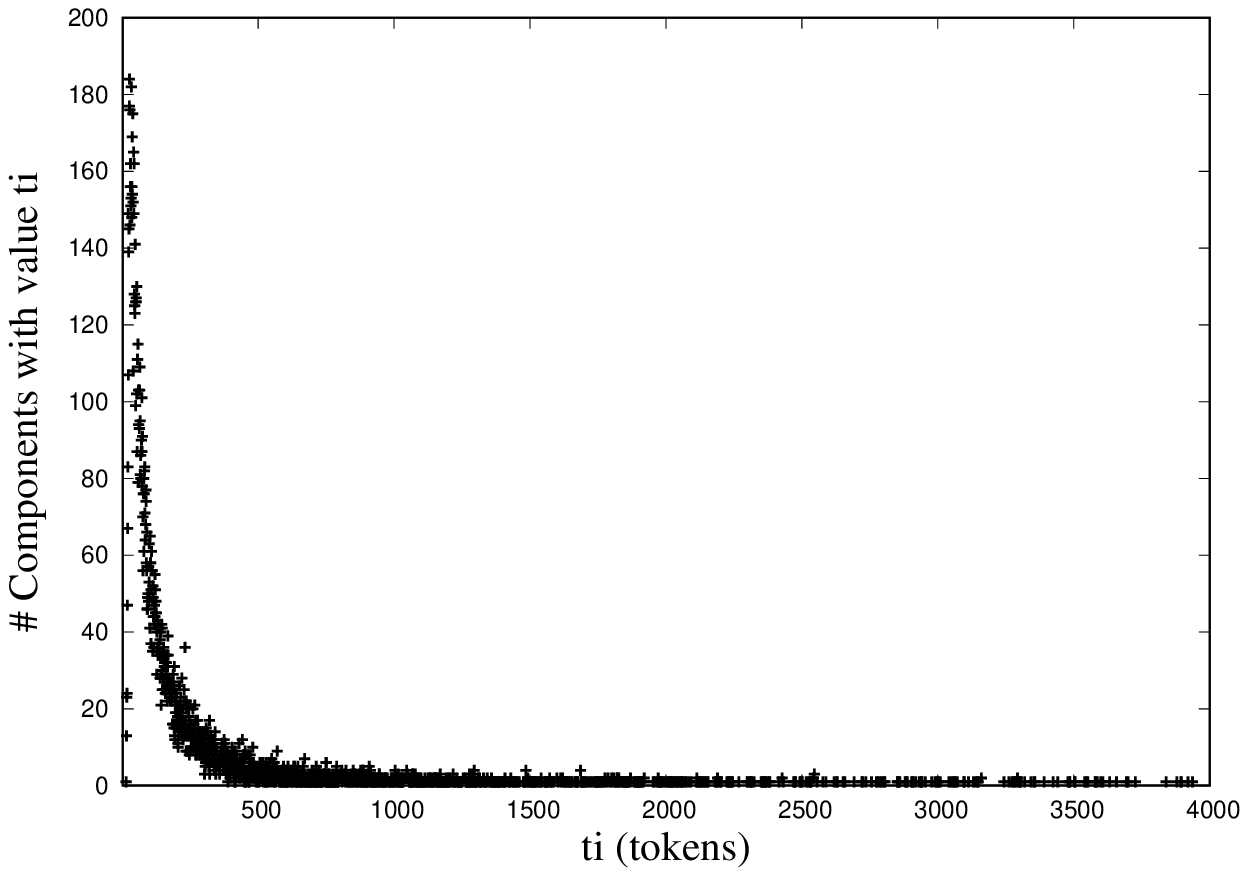}
        \label{fig:kdelibs_pdf}
    \end{subfigure}%

    \begin{subfigure}[t]{0.5\textwidth}
        \centering
        \caption{C}
        \includegraphics[width=6cm]{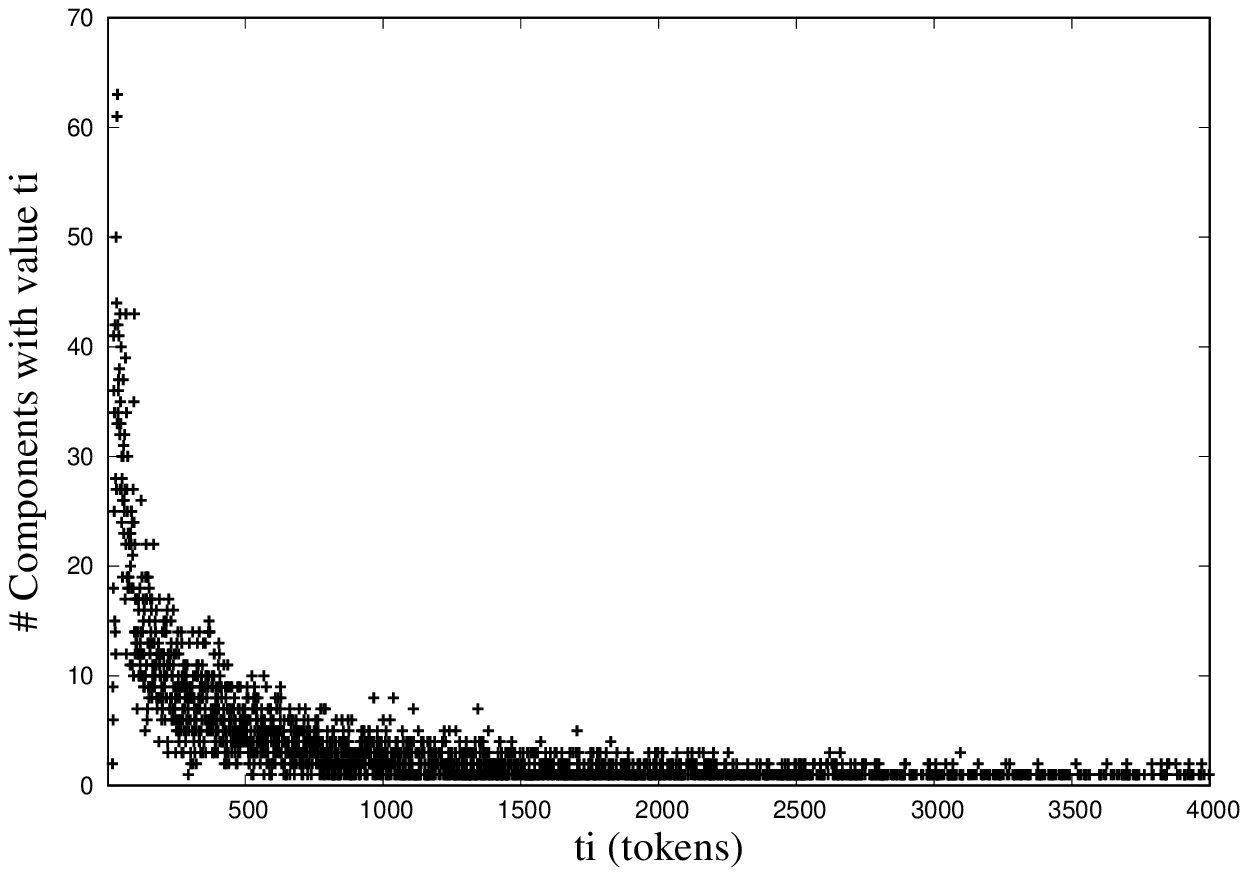}
        \label{fig:eclipse_pdf}
    \end{subfigure}%
    ~ 
    \begin{subfigure}[t]{0.5\textwidth}
        \centering
        \caption{D}
        \includegraphics[width=6cm]{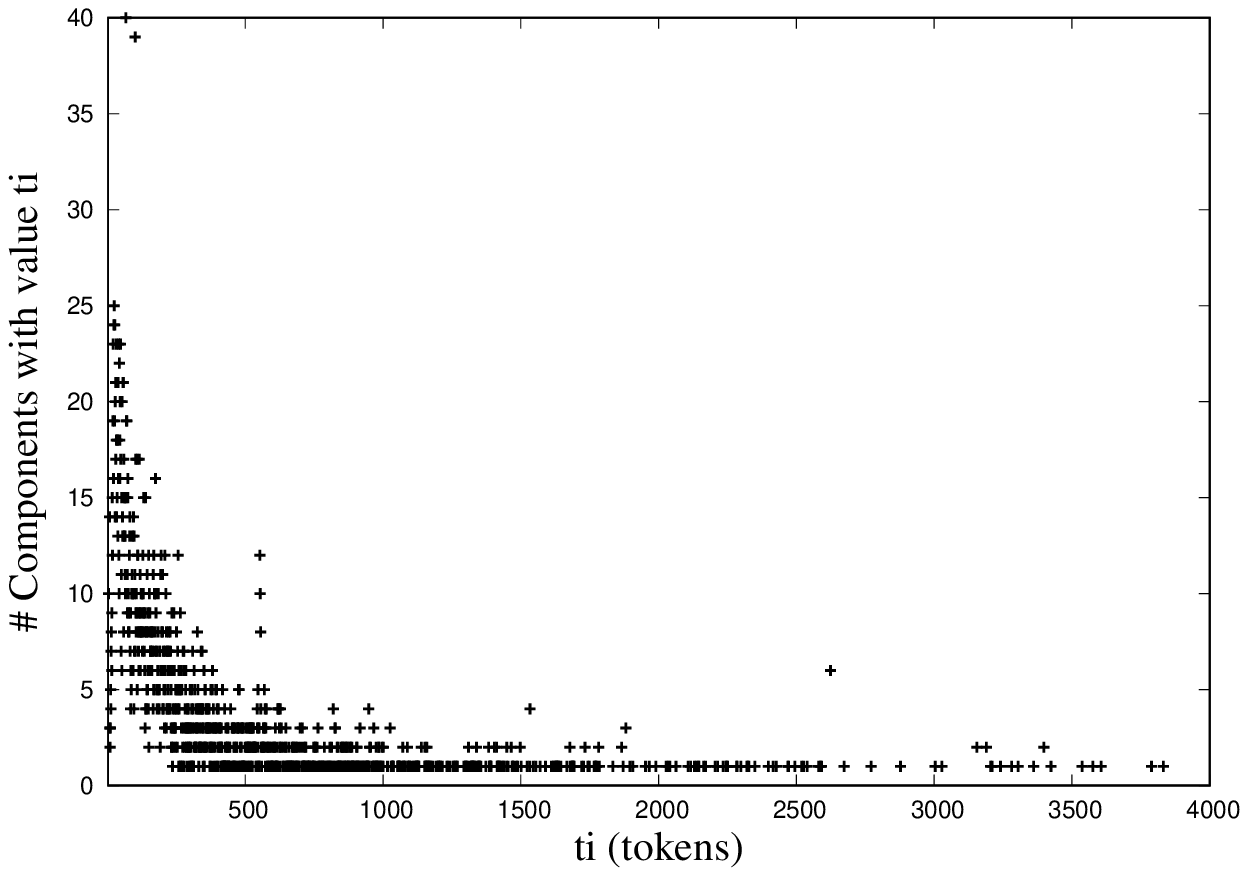}
        \label{fig:gcc_pdf}
    \end{subfigure}%
    \caption{The length distributions of functions in four packages each in different programming languages, (A): Gimp (ISO C), (B): kdelibs (C++), (C): Java (Eclipse) and (D): Ada (GCC).}
\end{figure}

Again, there are considerable differences in scale compared with that of proteins, but we can still identify in software collections equivalent to species on the basis of software written for individual applications.  Figs. \ref{fig:gimp_pdf}-\ref{fig:gcc_pdf} illustrates the length distributions of collections of software functions in four applications of very different sizes, in four different programming languages  (Fig. \ref{fig:gimp_pdf}, The Gimp image manipulation program (ISO C) 18,693 components; Fig. \ref{fig:kdelibs_pdf}, The KDE desktop libraries (C++) 16,241 components; Fig. \ref{fig:eclipse_pdf}, The Eclipse interactive Development Environment (Java) 9,588 components and Fig. \ref{fig:gcc_pdf}, The gcc Ada compiler (Ada) 3,765 components).  Once again, despite the disparity in their sizes, each of these collections again shows striking similarity to the canonical form of length distributions exhibited by Figs. \ref{fig:trembl_pdf}, \ref{fig:c_pdf}.  In this case, the x-axis scale is the same whilst the y-axis is again scaled according to the size of the packages to normalise the size of the peaks approximately.

\subsection*{Music}
CoHSI also predicts the length distributions of musical compositions.  Much of the theory and discussion is deferred to Appendix C p. \pageref{app:musicalphabets}.  We will simply point out here that modern digital formats for musical annotation such as MusicXML\footnote{\url{https://en.wikipedia.org/wiki/MusicXML}, accessed 07-Jul-2017} allow us to apply the heterogeneous theory described in Appendix A p. \pageref{app:heterogeneous}, to yet another distinct discrete system where, in this case, the components are pieces of music and the unique alphabet comprises of notes as shown in Table \ref{tab:entity}.

Extracting the appropriate data is fortunately relatively simple compared with the daunting task of extracting possibly post-translationally modified amino acids in proteins or programming language tokens in computer programs, as the following XML snippet\footnote{\url{https://hymnary.org/media/fetch/99378}, accessed 07-Jul-2017} taken from ``Nun danket alle Gott'', (Words Rinkart 1636, Music Cr\"{u}ger, 1647) shows.

\begin{verbatim}
 <part id="P1">
    <measure number="1">
    ...
     <note>
        <pitch>
          <step>E</step>
          <alter>-1</alter>
          <octave>4</octave>
        </pitch>
        <duration>480</duration>
        <voice>1</voice>
        ...
\end{verbatim} 

This snippet refers to the note Eb in the 4th octave (middle C is annotated C4, so this is a minor third above middle C).  The duration must be determined from other parameters in the XML but this note actually corresponds to a 1/4 note or crotchet.

Arguably the most beneficial aspect of studying music from the point of view of this paper, however, is that \textit{it provides a simple example of when there are multiple candidate unique alphabets}, for example, whether or not to include musical note \textit{duration} as well as \textit{pitch} in defining the alphabet.  When we first considered this aspect, the potential ambiguity worried us until we eventually realised that it led naturally and elegantly to the important conclusion, proved in Appendix C p. \pageref{app:musicalphabets} and verified experimentally, that \textit{all consistent unique alphabets are themselves related by a power-law.}

\begin{figure}[H]
    \captionsetup[subfigure]{labelformat=empty}
    \centering
    \begin{subfigure}[t]{0.5\textwidth}
        \centering
        \caption{A}
        \includegraphics[width=6cm]{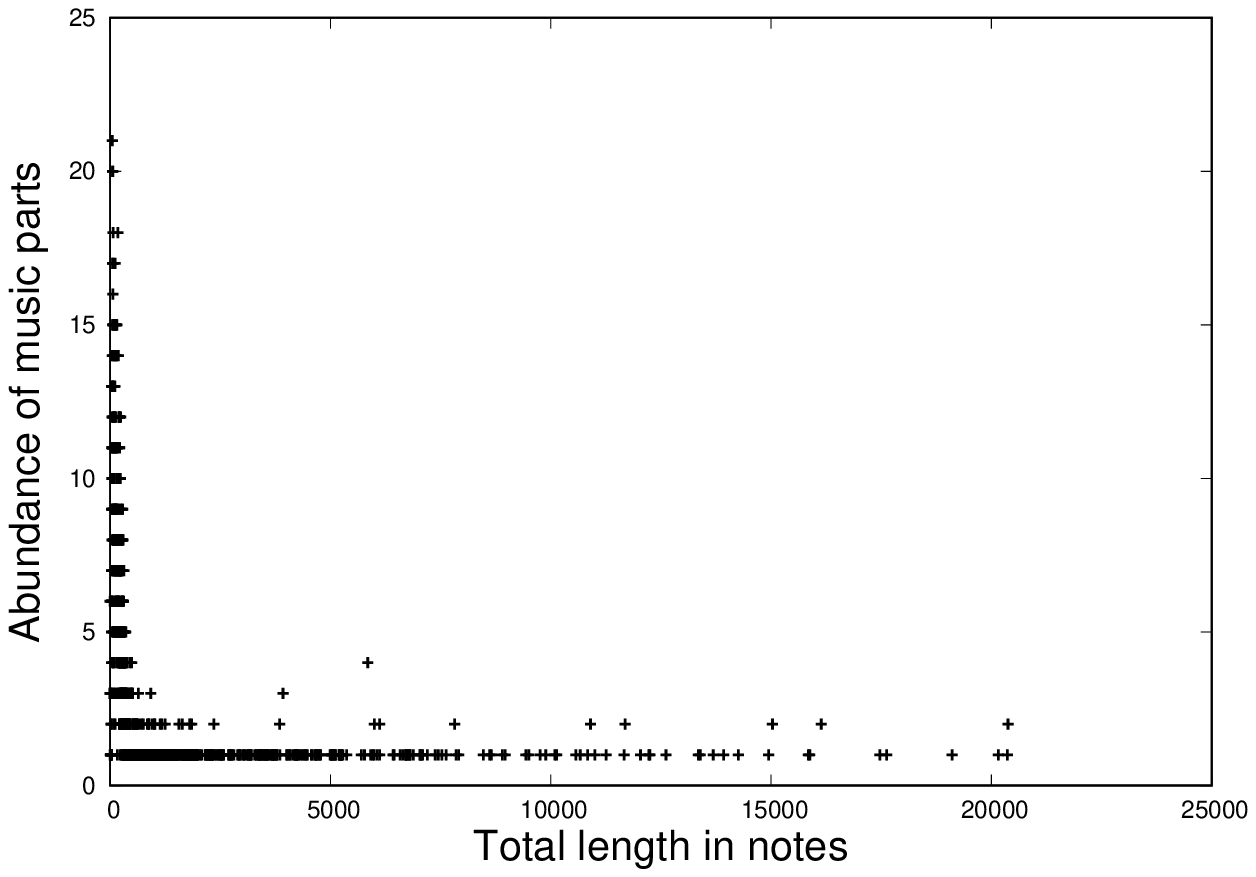}
        \label{fig:music_ti_pdf}
    \end{subfigure}%
    ~ 
    \begin{subfigure}[t]{0.5\textwidth}
        \centering
        \caption{B}
        \includegraphics[width=6cm]{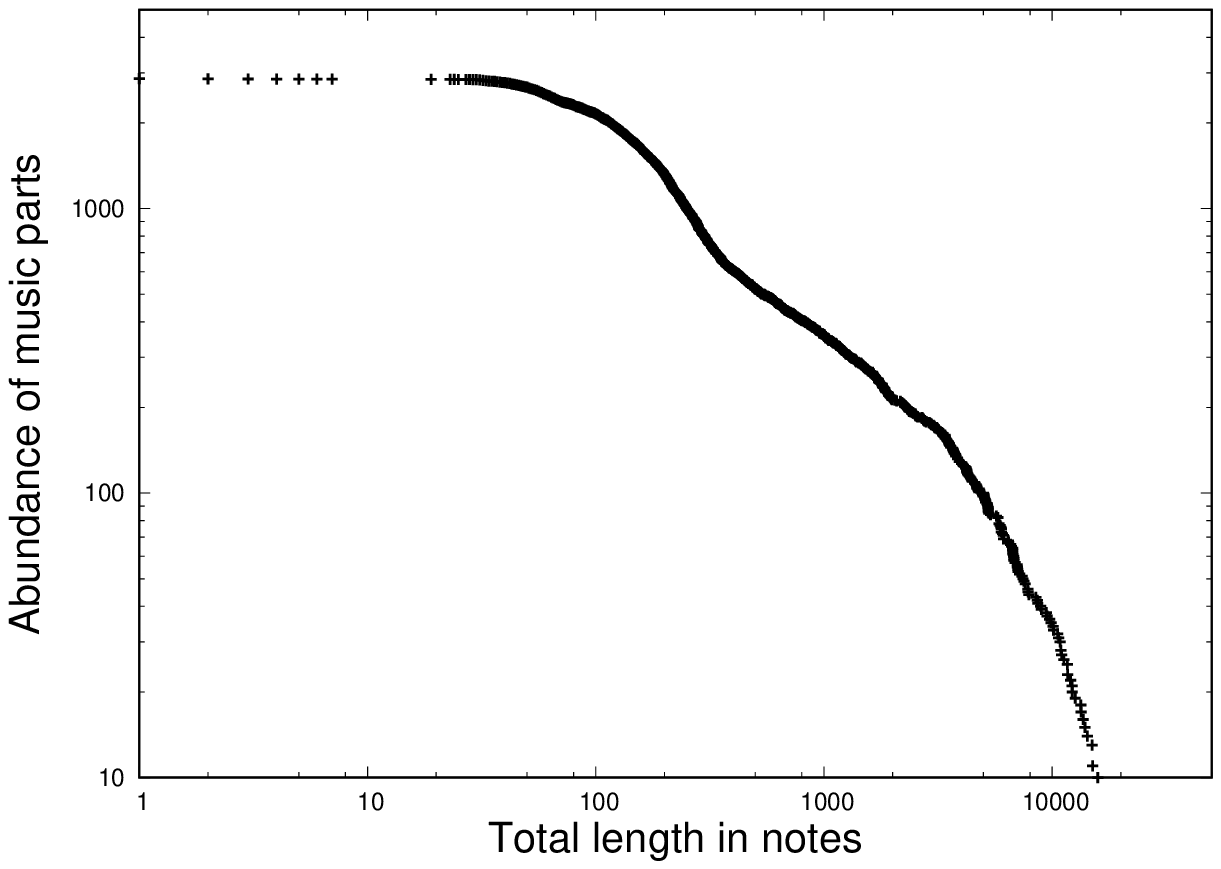}
        \label{fig:music_ti_ccdf}
    \end{subfigure}%

    \caption{The length distribution of the 883 pieces of music analysed shown as a pdf (A) and a ccdf (B).}
\end{figure}

In this study, we used 883 pieces of music, mostly classical but a very eclectic mix of chorales, piano concertos, horn duets, blue-grass music and indeed almost anything in an XML format we could get our hands on.  This process was not as simple as accumulating large amounts of open source code unfortunately and took some considerable time and manual effort.  As a result, this is by far the smallest system we consider and does not therefore allow us much scope to demonstrate different sized aggregations.  Even so, the length distribution of these 883 pieces of music is still gratifyingly suggestive of the presence of the predicted canonical distribution as shown in Figs. \ref{fig:music_ti_pdf}-\ref{fig:music_ti_ccdf}.

\paragraph{}
\fbox{%
\begin{minipage}{12cm}
\textbf{An R lm() analysis on the tail of Fig \ref{fig:music_ti_ccdf} reports that the associated p-value matching the power-law tail linearity in the ccdf of Fig. \ref{fig:elements} is $< 2.2 \times e^{-16}$ over the range $100.0-10000.0$, with an adjusted R-squared value of $0.9936$.  The slope is $-1.66 \pm 0.08$.}
\end{minipage}}
\paragraph{}

\subsection*{The written word}
The pioneering work which first suggested the ubiquity of power-laws in texts was that by George K. Zipf \cite{Zipf35}.  Zipf showed \textit{empirically} that if the frequency of occurrence of words in a text were plotted in rank order on a $\log-\log$ ccdf, a power-law in frequency was observed.  This is an example of a system of homogeneous boxes, Appendix A p. \pageref{app:homogeneous}, where we give a proof of Zipf's law using the methodology of this paper.

\begin{figure}[H]
    \captionsetup[subfigure]{labelformat=empty}
    \centering
    \begin{subfigure}[t]{0.5\textwidth}
        \centering
        \caption{A}
        \includegraphics[width=6cm]{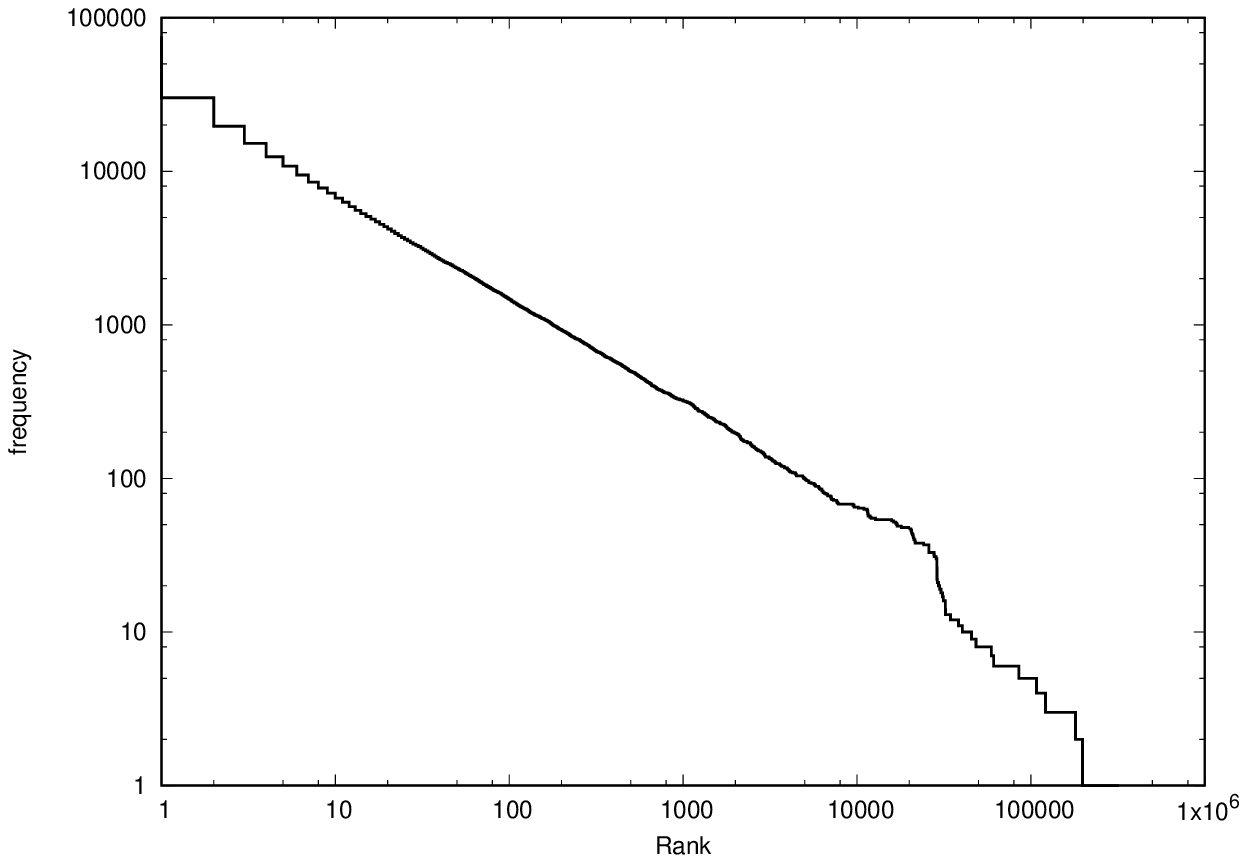}
        \label{fig:cv_ccdf}
    \end{subfigure}%
    ~ 
    \begin{subfigure}[t]{0.5\textwidth}
        \centering
        \caption{B}
        \includegraphics[width=6cm]{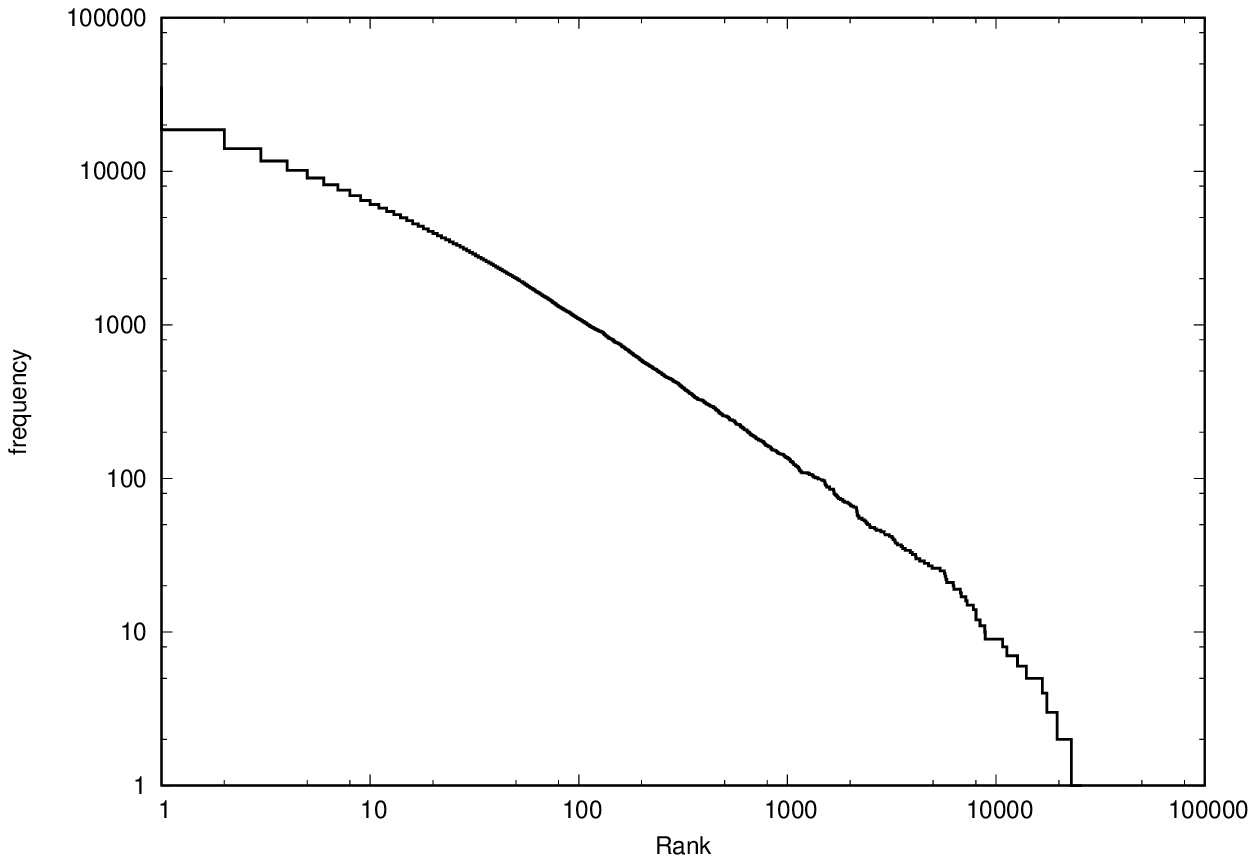}
        \label{fig:shakespeare_ccdf}
    \end{subfigure}%

    \begin{subfigure}[t]{0.5\textwidth}
        \centering
        \caption{C}
        \includegraphics[width=6cm]{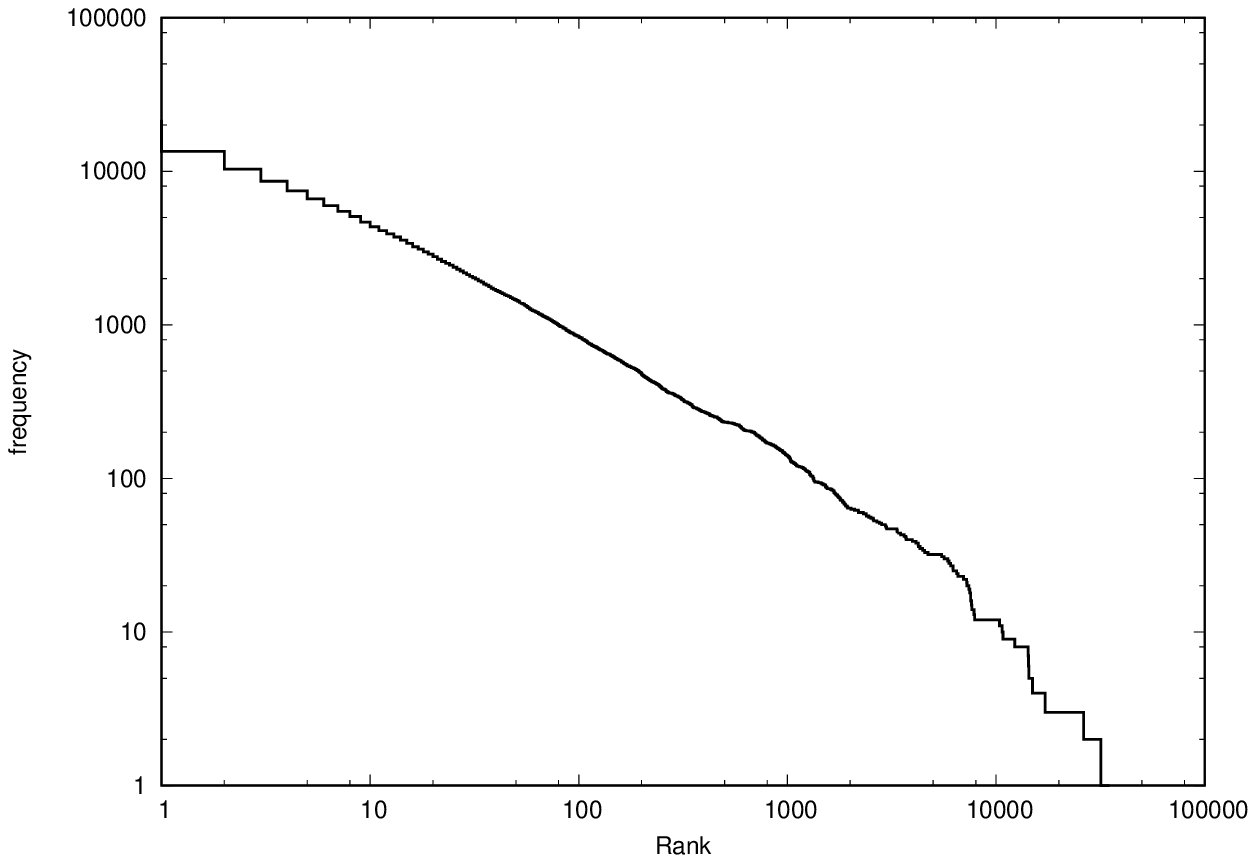}
        \label{fig:bibel_ccdf}
    \end{subfigure}%
    ~ 
    \begin{subfigure}[t]{0.5\textwidth}
        \centering
        \caption{D}
        \includegraphics[width=6cm]{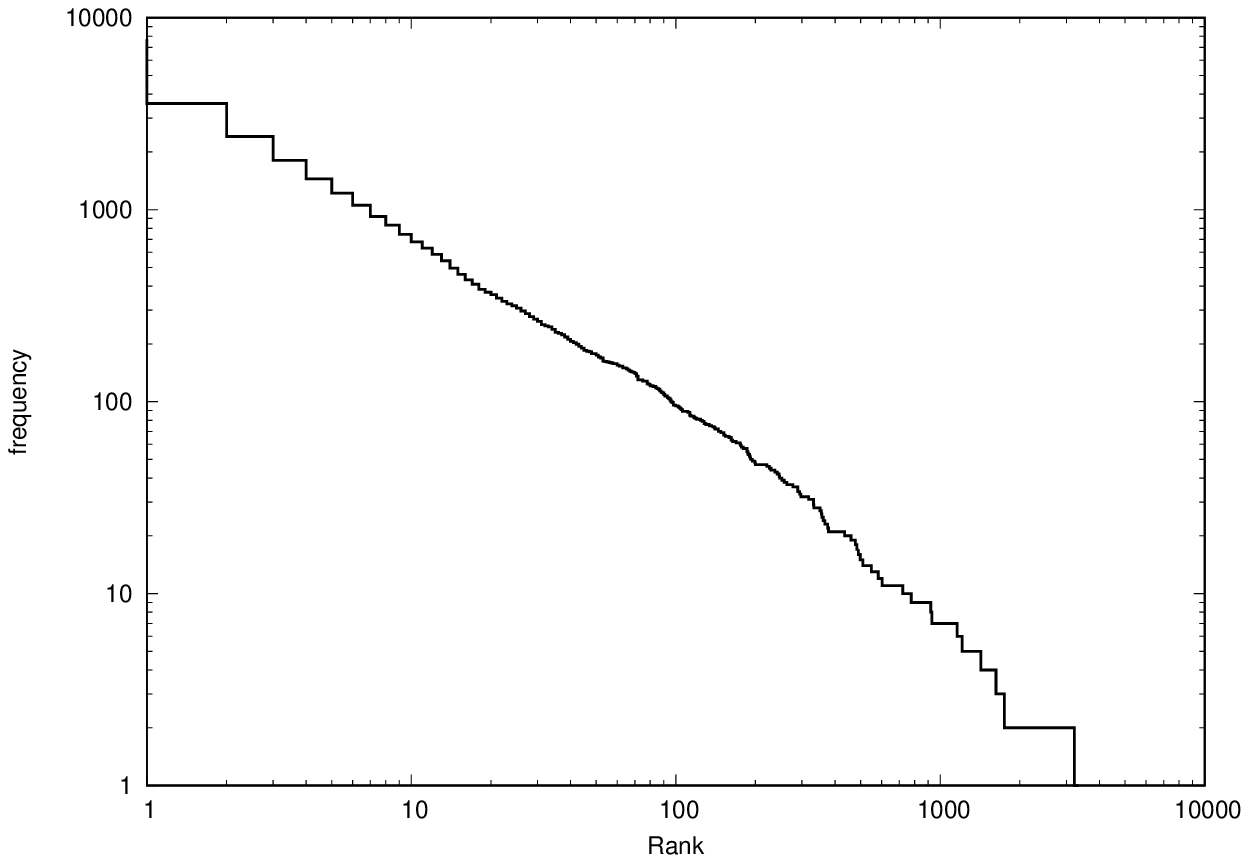}
        \label{fig:threemen_ccdf}
    \end{subfigure}%
    \caption{The rank ordered word distributions in, (A): The Common Vulnerabilities database, (B): The complete works of Shakespeare, (C): The King James Bible in Swedish and (D): Jerome K. Jerome's ``Three Men in a Boat''.}
\end{figure}

Archetypal examples of this at different scales are shown in Figs. \ref{fig:cv_ccdf}-\ref{fig:threemen_ccdf}.  These show respectively, (Fig. \ref{fig:cv_ccdf} The Mitre Common Vulnerabilities database 2,410,350 words\footnote{\url{https://cve.mitre.org/, accessed 01-May-2015}}; Fig. \ref{fig:shakespeare_ccdf} The complete works of Shakespeare 948,516 words; Fig. \ref{fig:bibel_ccdf} The King James Bible in Swedish 807,969 words and Fig. \ref{fig:threemen_ccdf} the classic English text ``Three Men in a Boat'' published by Jerome K. Jerome in 1889 67,435 words)\footnote{\url{https://www.gutenberg.org/, accessed 01-Jul-2017}}.  The classic straight line signature of the power-law is evident in each case even though the datasets are different in size by a factor of 40 from largest to smallest.

Before leaving this section, we point out that some systems can in fact be characterised both by using the \textit{homogeneous} model, Appendix A p. \pageref{app:homogeneous}, and also using the \textit{heterogeneous} model, Appendix A p. \pageref{app:heterogeneous}.  It would therefore provide substantial additional support for the generality of the information model we propose in this paper if \textit{both} the homogeneous model predictions and the heterogeneous model predictions held for a system in which \textit{both} could be used.  There is no conflict between information measures here even though they are different, provided they are consistently applied.

Treating the words of a text as indivisible as we have done above yields the homogeneous model where word frequency follows the predicted Zipf power-law in rank, and we have already seen that the homogeneous model predicts exactly this long-established behaviour, Appendix A p. \pageref{app:homogeneous}.

\textit{It is also possible to consider the individual words of a text as being further sub-divided into their letters as a heterogeneous model, just as if each word were a protein built from a unique alphabet, which in English, is 26 letters.}  Word length has been studied extensively for various languages, for example \cite{Smith2012}, however for our purposes, we expect by analogy with our protein studies, that applying the heterogeneous model to word-length frequency will yield the canonical distribution seen in Fig. \ref{fig:trembl_pdf} for example.

\begin{figure}[H]
    \captionsetup[subfigure]{labelformat=empty}
    \centering
    \begin{subfigure}[t]{0.5\textwidth}
        \centering
        \caption{A}
        \includegraphics[width=6cm]{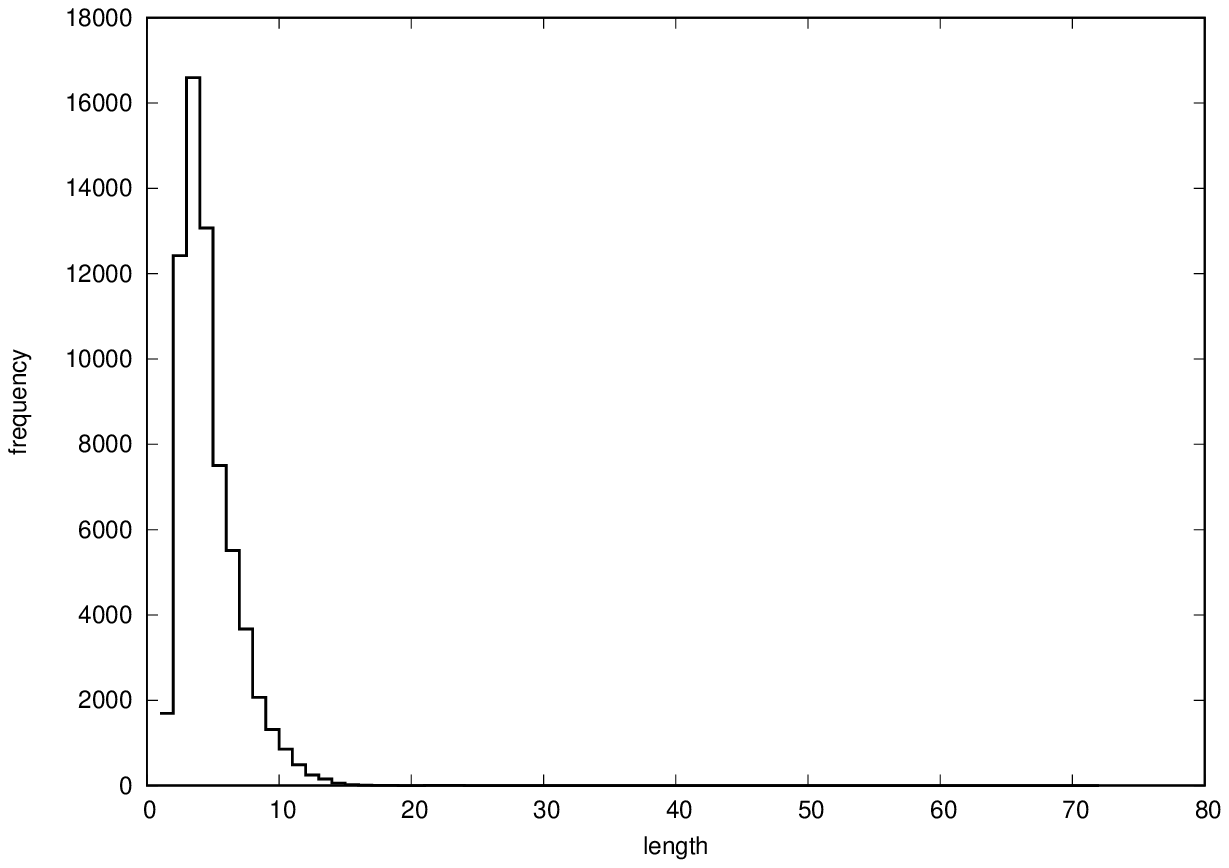}
        \label{fig:threemen_ti_pdf}
    \end{subfigure}%
    ~ 
    \begin{subfigure}[t]{0.5\textwidth}
        \centering
        \caption{B}
        \includegraphics[width=6cm]{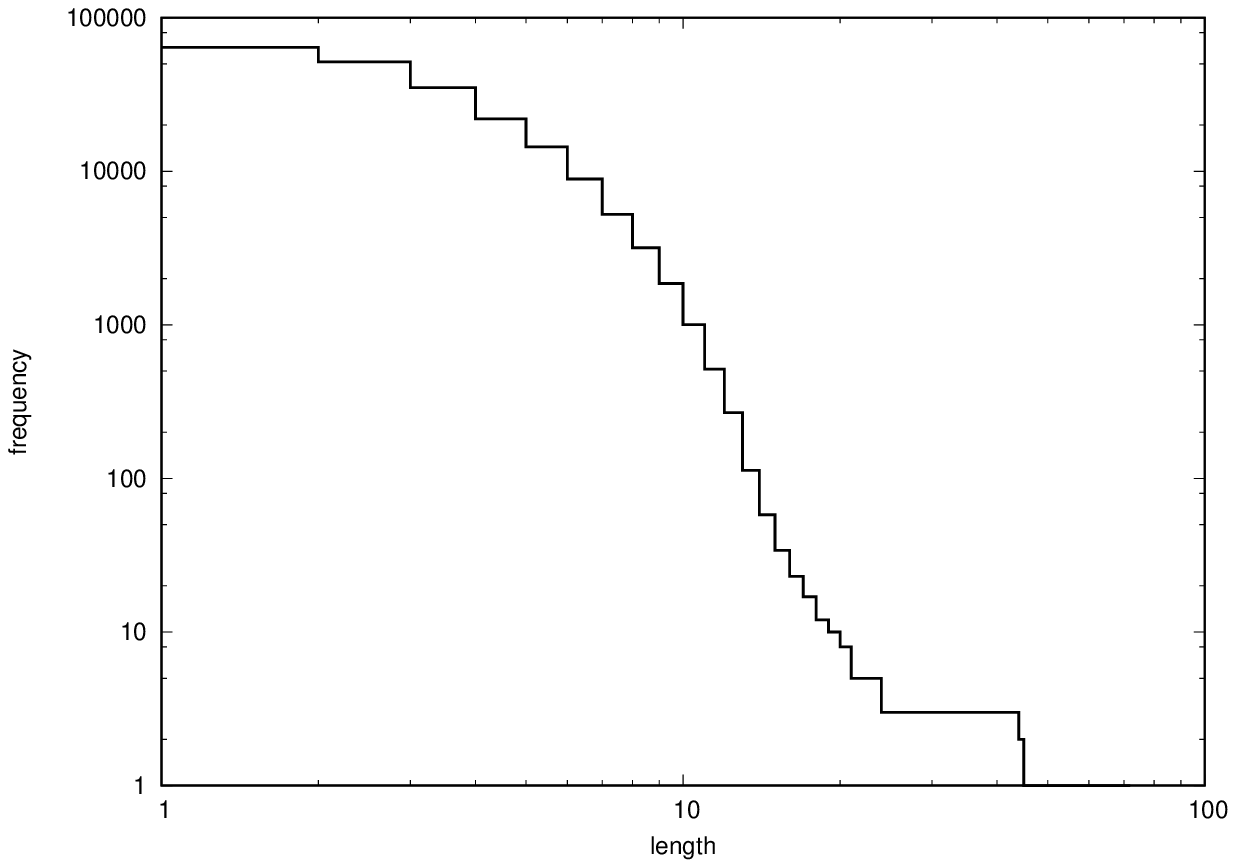}
        \label{fig:threemen_ti_ccdf}
    \end{subfigure}%

    \caption{The text ``Three Men in a Boat'' with each word considered as a component and the frequency of occurrence of length of word plotted as a pdf (A) and as a ccdf (B).}
\end{figure}

Fig. \ref{fig:threemen_ti_pdf} shows the word\textbf{-length} frequency for the text ``Three Men in a Boat'' whose word frequency is shown in Fig. \ref{fig:threemen_ccdf}.  Even though the x-axis is limited to a maximum word length of around 40 in this novel (it includes single- and, unusually, double- hyphenated words such as the archaic currency reference ``two-pounds-ten''), the canonical shape is once again evident with a sharp unimodal peak and as can be seen in Fig. \ref{fig:threemen_ti_ccdf}, good evidence of the predicted power-law tail.  The steep slope is associated with a relatively small alphabet as described in Appendix D, p. \pageref{app:powerlaws}.

\paragraph{}
\fbox{%
\begin{minipage}{12cm}
\textbf{An R lm() analysis on the tail of Fig. \ref{fig:threemen_ti_ccdf} reports that the associated p-value matching the power-law tail linearity in the ccdf of Fig. \ref{fig:elements} is $2.303 \times e^{-15}$ over the range starting at the mode $5.0-30.0$, with an adjusted R-squared value of $0.9853$.  The slope is $-6.40 \pm 0.32$.}
\end{minipage}}
\paragraph{}

\paragraph{}
\fbox{%
\begin{minipage}{12cm}
\textbf{We believe this adds considerable weight to the information-theoretic arguments of this paper.  In this case, the same system, treated using two different models of information (the homogeneous case of word frequencies and the heterogeneous case of letter frequencies and word lengths) obeys the predictions of \textit{both} models, showing that consistent but different measures of information within the same system lead to valid predictions for the different distributions.  In both cases, conserving Hartley-Shannon information in an ergodic system is the underlying mechanism.}
\end{minipage}}
\paragraph{}

\subsection*{The atomic elements}
\begin{figure}[H]
\centering
\includegraphics[width=8cm,height=6cm]{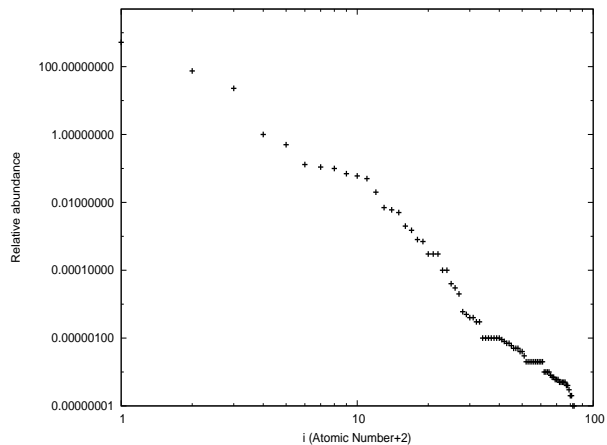}
\caption{The frequency of occurrence of the elements in the universe supplemented by estimates of dark energy (data point 1) and dark matter (data point 2) shown as a $\log-\log$ ccdf.}
\label{fig:elements}
\end{figure}

The distribution of atomic elements in the universe is a similar system to that of word frequencies and we use the homogeneous model, Appendix p. \pageref{app:homogeneous}.  Again, the components each consist of one type only, in this case atoms of each element and, intrigued by its apparent preponderance we chose to include current estimates of dark material, i.e. energy and matter. The frequencies of occurrence, Fig. \ref{fig:elements}, have been taken from NASA\footnote{\url{https://map.gsfc.nasa.gov/universe/uni_matter.html}, accessed 29-Jun-2017} and Wikipedia\footnote{,\url{https://en.wikipedia.org/wiki/Abundance_of_the_chemical_elements}, accessed 29-Jun-2017}.

The distribution of elements fits well on the predicted power-law distribution for homogeneous boxes, where the rank ordering turns out to follow the atomic number.  (The relationship with atomic number is outwith the theory described here.)

\paragraph{}
\fbox{%
\begin{minipage}{12cm}
\textbf{An R lm() analysis on this tail reports that the associated p-value matching the power-law tail linearity in the ccdf of Fig. \ref{fig:elements} is $< 2.2 \times e^{-16}$ over the range $1.0-85.0$, with an adjusted R-squared value of $0.9779$.  The slope is $-6.80 \pm 0.94$.}
\end{minipage}}
\paragraph{}

Intriguingly, the observed amount of dark energy (first point Fig. \ref{fig:elements}) and dark matter (second point Fig. \ref{fig:elements}) fit the predicted homogeneous box distribution well, suggesting that from the point of view of CoHSI, \textit{dark matter corresponds to something with an atomic number of zero (a re-generating sea of neutrons ?) and dark energy corresponds to something with an atomic number of -1 (?);} so that's something for the theorists to chew on.

\begin{figure}[H]
\centering
\includegraphics[width=8cm,height=6cm]{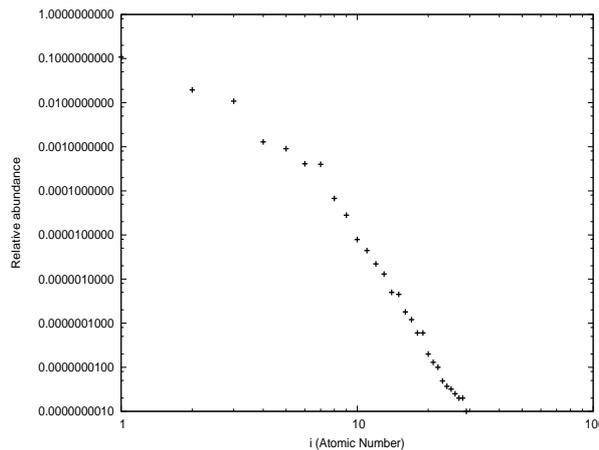}
\caption{The frequency of occurrence of the elements in sea water shown as a $\log-\log$ ccdf.}
\label{fig:seawater}
\end{figure}

On a much smaller scale than Fig. \ref{fig:elements}, the characteristic straight line signature is again visible in the distribution of the elements in seawater\footnote{\url{https://en.wikipedia.org/wiki/Abundances_of_the_elements_(data_page)}, accessed 29-Jun-2017}, Fig. \ref{fig:seawater}, \textit{not} in this case including dark material.

\paragraph{}
\fbox{%
\begin{minipage}{12cm}
\textbf{Even though the tail is short, an R lm() analysis on this tail reports that the associated p-value matching the power-law tail linearity in the ccdf of Fig. \ref{fig:elements} is $< 2.2 \times e^{-16}$ over the range $10.0-72.0$, with an adjusted R-squared value of $0.9963$.  The slope is $-8.82 \pm 0.29$.}
\end{minipage}}
\paragraph{}

\section*{Conclusions}
This paper, through theory and testing against multiple datasets of different provenance and levels of aggregation, makes the case that a conservation principle derived from information theory (Conservation of Hartley-Shannon Information) operates within all discrete systems to impose important and common structural properties on length and unique alphabet size distributions.  By development of a statistical mechanics argument in which we consider ergodic ensembles with a fixed number of tokens and a fixed total H-S information content, independent of the meaning of the tokens chosen without bias, we demonstrate that the length and unique token alphabet for components of discrete systems are inextricably linked by a canonical distribution - the heterogeneous CoHSI distribution (\ref{eq:cohsi}) - visible in all aggregations and at all scales where numbers are sufficiently large for statistical mechanics to operate.  The two Lagrange multipliers which naturally emerge $\alpha, \beta$ are undetermined and simply parameterise the range of possible solutions.

This has a number of interesting implications which we will divide into levels of confidence based on the data analysis, distinguishing in each case whether this is a \textit{heterogeneous} system, Appendix A p. \pageref{app:heterogeneous} (which has a canonical distribution (\ref{eq:cohsi})) or a \textit{homogeneous} system, Appendix p. \pageref{app:homogeneous} (which has a canonical distribution (\ref{eq:powerrank}) corresponding to Zipf's law) or both, recalling that such systems differ only in the relevant definition of Hartley-Shannon information.  

\subsection*{Very confident}
These conclusions are strongly supported by statistical analysis using R and documented individually in the body of the paper.  Where linear analysis was done on a power-law tail, this is noted below as \textbf{(R)}.  \textit{All} $14$ such analyses gave an adjusted $R^{2}$ within the range $0.970 - 0.999$ with values of $p < e^{-14}$.
\begin{itemize}
 \item 

 \fbox{%
\begin{minipage}{12cm}
All \textit{heterogeneous} systems will tend to the canonical frequency distribution (\ref{eq:cohsi}) as total size grows, with a sharp unimodal peak and a power-law tail.
\end{minipage}}

 \begin{description}
  \item[Justification] Equation (\ref{eq:cohsi}).  Development starting Appendix A p. \pageref{app:heterogeneous}.
  \item[Evidence] Fig. \ref{fig:trembl_pdf} (Proteins); Fig. \ref{fig:c_pdf} (Software); Fig. \ref{fig:music_ti_pdf} (Music); Fig. \ref{fig:threemen_ti_pdf} (Texts) \textbf{(R)}).
 \end{description}
 
 \item
  
\fbox{%
\begin{minipage}{12cm}
All \textit{heterogeneous} discrete systems made up from components, themselves comprising indivisible tokens chosen from an alphabet, will exhibit a precise power-law tail in both their length distribution and in the distribution of their unique alphabet.
\end{minipage}}

 \begin{description}
  \item[Justification] \textit{Length:} Eq. (\ref{eq:pwrlaw2}), \textit{Alphabet:} Eq. (\ref{eq:pwrlaw}).  Development starting at Appendix A p. \pageref{app:heterogeneous} and also p. \pageref{app:dualdistrib}.
  \item[Evidence] \textit{Length:} Fig. \ref{fig:trembl_cdf} (Proteins); Fig. \ref{fig:software_ti_cdf} (Software); Fig. \ref{fig:music_ti_ccdf} (Music), \textbf{(R)}.  \textit{Alphabet:} Fig. \ref{fig:selenealphabets} (Proteins); Fig \ref{fig:software_ai_cdf} (Software); Fig. \ref{fig:musicpowerlaws_pdf} (Music), \textbf{(R)}.
 \end{description}
 
 \item
  
\fbox{%
\begin{minipage}{12cm}
The canonical CoHSI distribution (\ref{eq:cohsi}) tends to the asymptotic solution found in \cite{HattonWarr2015}.
\end{minipage}}

\begin{description}
 \item[Justification] Appendix A p. \pageref{app:chocolatebox} onwards.
 \item[Evidence] Figs. \ref{fig:cohsiclose}, \ref{fig:cohsifar}, example solutions of equation (\ref{eq:cohsi}).
\end{description}

\item

\fbox{%
\begin{minipage}{12cm}
  The canonical frequency distribution will appear in all aggregations of a system.
\end{minipage}}

\begin{description}
 \item[Justification] The theory for both \textit{heterogeneous} systems and \textit{homogeneous} systems is scale independent.  It finds the most likely length distribution for a given total size $T$ and a given total heterogeneous or homogeneous Information $I$, but other than requiring a reasonably large system in the general sense of Statistical Mechanics \cite{GlazerWark2001}, does not depend on the values of $T$ or $I$.  Appendix A p. \pageref{app:chocolatebox} onwards.
 \item[Evidence] \textit{Heterogeneous} systems, Figs. \ref{fig:archaea_pdf}-\ref{fig:viruses_pdf},  Figs. \ref{fig:human_pdf}-\ref{fig:halma_pdf} (Proteins); \textit{Homogeneous} systems, Figs. \ref{fig:cv_ccdf}-\ref{fig:threemen_ccdf} (Texts); Figs. \ref{fig:elements}, \ref{fig:seawater} (Atomic elements).
\end{description}

\item

\fbox{%
\begin{minipage}{12cm}
  The canonical frequency distribution will appear in all qualifying discrete systems independently of their provenance.
\end{minipage}}

\begin{description}
 \item[Justification] At the heart of the definition of Hartley-Shannon Information is the original prescient advice from Ralph Hartley that the meaning of the tokens is irrelevant.  
 \item[Evidence] \textit{Heterogeneous} systems, Fig. \ref{fig:trembl_pdf} (Proteins), Fig. \ref{fig:c_pdf} (Software), Fig. \ref{fig:music_ti_pdf} (Music),  Fig. \ref{fig:threemen_ti_pdf} (Text); \textit{Homogeneous} systems, Figs. \ref{fig:cv_ccdf}-\ref{fig:threemen_ccdf} (Text), Figs. \ref{fig:elements}, \ref{fig:seawater}  (Atomic Elements).
\end{description}

\item

\fbox{%
\begin{minipage}{12cm}
  If a system can be considered as both a \textit{heterogeneous} system \textit{and} a \textit{homogeneous} system, then the predictions for \textit{both} kinds of system will appear.
\end{minipage}}

\begin{description}
 \item[Justification] Only a different definition of information is needed.  The mechanism of statistical mechanics then automatically generates the relevant length distribution, Appendix A p. \pageref{app:heterogeneous} and p. \pageref{app:homogeneous}.  
 \item[Evidence] \textit{Heterogeneous} system, Fig. \ref{fig:threemen_ti_ccdf} (Lengths of words in a text); \textit{Homogeneous} systems, Fig. \ref{fig:threemen_ccdf} (Rank-ordered word frequency in text) \textbf{(R)}.
\end{description}

\item

\fbox{%
\begin{minipage}{12cm}
  The alphabet used to categorise a heterogeneous system is irrelevant provided it is consistent.
\end{minipage}}

\begin{description}
 \item[Justification] Consistent alphabets are power-laws of each other asymptotically for large components, Appendix C p. \pageref{app:musicalphabets} onwards.  
 \item[Evidence] Figs. \ref{fig:musicpowerlaws_pdf}, \ref{fig:musicalphabets_pdf} (Duration and no-duration notes in Music) \textbf{(R)}.
\end{description}

\item

\fbox{%
\begin{minipage}{12cm}
  Average component length is highly preserved across collections in \textit{heterogeneous} systems.
\end{minipage}}

\begin{description}
 \item[Justification] This is a direct consequence of the sharply unimodal shape of the canonical Distribution for heterogeneous systems.  
 \item[Evidence] Figs. \ref{fig:bacteria_average}, \ref{fig:eukaryota_average} (Proteins) \textbf{(R)}, from \cite{HattonWarr2015}).  This is also reported for software \cite{HatTSE14}.
\end{description}

\item

\fbox{%
\begin{minipage}{12cm}
  Unusually long components are inevitable in \textit{heterogeneous} systems.
\end{minipage}}

\begin{description}
 \item[Justification] Such components are inevitable because of the presence of the ubiquitous power-law tail, (noting the comments of \cite{Clauset2011}).  
 \item[Evidence] Fig. \ref{fig:trembl_cdf} (Proteins) which shows for example, approximately 10,000 proteins longer than 10 times the average length.).  This phenomenon has also been reported for software \cite{Louridas:2008:PLS:1391984.1391986}.
\end{description}
 
\end{itemize}

\subsection*{Confident}

\begin{itemize}
 \item We expect to see the numbers of known amino acids to expand with time to preserve the power-law tail already evident in Fig. \ref{fig:selene13-11_aa_ccdf}.  There were around $800$ structurally distinct Post Translationally Modified amino acids known in SwissProt 13-11.  There are already suspected to be thousands more \cite{KhouryBalibanFloudas2011}.  Our information argument strongly supports this thesis and may indeed help to quantify it.
\end{itemize}

\subsection*{Speculative}

\begin{itemize}
 \item We find it intriguing that the power-law slope of virus protein lengths is quite different from those observed in the three domains of life.  We speculate that this relates to their unique alphabet.
 \item The asymptotic behaviour for large components (\ref{eq:tloga}) in heterogeneous systems, implies that tokens carry a payload of the average information content of the component in which they appear.  For example, in proteins a particular amino acid might carry a different information payload in different proteins by virtue of the company it keeps.  We do not know if this has any useful physical interpretation.
 \item We (wildly) speculate that since both dark energy and dark matter lie on the same information theoretic distribution as the elements in order of atomic number, there is an undiscovered but intimate relationship between dark material and atomic number.
\end{itemize}

It is clear from the above results that important structural features of discrete systems are well-predicted by a single conservation principle applied to ergodic systems at all levels of aggregation and of all kinds.  Nothing more is asked of the theory than that the size of the systems should be sufficiently large that the methodology of Statistical Mechanics can be applied.

That the above features need not depend for explanation on any mechanism of natural selection in the case of proteins or anything to do with human volition in the case of either music or software, we find remarkable.  Instead, they simply manifest themelves as emergent properties of \textit{heterogeneous} or \textit{homogeneous} large systems of components, revealed when we consider an ergodic ensemble of the same size and H-S information content, from which we seek the most likely distribution of its component lengths using the methodology of this paper.

\section*{Acknowledgements}

\begin{itemize}
 \item \textbf{Competing Interests:} The authors declare that they have no
competing financial interests.
 \item \textbf{Correspondence:} Correspondence and requests for materials
should be addressed to Les Hatton ~(email: lesh@oakcomp.co.uk).
 \item This material is based on work supported in part by the National Science Foundation. Any opinion, finding, and conclusions or recommendations expressed in this material are those of the authors and do not necessarily reflect the views of the National Science Foundation.
\end{itemize}

Although some parts of this paper are published \cite{HatTSE08,HatTSE14,HattonWarr2015}, the main body of the paper, the lower and upper bounds on the implicit canonical pdf (the heterogeneous CoHSI distribution), the relationship between alphabets and most of the empirical work, appear here for the first time both to clarify and extend the theory with new results.  During review processes for various journals, the general thesis of this paper that a simple global principle appears to be responsible for some important aspects of protein evolution was a bridge too far for most of the reviewers given these restrictions.  This resistance has been noted before in biology research \cite{Frank2009} and publishing inter-disciplinary papers in general remains much more difficult than it should be.

However, there were some insightful responses for which we are both very grateful.  One anonymous reviewer pointed out an error with the combinatorics leading up to the chocolate box argument of Appendix p. \pageref{app:chocolatebox}, which opened up a real can of worms (the role of additive partitions) which was finally resolved by the recursive argument of Appendix A p. \pageref{app:sm}. Yet another reviewer made us really think about the application of this apparently abstract principle to real systems we can touch and feel, as well as suggesting we clarify the emergent role of $A$ in the dual distribution, which we hopefully have in Appendix A p. \pageref{app:dualdistrib}.

We would particularly like to thank Dan Rothman for a whole string of insights which really helped clarify some of the more opaque sections; Ken Larner who went through the document with a fine toothcomb suggesting numerous clarifications; Leslie Valiant for bringing to our attention Frank's excellent paper \cite{Frank2009}; Alex Potanin and colleagues in New Zealand for useful comments after an early seminar on this topic.

Any mistakes which remain are our responsibility of course but we hope that the theory is now sufficiently well explained and its novelty delineated and that the associated reproducibility packages will help others to verify the computational aspects and extend it in new directions.

\newpage
\section*{Appendix A: The Conservation of Hartley-Shannon Information: From Statistical Mechanics to the Canonical CoHSI Distribution}
\label{app:sm}
Statistical mechanics is a methodology for predicting component distributions of general systems made from discrete pieces or components subject to restrictions known as constraints.  Such systems include gases (made from molecules), proteins (made from amino acids), software (made from programming language tokens) and even boxes containing beads.  Conventionally, constraints are applied by fixing the total number of pieces and/or the total energy \cite{GlazerWark2001}.

\subsection*{Statistical Mechanics: Classical use in physics}
To illustrate the method, we describe a classical problem of determining the most likely distribution of particles amongst energy levels.

To see this, the following variational methodology is borrowed from the world of statistical physics, (\cite{Sommerfeld56} (p.217-); and for an excellent introduction, see \cite{GlazerWark2001}).  In the kinetic theory of gases, a standard application is to find the most common arrangement of molecules amongst energy levels in a gas subject to various constraints such as a fixed total number of molecules and fixed total energy.  For this, imagine that there are M energy levels, where the number of particles with energy level $\varepsilon_{i}$ is $t_{i}, i=1,..,M$.

For this system, the total number of ways $W$ of organising the particles amongst the M energy levels is given by:-

\begin{equation}
W=\frac{T!}{t_{1}!t_{2}!..t_{M}!}, \label{eq:comb}
\end{equation}

where

\begin{equation}
T = \sum_{i=1}^{M} t_{i}  \label{eq:con1}
\end{equation}

The total amount of energy in this system is just the sum of all the particle energies and is given by

\begin{equation}
E = \sum_{i=1}^{M} t_{i} \varepsilon_{i} \label{eq:con2}
\end{equation}

In a physical system, E corresponds to the total internal energy and the variational method to follow constrains this value to be fixed; i.e. solutions are sought in which energy is conserved.

Using the method of Lagrangian multipliers and Stirling's approximation as described in \cite{GlazerWark2001}, will give the most likely distribution satisfying equation (\ref{eq:comb}) subject to the constraints in equations (\ref{eq:con1}) and (\ref{eq:con2}).  This is equivalent to maximising the following variational derived by taking the natural log of (\ref{eq:comb}).  Just as in maximum likelihood theory, taking the log dramatically simplifies the proceedings, in this case the factorials, and allows the use of Stirling's theorem for large numbers.  Also, since it is monotonic, a maximum in log W is coincident with a maximum in W.  This leads to

\begin{equation}
\log W = T \log T - \sum_{i=1}^{M} t_{i} \log (t_{i}) + \lambda \lbrace T - \sum_{i=1}^{M}t_{i} \rbrace + \beta \lbrace U -  \sum_{i=1}^{M} t_{i} \varepsilon_{i} \rbrace    \label{eq:minclassical}
\end{equation}

where $\lambda$ and $\beta$ are the multipliers \cite{GlazerWark2001}.  In essence, the variational process envisages varying the contents $t_{i}$ of each of the components until a maximum of $\log W$ is found.  The maximum is indicated by taking $\delta (\log W) = 0$, (analogous to finding maxima in differential calculus).  Noting that

\begin{itemize}
 \item the variational operator $\delta$ acts on pure constants such as $T \log T$, $\lambda T$ and $\beta U$ to produce zero just as when differentiating a constant,
 \item the product rule of differentiation gives $\delta (t_{i} \log (t_{i})) = \delta t_{i} \log (t_{i}) + t_{i} \delta (\log (t_{i}) = \delta t_{i} (1 + \log(t_{i}))$,
 \item $\varepsilon_{i}$ is independent of the variation by assumption,
 \item T and the $t_{i}$ are $\gg 1$ (to satisfy Stirling's theorem, although it is surprisingly accurate even for relatively small values).
\end{itemize}

This leads to

\begin{equation}
0 = - \sum_{i=1}^{M} \delta t_{i} \lbrace \log (t_{i}) + \alpha + \beta \varepsilon_{i} \rbrace    \label{eq:var1}
\end{equation}

where $\alpha = 1 + \lambda$.  (Further elaboration of this standard technique can be found in Glazer and Wark \cite{GlazerWark2001}.)

Finally, (\ref{eq:var1}) must be true for all variations to the occupancies $\delta t_{i}$ and therefore implies

\begin{equation}
\log (t_{i}) = - \alpha - \beta \varepsilon_{i}    \label{eq:var2}
\end{equation}

\textbf{for all $i$}.

Using equation (\ref{eq:con1}) to replace $\alpha$, this can be manipulated into the most likely, i.e. the equilibrium distribution, of particles amongst the M components.

\begin{equation}
t_{i} = \frac{T e^{-\beta \varepsilon_{i}}}{\sum_{i=1}^{M} e^{-\beta \varepsilon_{i}}}    \label{eq:varn}
\end{equation}

Defining $p_{i} = \frac{t_{i}}{T}$ means that $p_{i}$ can be interpreted as a probability density function since it is non-negative everywhere and its sum everywhere is equal to 1.  Then (\ref{eq:varn}) yields

\begin{equation}
p_{i} = \frac{e^{-\beta \varepsilon_{i}}}{\sum_{i=1}^{M} e^{-\beta \varepsilon_{i}}}    \label{eq:varp}
\end{equation}

This is a result with a profound interpretation in physics.  It states

\paragraph{}
\fbox{%
\begin{minipage}{12cm}
\textbf{If we consider all possible systems which share the same number of particles T and the same total energy U (i.e. energy is \textit{conserved}), then given any one example of such a system, it is overwhelmingly likely to obey (\ref{eq:varp}).  In other words by considering all possible systems with these parameters and constraining them to have the same T and U, probability distribution (\ref{eq:varp}) is overwhelmingly likely, provided the $t_{i}$ are large enough for Stirling's approximation to hold.}
\end{minipage}}
\paragraph{}

In this case, the distribution is exponential and exactly as is found in nature - \textit{exponentially fewer particles occupy higher energy levels.}  All the above has been known for decades and is extremely successful at explaining classical systems such as gases and even quantum mechanical systems; the methodology of statistical mechanics however is exceedingly versatile, so let us consider a simple model just consisting of boxes of coloured beads.

\subsection*{Conservation of Hartley-Shannon Information}
\subsubsection*{Identical boxes of beads}
Let us put some flesh on the meaning of ``overwhelmingly likely'' as used earlier in this paper.  Consider now a system of M boxes of identical beads, where the $i^{th}$ box contains $t_{i}$ beads and M is reasonably large.  If we have T beads in total numbered by sequence, where $T = \sum_{i=1}^{M} t_{i}$, so that they are distinguishable by their order, the number of possible ways of arranging them in each of the M boxes is given by

\begin{equation}
 \Omega = \frac{T!}{\prod_{i=1}^{M} (t_{i}!)}		\label{eq:distinguishable}
\end{equation}

\begin{figure*}[h!]
\centerline{\includegraphics[width=6cm]{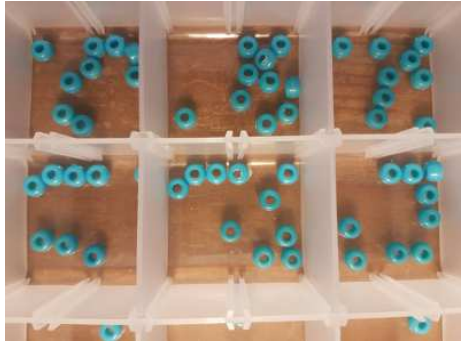}}
\caption{Boxes containing exactly the same number of exactly the same bead.}
\label{beads_same}
\end{figure*}

Suppose there are $M=10$ boxes and $T=100$ beads and we simply assign them one by one to a randomly chosen box.  We would be very surprised if the first box contained all the beads with the others empty, and the number of ways this can happen according to (\ref{eq:distinguishable}) is $100!/(100!\times0!\times0!..0!) = 1$.  If each box contains 10 beads however as shown in Fig. \ref{beads_same}, this situation can happen in $100!/(10!\times10!\times..10!)$ ways, which is approximately $10^{100}$, a gigantic number.

In other words, we are overwhelmingly more likely to see equal box populations than 1 single filled box.  In fact statistical mechanics allows us to prove that, in this case, equal population is by far the most likely distribution of contents simply by finding the maximum of (\ref{eq:distinguishable}) subject to a fixed number of T beads in the form

\begin{multline}
\log \Omega = T \log T - T - \sum_{i=1}^{M} \lbrace t_{i} \log (t_{i}) - t_{i} \rbrace \\
+ \alpha \lbrace T - \sum_{i=1}^{M}t_{i} \rbrace    \label{eq:mint}
\end{multline}

where the constraint on fixing T is controlled by the Lagrangian parameter $\alpha$.  Finding the maximum of (\ref{eq:mint}) using the standard $\delta()$ method \cite{GlazerWark2001}, gives the solution $t_{i} \sim constant$, corresponding to equal box populations.  Frank also demonstrates this in his maximum entropy formulation \cite{Frank2009}, p. 9.

\subsubsection*{Heterogeneous boxes of beads}
\label{app:heterogeneous}
\begin{figure*}[h!]
\centerline{\includegraphics[width=6cm]{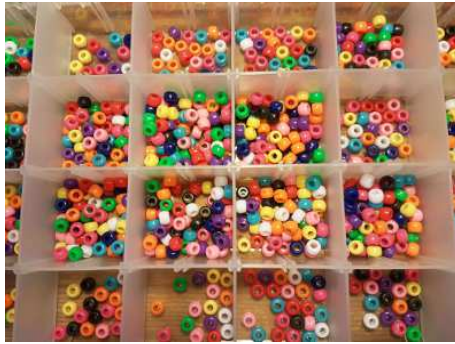}}
\caption{The heterogeneous case where each box contains mixed types and different numbers of beads.  This is relevant to proteins, computer program functions and the length distribution of words in texts.}
\label{beads}
\end{figure*}

We now describe an extension which is directly relevant to systems such as the known proteome or computer programs.  Consider Fig. \ref{beads}.

Here the boxes contain differently coloured beads.  We envisage this as the $i^{th}$ box containing $t_{i}$ beads selected randomly from a \textit{unique alphabet} of $a_{i}$ colours, ordered by sequence.  For the proteome, the “colours” correspond to different amino acids and for software functions they correspond to different programming language tokens.

Now we utilise the great generality of statistical mechanics by generalizing the payload to be Hartley-Shannon (H-S) information content instead of energy \cite{HatTSE14}.

The H-S information content of the $i^{th}$ box $I_{i}$, (Appendix p. \pageref{app:hsinfo}) is simply the log of the number of ways of arranging the beads in that box, \textit{so that it is guaranteed to contain at least one of each of the $a_{i}$ colours}.  H-S information is completely agnostic about what the colours actually mean, \textit{indeed Hartley specifically advised against attaching any meaning to a token \cite{Hartley1928}.}   The only thing that matters is that beads \textit{change} colour, so the actual colour is irrelevant and the total H-S information is just the sum of the information for each box.  Presented with such a system, we can ask what is the most likely distribution of contents for systems for which both the total number of beads and the total H-S information are conserved ?  We must also recall that proteins and software are both constructed sequentially so we are considering systems where beads are distinguishable by the order in which they appear, but the actual order is irrelevant.

The relevant variational form we must solve is therefore

\begin{multline}
\log \Omega = T \log T - T - \sum_{i=1}^{M} \lbrace t_{i} \log (t_{i}) - t_{i} \rbrace \\
+ \alpha \lbrace T - \sum_{i=1}^{M}t_{i} \rbrace + \beta \lbrace I -  \sum_{i=1}^{M} I_{i} \rbrace    \label{eq:mini}
\end{multline}

The only term which is different in this formulation from the classical solution derived above (\ref{eq:minclassical}), is the last term on the right hand side of (\ref{eq:mini}).  In the variational methodology, each term has the $\delta()$ operation applied in order to vary the $t_{i}$ and derive the distribution in (\ref{eq:var1}), so we are interested specifically in

\begin{equation}
 \delta \big (\beta \lbrace I -  \sum_{i=1}^{M} I_{i} \rbrace \big ) =  - \beta \sum_{i=1}^{M} \delta ( I_{i} )  = - \beta \sum_{i=1}^{M} \frac{d I_{i}}{d t_{i}} \delta t_{i}, \label{eq:deltai}
\end{equation} 

since I is being held constant.

Now consider what happens when boxes are very large compared with their unique alphabet, i.e. $t_{i} \gg a_{i}$.  In this case, \cite{HatTSE14}, the information content is 

\begin{equation}
 I_{i} = \log(a_{i}\times a_{i}\times ... \times a_{i}) = \log(a_{i}^{t_{i}}) = t_{i} \log a_{i}
 \label{eq:tloga}
\end{equation} 

In other words, we select $t_{i}$ times from a choice of $a_{i}$ colours secure in the knowledge that since $t_{i} \gg a_{i}$, it is very unlikely that any of the $a_{i}$ colours would be missed out and we therefore meet the requirement of having \textit{exactly} $a_{i}$ unique colours.

In this case, (\ref{eq:deltai}) becomes

\begin{equation}
- \beta \sum_{i=1}^{M} \frac{d I_{i}}{d t_{i}} \delta t_{i} = - \beta \sum_{i=1}^{M} \frac{d (t_{i} \log a_{i})}{d t_{i}} \delta t_{i} =  - \beta \sum_{i=1}^{M} (\log a_{i}) \delta t_{i} \label{eq:deltaiasymp}
\end{equation}

(\ref{eq:deltaiasymp}) fits perfectly into the variational methodology leading to (\ref{eq:mini}), modifying (\ref{eq:var2}) to give

\begin{equation}
\log (t_{i}) = - \alpha - \beta \log a_{i}    \label{eq:var2a}
\end{equation}

The analogue of (\ref{eq:varp}) is therefore

\begin{equation}
p_{i} \equiv \frac{t_{i}}{T}= \frac{a_{i}^{-\beta}}{\sum_{i=1}^{M} a_{i}^{-\beta}}    \label{eq:varpa}
\end{equation}

To summarize, maximising (\ref{eq:distinguishable}) subject to a fixed total number of beads T AND a fixed total H-S information $I = \sum_{i=1}^{M} I_{i}$ is directly analogous to maximising (\ref{eq:minclassical}) with $\log a_{i}$ replacing $\epsilon_{i}$.  Like its classical equivalent (\ref{eq:varp}),  (\ref{eq:varpa}) is also fundamental.  It states

\paragraph{}
\fbox{%
\begin{minipage}{12cm}
\textbf{In any discrete system satisfying the model described here, the tail (i.e. large $t_{i}$) of the distribution of unique alphabets is overwhelmingly likely to obey a power-law.}
\end{minipage}}
\paragraph{}

Note that by analogy with (\ref{eq:minclassical}), we can interpret $t_{i} \log a_{i}$ as each bead carrying a payload of $\log a_{i}$, so that even though H-S information is token agnostic, the beads in a particular box still carry a box-dependent payload which is a function of the unique alphabet of colours in that box, $a_{i}$.  This is exactly analogous to $t_{i} \varepsilon_{i}$ being interpreted as each particle carrying an energy $\varepsilon_{i}$ in classical statistical mechanics.  In other words, each box behaves as if it had a fixed \textit{information} level $\log a_{i}$ determined by its unique alphabet.  In a protein for example, this has the intriguing implication that even though H-S information is token-agnostic, a particular amino acid in one protein may carry a different information payload than when present in another protein, simply because its neighbours are different.

\subsubsection*{The asymptotic dual distribution}
\label{app:dualdistrib}
As pointed out by \cite{HattonWarr2015}, (\ref{eq:varpa}) has a dual solution.  With some algebra, it can be shown that

\begin{equation}
q_{i} \equiv \frac{a_{i}}{A} = \frac{t_{i}^{-1/\beta}}{\sum_{i=1}^{M} t_{i}^{-1/\beta}},	  \label{eq:varqa}
\end{equation} 

where

\begin{equation}
A = \sum_{i=1}^{M} a_{i}
\end{equation}

Note here that $A$ emerges naturally as the sum of the unique alphabets of each component.  It is \textit{not} the size of the unique alphabet across all components.  This is simply another manifestation of the token-agnosticism of Hartley-Shannon information - system-wide uniqueness of the alphabet simply does not emerge as a requirement.  The only requirements for a pdf are that it be positive definite and normalisable so this in no way detracts from the fact that (\ref{eq:varqa}) is also a power-law.  In other words, 

\paragraph{}
\fbox{%
\begin{minipage}{12cm}
\textbf{The length distribution of large proteins or software functions for which $t_{i} \gg a_{i}$ will also be a power-law.}
\end{minipage}}
\paragraph{}

Note also the natural appearance of the reciprocal slope $1/\beta$.  This value is not found in the datasets here but this difference is discussed and we think resolved in the discussion of alphabets in music in the Appendices p. \pageref{app:musicalphabets}.

\subsubsection*{The chocolate box analogy and additive partitions: the CoHSI distribution}
\label{app:chocolatebox}
For smaller boxes containing fewer beads, the above value of $I_{i}$ (\ref{eq:tloga}) is not correct.  If $t_{i}$ is closer in size to $a_{i}$, (it cannot be smaller since the length must be at least equal to the unique alphabet), there is an increasingly high probability that we might miss out one of the colours in the unique alphabet as we select our $t_{i}$ beads, negating the fundamental assumption that each box contains a unique alphabet of \textbf{exactly} $a_{i}$.  We must therefore make different provisions as the boxes get smaller.

\begin{figure}[H]
\centerline{\includegraphics[width=6cm]{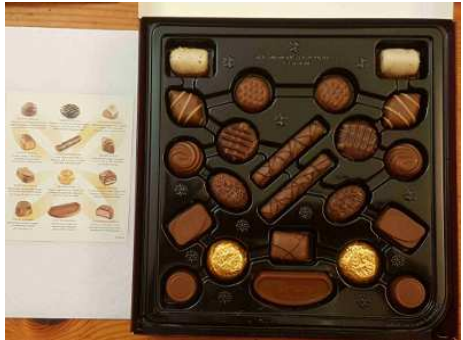}}
\caption{A box of 22 chocolates chosen from 12 different types as shown on the left.}
\label{fig:chocolates}
\end{figure}

The situation is akin to boxes of mixed chocolates, Fig. \ref{fig:chocolates}.  Such boxes are constructed from a fixed set of chocolates advertised on the lid, and every box must contain at least one of each.  Larger boxes simply contain more than one of some kinds.  In how many ways can such boxes be created ?

Note that it is simple to find an algorithm to guarantee that the unique alphabet is exactly $a_{i}$.  All that is necessary is to fill any $a_{i}$ places with one chocolate of each type and then fill the remaining $t_{i} - a_{i}$ at random from the available types.  The number of ways of doing this is

\begin{equation}
 \big((a_{i}!) . (\Comb{t_{i}}{a_{i}})) . (a_{i}^{(t_{i} - a_{i})}\big),   \label{eq:forcechocs}
\end{equation} 

where $\Comb{n}{r} = n!/((n-r)!r!)$ is the combination operator.  This however, is not the same as counting \textit{all} the possible ways of filling the box such that it contains exactly $a_{i}$ chocolates.

We are trying to find the number of different ways of filling the $i^{th}$ box with $t_{i}$ chocolates chosen from a unique set of exactly $a_{i}$ chocolates and we must do this in a way which fits into the statistical mechanical framework so we can use its methodology.

To explore this, suppose we have a box of $t_{i} = 5$ chocolates such that it contains \textit{exactly} $a_{i} = 2$ different chocolates of types A and B.  The total number of ways this can be done $N(t_{i},a_{i})$, is given by

\begin{equation}
 N(5,2) = \frac{5!}{1!4!} + \frac{5!}{4!1!} + \frac{5!}{3!2!} + \frac{5!}{2!3!}			\label{eq:chocs52}
\end{equation} 

Note

\begin{itemize}
 \item The first term on the right hand side of (\ref{eq:chocs52}) is the total number of ways of selecting 5 chocolates by using 1 chocolate of type A and 4 chocolates of type B.  This is equal to 5 (ABBBB, BABBB, BBABB, BBBAB, BBBBA).
 \item The second term corresponds to 4 chocolates of type A and 1 of B and is also equal to 5 (BAAAA, ABAAA, AABAA, AAABA, AAAAB).
 \item The third term corresponds to taking 3 of type A and 2 of type B.  This is equal to 10, (AAABB, AABAB, AABBA, ABAAB, ABABA, ABBAA, BBAAA, BABAA, BAABA, BAAAB).
 \item The fourth term corresponds to taking 2 of type A and 3 of type B.  This is also equal to 10, (BBBAA, BBABA, BBAAB, BABBA, BABAB, BAABB, AABBB, ABABB, ABBAB, ABBBA).
\end{itemize}

There are no other ways of arranging the box such that there are exactly 2 colours and exactly 5 chocolates altogether.  There are therefore 5 + 5 + 10 + 10 = 30 different such boxes in total.  Note that (\ref{eq:forcechocs}) gives $(2!) . (\Comb{5}{2})) . (2^{(5 - 2)}) = 160$ boxes.  This over-counting is because a box such as ABBAB could be generated several times by that algorithm, for example, by filling the first two places with AB and then the rest at random or by filling the first and third places with AB and the rest at random.).

The denominators of (\ref{eq:chocs52}) correspond to elements of the \textit{additive compositions}\footnote{\url{https://en.wikipedia.org/wiki/Partition_(number_theory)}, accessed 02-Jun-2017.} of size 2 of the number 5.  These are

\begin{equation}
 5 = 1 + 4; 5 = 4 + 1; 5 = 3 + 2; 5 = 2 + 3
\end{equation} 

There are other additive compositions such as $2+2+1$, but this corresponds to three different kinds of chocolate so must be excluded.

The fact that the compositions are \textit{additive} presents a real complication when merging with the methodology of statistical mechanics because it breaks the steps leading from (\ref{eq:deltai})-(\ref{eq:varpa}) by introducing the log of a recursive definition as we shall see.  Prior to discovery of this recursive method, the solution was simply trapped between a lower and upper bound.  The lower bound consisted of just one of the terms leading to the recursive definition and the upper bound was the pure power-law (\ref{eq:varpa}).  The recursive method is however far more compelling.

First we slightly modify the definition in (\ref{eq:chocs52}) by letting $N(t_{i}, a_{i}; a'_{i})$ be the number of ways of producing a chocolate box with $t_{i}$ chocolates containing exactly $a'_{i}$ unique types chosen from a total unique number of types of $a_{i}$.  In this notation, for example, $N(5,2;1) = 2$ and $N(5,2;2) = 30$.  The distinction between $a_{i}$ and $a'_{i}$ is to make way for the use of recursion.

It can be easily verified that the following recursion then generates the desired total number of ways $N(t_{i}, a_{i}; a_{i})$ of generating a chocolate box of $t_{i}$ chocolates from a unique set of chocolates $a_{i}$.

\begin{equation}
 N(t_{i}, 1; 1) = 1; \hspace{0.5cm} N(t_{i}, a_{i}; i) = \Comb{a_{i}}{i} N(t_{i}, i; i), i = 1,..,a_{i} - 1, a_{i} = 1,..,t_{i}
\end{equation} 

completed by

\begin{equation}
 N(t_{i}, a_{i}; a_{i}) = a_{i}^{t_{i}} - \sum_{i=1}^{a_{i}-1} N(t_{i},a_{i}; i)
\end{equation}

The corresponding Hartley-Shannon information content for a box containing $t_{i}$ chocolates chosen from a unique alphabet of $a_{i}$ chocolates is therefore given by

\begin{equation}
 I_{i} = \log \big ( N(t_{i}, a_{i}; a_{i}) \big ) 	\label{eq:chocs52ta}
\end{equation}

In contrast, the equivalent form for the pure power-law (\ref{eq:tloga}) is

\begin{equation}
 I_{i} \lvert_{P} = \log \big ( a_{i}^{t_{i}} \big ) = \log \big ( t_{i} \log a_{i} \big ) \label{eq:purepower}
\end{equation}

We can now see the problem posed by (\ref{eq:chocs52ta}) when we apply the $\delta()$ operator to $I_{i}$ in the statistical mechanical framework leading from (\ref{eq:deltai})-(\ref{eq:varpa}).  The presence of the recursion prevents the clean separation of factors by the $\log$ operation.  If we simply solved this computationally, that would ordinarily be no problem but (\ref{eq:chocs52ta}) is computationally difficult for the large factorial values which arise even for modest values of $(t_{i},a_{i})$.  \textit{(Whatever method we choose, however, it must have the property of producing the power-law form (\ref{eq:purepower}) in the asymptotic limit $t_{i} \gg a_{i}$.)}

Applying the $\delta()$ operator to (\ref{eq:mini}) using (\ref{eq:deltaiasymp}) leads to the pure power-law equation

\begin{equation}
\log t_{i} = -\alpha -\beta ( \log a_{i} ),    \label{eq:minip}
\end{equation}

whereas applying the $\delta()$ operator to (\ref{eq:mini}) using (\ref{eq:deltai}), (\ref{eq:chocs52ta}) leads to the full equation

\begin{equation}
\log t_{i} = -\alpha -\beta ( \frac{d}{dt_{i}} \log N(t_{i}, a_{i}; a_{i} ) ),    \label{eq:minif}
\end{equation}

Here, the unique alphabet $a_{i}$ is playing a dual role as the frequency in a pdf by analogy with (\ref{eq:varpa}) using (\ref{eq:varqa}).  To complete our chocolate box analogy, if we are presented with a system of boxes with a total number of chocolates $T$ chosen from a fixed alphabet of chocolates and a total H-S information $I$, then by far the most likely distribution of numbers of chocolates in any box will be given by a pdf which is the solution of (\ref{eq:minif}).

Finally we note that following the argument that led up to (\ref{eq:tloga}), for $t_{i} \gg a_{i}$,

\begin{equation}
 \log N(t_{i}, a_{i}; a_{i} ) \rightarrow t_{i} \log a_{i},
\end{equation}

so the full solution correctly asymptotes to the pure power-law.  This will be confirmed during the computation with both forms being displayed together.

\subsubsection*{Computational aspects of the CoHSI distribution}
Before we proceed with this, there is a technical limitation to overcome since the equation for the pdf which results from applying the variational method to (\ref{eq:minif}) is implicit. As we pointed out in the text, there is a precedent for this in the definition of Tsallis entropy \cite{Tsallis1988,Tsallis1999}, although in our case, the implicit nature of the pdf arises naturally from CoHSI.  (In Tsallis entropy, the entropy term is adjusted using an additional parameter and this adjustment can lead to an implicit pdf.)

We must therefore generalise the argument from integer values of ($t_{i}, a_{i}$), to the real line.  This will not affect our computation of factorials however, which are done at integer values of $t_{i}, a_{i}$, with interpolation for non-integer values.

\paragraph{}
\fbox{%
\begin{minipage}{12cm}
\textbf{(\ref{eq:minif}) defines the canonical implicit pdf with solutions $(t_{i},a_{i})$ which a) conserves H-S information and b) asymptotes to the pure power-law (\ref{eq:pwrlaw}) for $t_{i} \gg a_{i}$ as required for any heterogeneous system at all scales.}
\end{minipage}}
\paragraph{}

Solving (\ref{eq:minif}) proved challenging.  The one thing we do know, however, is that it asymptotes to the simple explicit solution $I_{i} \sim t_{i} \log a_{i}$ leading to (\ref{eq:pwrlaw}) when $t_{i} \gg a_{i}$.  This suggested the following procedure.

We start with large $t_{i}$ solving (\ref{eq:minip}) explicitly for $a_{i}$, followed by the use of this value as the starting value for the full solution of (\ref{eq:minif}).  This was found by searching a pre-computed grid of integer $(t_{i},a_{i})$ values for the appropriate value of $d/dt_{i} (\log N(t_{i}, a_{i}; a_{i}))$ interpolating as necessary.  When the solutions are found, we decrement $t_{i}$ and start again.  This process is somewhat akin to shooting methods in boundary layer solutions in fluid dynamics, \cite{vdyke70,Hatton1975}.  All code used is in the reproducibility package.

Figs. \ref{fig:trembl_pdf}, \ref{fig:c_pdf} indicate the behaviour we expect to find if CoHSI is indeed controlling these distributions.  The power-law behaviour for larger components is already modelled to a high degree of precision by the large component approximation (\ref{eq:tloga}).  We focus now on the behaviour for all component sizes.

\begin{figure}[H]
    \captionsetup[subfigure]{labelformat=empty}
    \centering
    \begin{subfigure}[t]{0.5\textwidth}
        \centering
        \caption{A}
        \includegraphics[width=6cm]{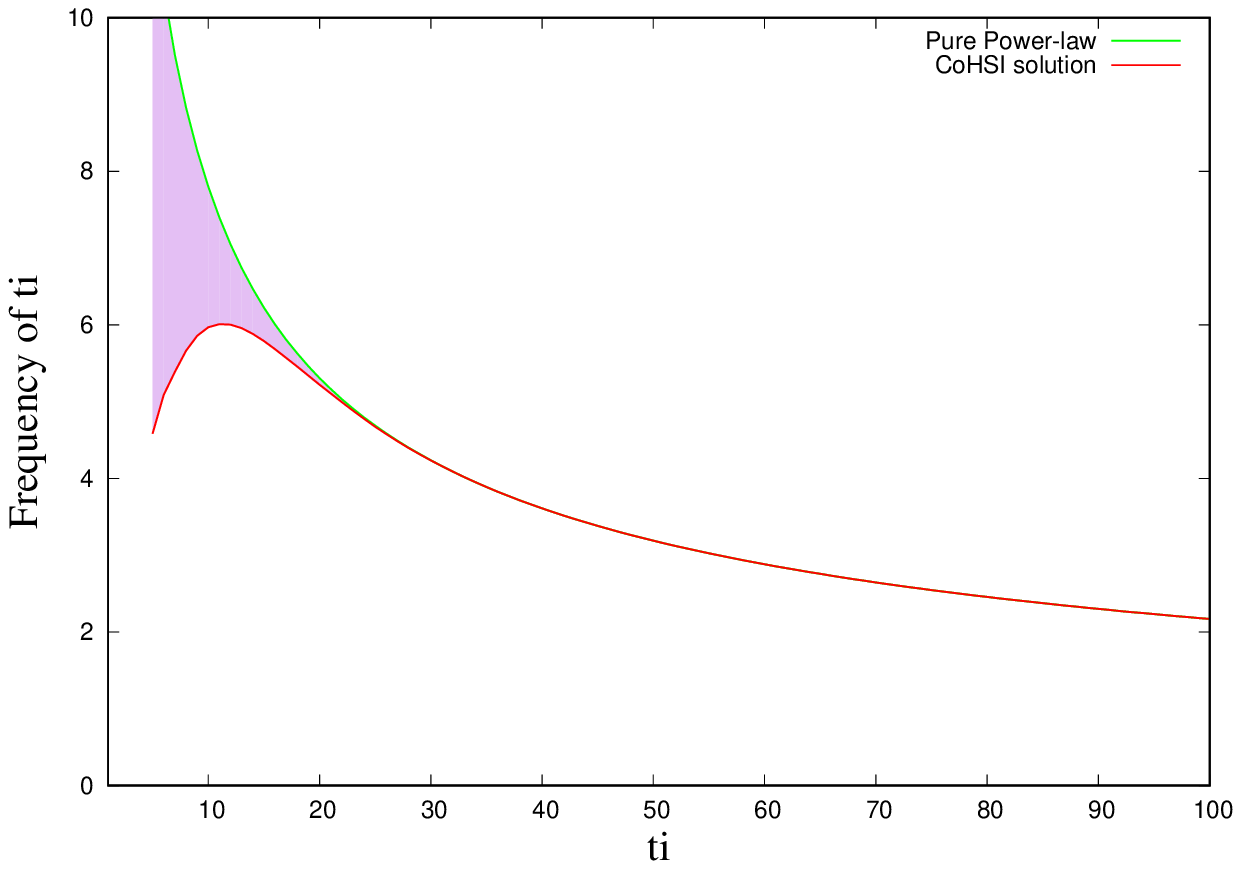}
        \label{fig:cohsiclose}
    \end{subfigure}%
    ~ 
    \begin{subfigure}[t]{0.5\textwidth}
        \centering
        \caption{B}
        \includegraphics[width=6cm]{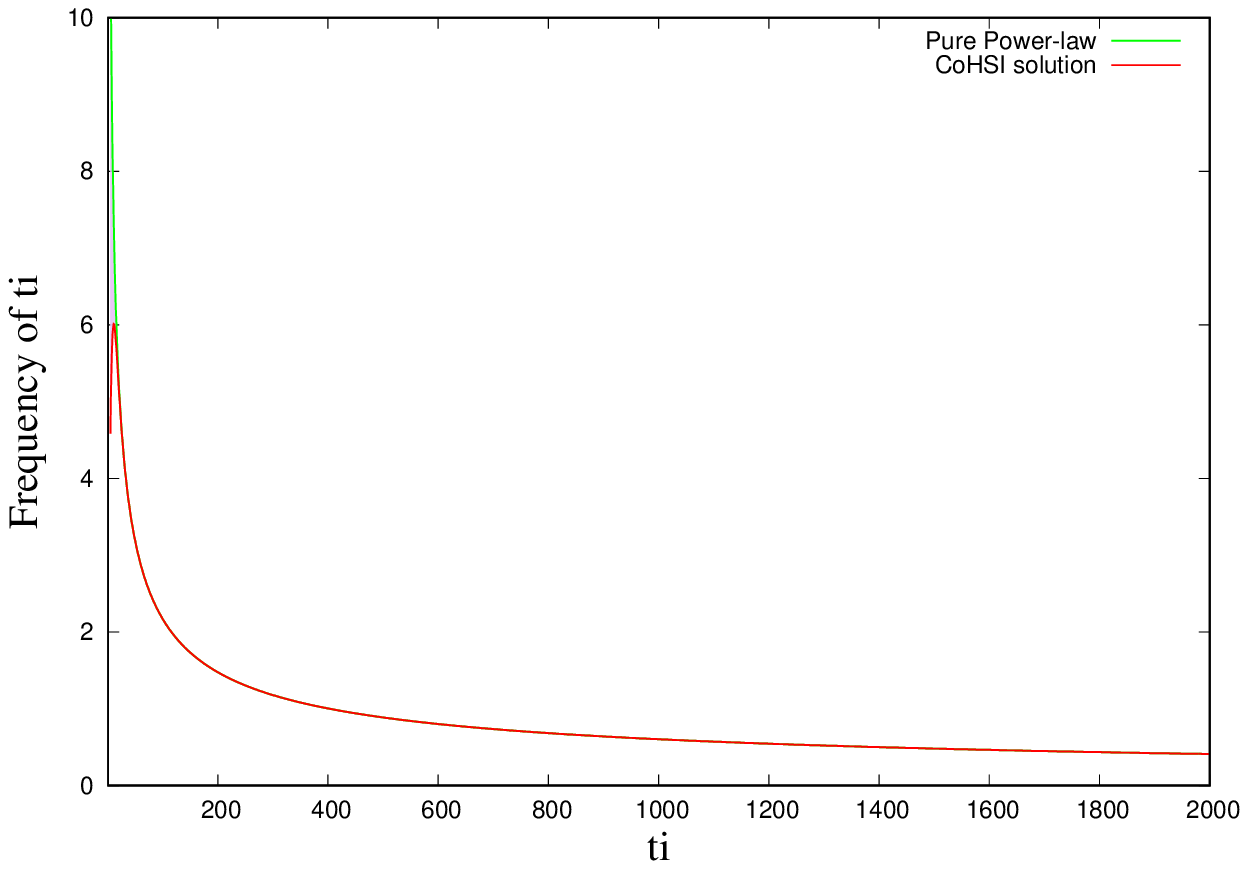}
        \label{fig:cohsifar}
    \end{subfigure}%

    \caption{The length distributions using the same modelling parameters for (A) the full CoHSI solution (\ref{eq:minif}), and the pure (asymptotic) power-law (\ref{eq:minip}) for components smaller than 100 tokens and (B), the same data for components up to 2,000 tokens long.}
\end{figure}

The full solution - the solution of (\ref{eq:minif}), and the pure power-law solution - the solution of (\ref{eq:minip}), using the same parameters, are shown together at two different scales as Figs. \ref{fig:cohsiclose} and \ref{fig:cohsifar}.  The shaded zone corresponds to the region where the full solution departs from the pure power-law solution.  Note that the the first order approximation for numerical differentiation means that the first few points for the full CoHSI solution do not converge.  The behaviour is clear from the remaining points however.

We make the following observations about Figs. \ref{fig:cohsiclose} and \ref{fig:cohsifar} with respect to (\ref{eq:minif}) and (\ref{eq:minip}).

\begin{figure}[H]
\centering
\includegraphics[width=8cm,height=6cm]{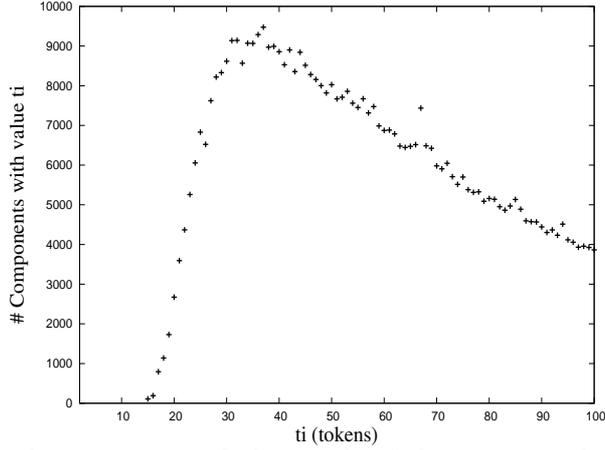}
\caption{A close-up around the peak of the measured dataset shown as Fig. \ref{fig:c_pdf}.}
\label{fig:c_pdf_close}
\end{figure}

\begin{itemize}
 \item In (\ref{eq:minip}) as the left hand side decreases with decreasing $t_{i}$, the value of $a_{i}$ must increase to give a solution.  This gives the pure power-law behaviour shown which continues to increase as $t_{i}$ decreases as shown clearly in Fig. \ref{fig:cohsiclose}.
 \item In (\ref{eq:minif}), $d/dt_{i}(N(t_{i}, a_{i}; a_{i})$ naturally decreases without having to keep increasing $a_{i}$ as was the case in (\ref{eq:minip}).  This is a natural consequence of CoHSI.
 \item As can be seen in Figs. \ref{fig:cohsiclose}, \ref{fig:cohsifar}, the qualitative behaviour of the full CoHSI solution around the unimodal peak remains sharp but is more rounded than the pure power-law solution and is qualitatively similar to the software data close up of Fig. \ref{fig:c_pdf} shown as Fig. \ref{fig:c_pdf_close}.  The sharpness of the peak is related to the boundary condition naturally emerging in this theory that $t_{i} \ge a_{i}$, i.e., no component can be shorter than its unique alphabet.
 \item The peak of the full solution differs slightly in position as well as their amplitudes as the power-law parameter $\beta$ changes.  A value of $\beta =1.8$ was used.  As $\beta$ increases, the peak moves left and the amplitude diminishes.
 \item The full solution naturally asymptotes to the pure power-law behaviour as required.
\end{itemize}

We can compare the behaviour around the peak with a close-up of the dataset of Fig. \ref{fig:c_pdf}, as shown as Fig. \ref{fig:c_pdf_close}.  Even on observed data, the transition from power-law to near linearity is abrupt, taking place over perhaps 10 tokens, and is qualitatively very similar to the full CoHSI solution.  

As was noted in the body of the paper, the juxtaposition of dual regions, one matched by a lognormal distribution and the other by a power-law has been described in the past, \cite{Montroll1982,Mitzenmacher2002}.  In our theory, this transition emerges completely naturally as the implicit solution of (\ref{eq:minif}).

To summarise, these results strongly support the thesis of this paper that the Conservation of Hartley-Shannon Information (CoHSI) acts as a constraint on how the length and alphabet size distributions of systems of a given size $T$ and total Hartley-Shannon information $I$, can evolve at all scales giving an excellent qualitative match which does not require juxtaposing existing pdfs of known properties.

\subsubsection*{Approximate properties of the heterogeneous CoHSI distribution}
\label{app:properties}
From the shape of Figs. \ref{fig:trembl_pdf}, \ref{fig:c_pdf} and the theory which led up to Figs. \ref{fig:cohsiclose}, \ref{fig:cohsifar}, we can approximate the distribution satisfactorily by glueing together a right-angled triangle up to the modal value $a_{max}$ at $t = t_{max}$, say and a power-law afterward because the solution corresponding to (\ref{eq:chocs52ta}) transitions from power-law to almost linear behaviour so quickly.  In other words, we can define the approximate canonical distribution $c(t)$ as follows

\[
 c(t) = \left \{ \begin{array} {ll}
                  (\frac{2(\beta - 1)}{(t_{max}^{2} (\beta + 1))}) t  & \mbox{$0 \le t \le t_{max}$} \\
                  (\frac{2(\beta - 1)}{(t_{max} (\beta + 1))}) (\frac{t}{t_{max}})^{-\beta}	& \mbox{$t_{max} < t < \infty$}
                 \end{array}
                 \right.
\]

We require $\beta >1$ for this to be positive definite.  This has been normalised so as to integrate to unity over it's support $[0,\infty]$.  This approximation will allow us to make useful inferences.  First we will calculate the mean location and spread of this distribution.

The mean location is given by

\[
 \langle c \rangle = \frac{2(\beta - 1)}{(t_{max}^{2} (\beta + 1))} \bigg [ \int_{s=0}^{t_{max}} s^{2} ds + t_{max} (\frac{1}{t_{max}})^{-\beta} \int_{s=t_{max}}^{\infty} s^{-\beta + 1} ds \bigg ],
\] 

which is

\[
 \langle c \rangle = \frac{2(\beta - 1)}{(t_{max}^{2} (\beta + 1))} \bigg [ \big [ \frac{s^{3}}{3} \big]_{s=0}^{t_{max}} + t_{max} (\frac{1}{t_{max}})^{-\beta} \big [ \frac{s^{-\beta + 2}}{-\beta + 2} \big]_{s=t_{max}}^{\infty}  \bigg ],
\]

and, provided $\beta > 2$, gives

\[
 \langle c \rangle = \frac{2(\beta - 1)}{(t_{max}^{2} (\beta + 1))} \bigg [ \big [ \frac{t_{max}^{3}}{3} \big] + ((t_{max})^{\beta + 1} \big [ - \frac{t_{max}^{-\beta + 2}}{-\beta + 2} \big]  \bigg ],
\]

and finally

\begin{equation}
 \langle c \rangle = \frac{2 (\beta - 1)t_{max}}{(\beta + 1)} \bigg [ \frac{1}{3} + \frac{1}{\beta - 2} \bigg ] = \frac{2 (\beta - 1)t_{max}}{3(\beta - 2)} ; \hspace{0.2cm} \beta > 2
 \label{eq:meanc}
\end{equation}

There is little point in computing higher moments because they place even greater constraints on the value of $\beta$ (they diverge unless the support for the distribution is a finite interval), and will not apply to our examples for which $2 \le \beta \le 4.5$, (recall that if the pdf has slope $-\beta$, the ccdf will have slope $-\beta + 1$ and we are measuring from the ccdf following \cite{Newman2006}.).

Applying the above estimate for $\langle c \rangle$ to the full Trembl distribution, Fig. \ref{fig:trembl_cdf}, for which $\beta = 4.14$ suggests that $\langle c \rangle / t_{max} \approx 1$, where as analysis of the data itself in R gives a ratio of around $1.5$ which is reasonable given the nature of the approximation.  The actual values from Trembl are given in Table \ref{tab:averagetrembl}.

\begin{table}[H]
\centering
\caption{Different measures of average length of proteins in amino acids in the full Trembl 17-03 distribution, Fig. \ref{fig:trembl_pdf}}
\begin{tabular}{@{\vrule height 10.5pt depth4pt  width0pt}cccc}
\hline
 & Mean & Median & Mode \\
\hline
Trembl & 335 & 268 & 219 \\
\hline
\end{tabular}
\label{tab:averagetrembl}
\end{table}

\subsubsection*{Homogeneous boxes of beads: CoHSI and Zipf's law}
\label{app:homogeneous}
\begin{figure*}[h!]
\centerline{\includegraphics[width=6cm]{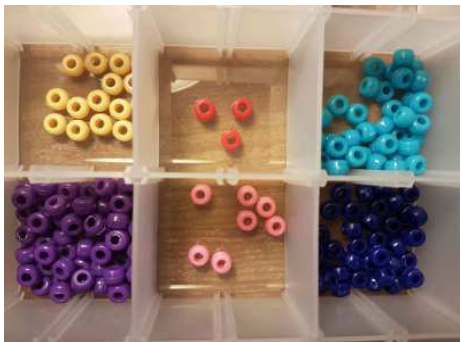}}
\caption{The homogeneous case.  In each box, all the beads are the same but different boxes contain different types and numbers of beads.  This is relevant to the distribution of atomic elements and to the rank ordering of frequency occurrence of words in texts.}
\label{boxes_same}
\end{figure*}

There are some kinds of discrete system for which the above information model does not apply.  Consider the case of homogeneous components.  Here, each bead carries a payload such that each box contains \textit{only beads with the same payload, unique to that box}.  We represent this by assembling beads of the same colour in the appropriate box, Fig. \ref{boxes_same}.

We could of course simply set $a_{i} = 1$ in the heterogeneous case above.  However, this immediately causes problems because the asymptotic Hartley-Shannon information content of any box in this case would be $t{_i} \log 1 = 0$ and is simply degenerate.  However, because Hartley-Shannon information is simply the $\log$ of the number of ways of arranging the beads of a box,  in the absence of an alphabet of choices in each box we can still find a suitable non-degenerate definition as follows.

Suppose we have a unique alphabet of beads $a'_{i}, i=1,..,M$ \textit{for the system as a whole}.  This is in contrast to the heterogeneous case where the unique alphabet $a_{i}$ was relevant only to the $i^{th}$ box.  Suppose from this system-wide alphabet, we seek to fill the M boxes each with $t_{i}$ of the $a'_{i}$ beads such that each box contains only one type.  The total population of the $M$ boxes is as before $T = \sum_{i=1}^{M} t_{i}$.  We will renumber them without loss of generality so that $t_{1} \ge t_{2} \ge .. \ge t_{M}$.

We proceed as follows.  Select any box and then fill it by selecting $t_{1}$ beads of the same colour.  Since we are selecting from $M$ different beads, the probability that we will achieve this selecting at random is $( 1/M )^{t_{1}}$.  For the second box, we then have an alphabet available of $M-1$, so the probability of filling this box with only one colour of the remaining colours is $( 1/(M-1) )^{t_{2}}$ and so on.

The total number of ways $N_{h}$ this can be done is then given by this probability multiplied by the total number of ways in which $T$ beads can be selected, which is $T!$.

\begin{equation}
 N_{h} = T! \bigg [ \big ( \frac{1}{M} \big )^{t_{1}} \times \big ( \frac{1}{M-1} \big )^{t_{2}} \times .. \times \big ( \frac{1}{1} \big )^{t_{M}} \big ] = T! \prod_{i=1}^{M} \big ( \frac {1}{i} \big )^{t_{i}}
 \label{eq:homogi}
\end{equation} 

Rewriting (\ref{eq:homogi}) then, the information content of this system is

\begin{equation}
 \log N_{h} = \log T! + \sum_{i=1}^{M} t_{i} \log \big ( \frac{1}{i} \big ) = \log T! - \sum_{i=1}^{M} t_{i} \log i
\end{equation} 

The development (\ref{eq:deltai})-(\ref{eq:varpa}) then follows but with $\log i$ replacing $\log a_{i}$.  The end result is the equivalent of (\ref{eq:varpa}) and amounts to

\begin{equation}
 t_{i} \sim i^{-\eta},
 \label{eq:powerrank}
\end{equation} 

where $\eta$ is some constant.

\paragraph{}
\fbox{%
\begin{minipage}{12cm}
(\textbf{\ref{eq:powerrank}) states that if we organise these homogeneous boxes in rank order of contents, (i.e. fullest first), then it is overwhelmingly likely that they will be distributed as a power-law in that \textit{rank}.  This is a famous law known as Zipf's law \cite{Zipf35}.  Zipf's law is empirical although others have produced statistical derivations \cite{Simon1955,WentianLi1992}.  The above derivation therefore serves as an alternative theoretical justification which places it nicely amongst those distributions which can be explained by the approach taken in this paper.}
\end{minipage}}
\paragraph{}

\newpage
\section*{Appendix B: CoHSI and Implications for Average Component Length and Long Components}
\subsection*{Average component length}
\label{app:averagecomplength}
It has been observed experimentally on several occasions \cite{Wang2005,XuJune2006,HattonWarr2015} that proteins appear to preserve their average length across aggregations within relatively tight bounds. The sharply unimodal peak of Figs. \ref{fig:trembl_pdf}, \ref{fig:c_pdf} as predicted by the theoretical development in this paper suggests that we should not be surprised at this.  Indeed at all scales and ensembles the estimates of average protein length will be highly conserved within collections as a result, even though the position of the peak may move a little.

\begin{figure}[H]
    \captionsetup[subfigure]{labelformat=empty}
    \centering
    \begin{subfigure}[t]{0.5\textwidth}
        \centering
        \caption{A}
        \includegraphics[width=6cm]{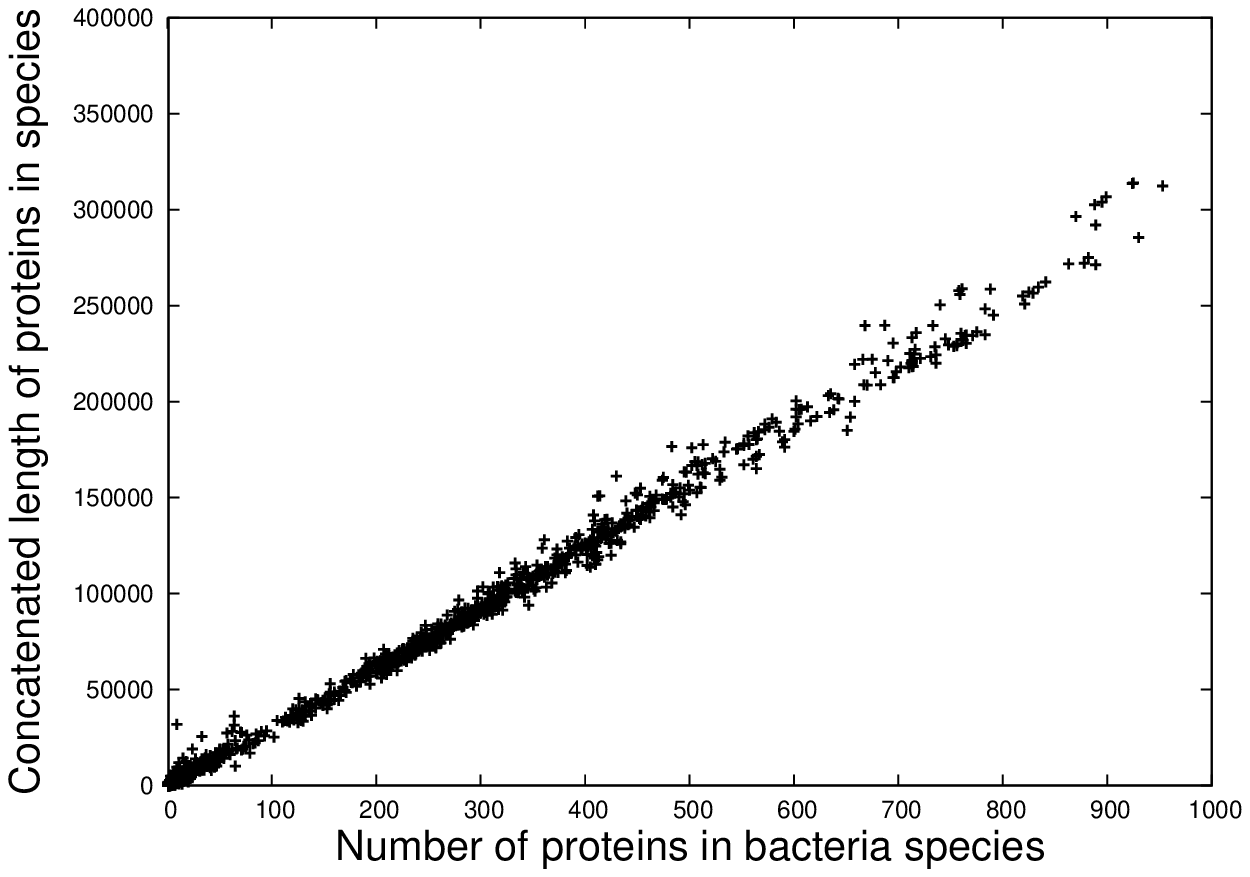}
        \label{fig:bacteria_average}
    \end{subfigure}%
    ~ 
    \begin{subfigure}[t]{0.5\textwidth}
        \centering
        \caption{B}
        \includegraphics[width=6cm]{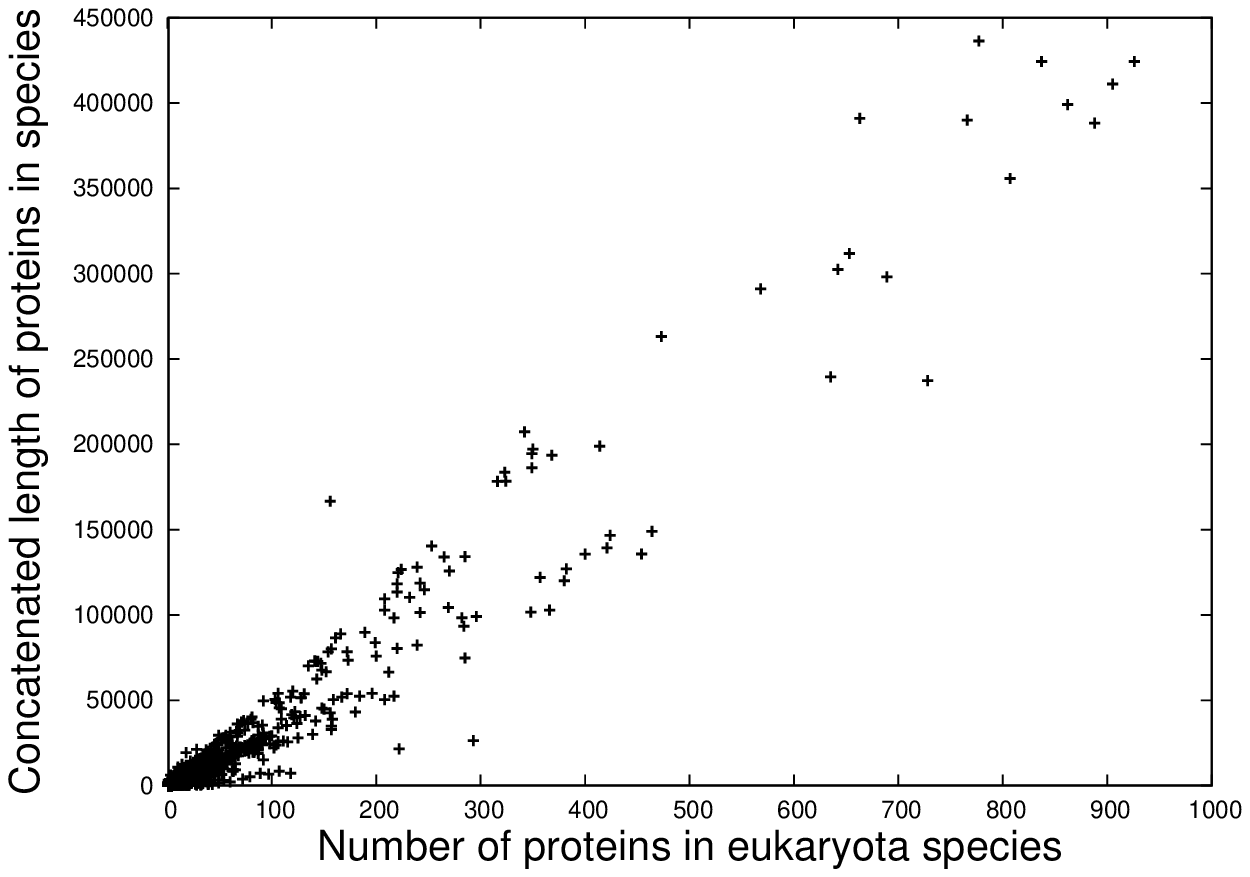}
        \label{fig:eukaryota_average}
    \end{subfigure}%

    \caption{A plot of the total concatenated length of proteins against the total number of proteins for each species in (A) Bacteria and (B) Eukaryota.  Each data point is a species.  The gradient of the linearity evident in both plots effectively defines the average protein length for that collection, from \cite{HattonWarr2015}.}
\end{figure}

In some aggregations, the degree to which the average length is preserved is quite remarkable, for example in Bacteria (Fig. \ref{fig:bacteria_average}), whilst in Eukarya, there is evidence of some fine structure \cite{HattonWarr2015} which invites further analysis Fig. \ref{fig:eukaryota_average}.

We also note that preservation of the average component length has also been reported for software \cite{HatTSE14}.

\subsubsection*{Measuring average protein length}
The skewed nature of the distribution of Figs. \ref{fig:trembl_pdf}, \ref{fig:c_pdf} suggests that the use of the mean as a measure of average length alone may be misleading and should be accompanied by other more robust measures such as the median and mode.  Table \ref{tab:average} demonstrates this by calculating them for the three domains of Archaea, Bacteria, Eukaryota, along with Viruses, as shown in Figs. \ref{fig:archaea_pdf}-\ref{fig:viruses_pdf}.  As expected, the more robust measure of median is less affected by the skew and the medians are therefore considerably less spread out than the means.  This is particularly true of viruses which although they have an anomalously large mean in comparison, their median is much closer aligned with those of Archaea and Bacteria.  The modes are subject to considerable noise.

\begin{table}[H]
\centering
\caption{Different Measures of average length of proteins in the domains of life and viruses}
\begin{tabular}{@{\vrule height 10.5pt depth4pt  width0pt}cccc}
\hline
Domain & Mean & Median & Mode \\
\hline
Archaea & 287 & 246 & 130 \\
Bacteria & 312 & 272 & 156 \\
Eukaryota & 435 & 350 & 379 \\
Viruses & 451 & 289 & 252 \\
\hline
\end{tabular}
\label{tab:average}
\end{table}

Table \ref{tab:average} shows that the mean is around $1-3$ times the modal value but we do not compare this using the approximate distribution, Appendix p. \pageref{app:properties}, as the data are rather noisier than for the full Trembl distribution (compare Fig. \ref{fig:cdf_domainsoflife} with Fig. \ref{fig:trembl_cdf}).

\subsection*{Long components}
\label{app:verylongcomponents}
One of the most important features of power-laws compared with any kind of exponential distribution such as the normal distribution is that ``events that are effectively 'impossible' (negligible probability under an exponential distribution) become practically commonplace under a power-law distribution.'' \cite{Clauset2011}.  The emphatic power-law in both the protein lengths and in software function lengths leads to large ratios when comparing the longest components with the average.  For example, proteins of around $36,000$ amino acids have been found and this is $100 \times$ the average.

\paragraph{}
\fbox{%
\begin{minipage}{12cm}
\textbf{In terms of the theory developed here, there is no need for any biological reason for very long proteins - they exist simply because of the naturally emerging power-law resulting from consideration of Information-conserving ergodic systems.}
\end{minipage}}
\paragraph{}

We note that precisely the same thing has been observed in software \cite{Louridas:2008:PLS:1391984.1391986}.

\section*{Appendix C: CoHSI and Token Alphabets}
\label{app:alphabets}
The definition of alphabets, i.e., unique sets of tokens from which choices can be made, poses interesting questions.  First of all, we must point out that there is generally no obvious definitive unique alphabet for any system.  Alphabets are partly subjective and partly objective because at their heart, they are about how humans categorise systems.  Take a simple example of a normally sighted person and a colour blind person both counting the number of differently coloured beads in a collection.  Barring counting errors, they will both find the same number of beads in total, however, \textit{they will not necessarily agree on the number of beads of each colour.}  In particular, red-green confusion is likely\footnote{\url{https://en.wikipedia.org/wiki/Color_blindness}}. 

How does this affect the theory we describe here ?  It might be thought that by linearly increasing the \textit{size} of the alphabet, the distribution of the two alphabets are themselves linearly related, i.e. $alphabet1_{i} = constant \times alphabet2_{i}$  However, this turns out to be \textit{not} the case and to understand what is happening, we must return to the duality of the asymptotic behaviour (\ref{eq:pwrlaw}) and (\ref{eq:pwrlaw2}), which we repeat here,

\begin{equation}
p_{i} \equiv \frac{t_{i}}{T} = \frac{a_{i}^{-\beta}}{Q(\beta)}					\label{eq:apwrlaw}
\end{equation}

and its algebraic dual given by

\begin{equation}
q_{i} \equiv \frac{a_{i}}{A} = \frac{t_{i}^{-1/\beta}}{\sum_{i=1}^{M} t_{i}^{-1/\beta}}		\label{eq:apwrlaw2}
\end{equation}

Since our normally-sighted person and our colour-blind person will count the same numbers but with different alphabets, we can say that for the normally sighted person,

\begin{equation}
p_{i} \equiv \frac{t_{i}}{T} = \frac{(a'_{i})^{-\beta'}}{Q(\beta')}					\label{eq:apwrlawprime}
\end{equation}

and for our colour blind person

\begin{equation}
p_{i} \equiv \frac{t_{i}}{T} = \frac{(a''_{i})^{-\beta ''}}{Q(\beta '')}					\label{eq:apwrlawprime2}
\end{equation}

where $a'_{i}, a''_{i}$ are the two unique alphabets they use and $\beta', \beta''$ their slopes.  Since the lengths are unchanged, we can see straight away from (\ref{eq:apwrlawprime}) and (\ref{eq:apwrlawprime2}), that the two unique alphabets will themselves be power-law related asymptotically

\begin{equation}
(a'_{i})^{-\beta'} \sim (a''_{i})^{-\beta''}	\Rightarrow a'_{i} \sim (a''_{i})^{-\beta'''},	\label{eq:pwralphabets}
\end{equation}

where $\beta''' = - \beta'' / \beta'$.  This leads us to predict a general rule

\paragraph{}
\fbox{%
\begin{minipage}{12cm}
\textbf{In any consistent categorisation of the same system with different unique alphabets, the distributions of the unique alphabets will also be related by a power-law.}
\end{minipage}}
\paragraph{}

\subsection*{Music alphabets}
\label{app:musicalphabets}
Consider an example from the world of music.  Music is also a system of discrete components in the sense described here, Table \ref{tab:entity}.  In recent years, discrete formats representing the notes and structure of a musical composition have appeared, for example MusicXML as referenced in the main text.

If we consider the 88 notes of a full-scale piano as defining the possible notes in the equal-tempered scale used in the vast majority of published music, then we have a candidate unique alphabet $a'_{i}$ of 88 \textit{(no-duration alphabet)}.  However, we can subdivide this alphabet quite naturally and consistently into notes \textit{and} duration.  The standard durations are divided into fractions of a whole note as breve (2), semi-breve (1), minim (1/2), crotchet (1/4), quaver (1/8), semiquaver (1/16) and demisemiquaver (1/32).  There are others defined off either end of this list but they are obviously rare as there were no occurrences in the body of music studied here.  This gives seven flavours of each note and expands the unique alphabet considerably to $88 \times 7 = 616$ items, \textit{(duration alphabet)}.

\begin{figure}[H]
    \captionsetup[subfigure]{labelformat=empty}
    \centering
    \begin{subfigure}[t]{0.5\textwidth}
        \centering
        \caption{A}
        \includegraphics[width=6cm]{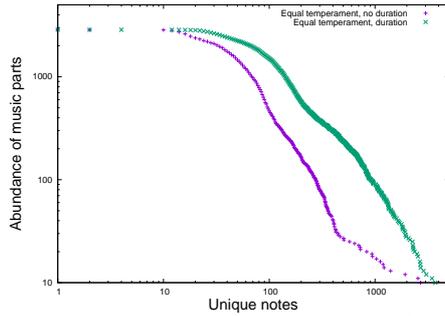}
        \label{fig:musicpowerlaws_pdf}
    \end{subfigure}%
    ~ 
    \begin{subfigure}[t]{0.5\textwidth}
        \centering
        \caption{B}
        \includegraphics[width=6cm]{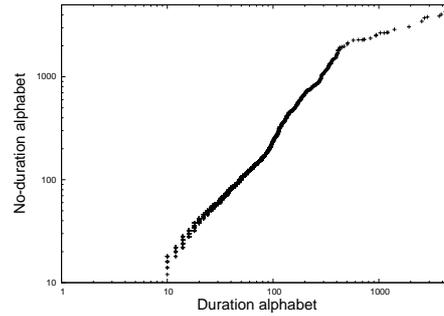}
        \label{fig:musicalphabets_pdf}
    \end{subfigure}%

    \caption{The $\log-\log$ ccdf of the duration and no-duration alphabets measured on the same body of music used in this study (A), and a comparison of the two alphabets as $\log-\log$ showing their clear linear power-law relationship (B).}
\end{figure}

Figs. \ref{fig:musicpowerlaws_pdf} shows the distribution of the two alphabets \textit{no-duration} and \textit{duration}, measured on the same body of music.  As expected from (\ref{eq:pwrlaw}) they both exhibit power-law behaviour.

\paragraph{}
\fbox{%
\begin{minipage}{12cm}
\textbf{For the \textit{no-duration} alphabet R reports that the associated p-value matching the power-law tail linearity in the ccdf of Fig. \ref{fig:musicpowerlaws_pdf} is $< (2.2) \times e^{-16}$ over the range $40.0-400.0$, with an adjusted R-squared value of $0.9937$.  The slope is $-1.66 \pm 0.01$.  For the \textit{duration} alphabet R reports that the associated p-value matching the power-law tail linearity in the ccdf of Fig. \ref{fig:musicpowerlaws_pdf} is $< (2.2) \times e^{-16}$ over the range $40.0-4000.0$, with an adjusted R-squared value of $0.9951$.  The slope is $-1.44 \pm 0.02$.}
\end{minipage}}
\paragraph{}

These are emphatic results.

\paragraph{}
\fbox{%
\begin{minipage}{12cm}
\textbf{Moreover in Fig. \ref{fig:musicalphabets_pdf} which compares the two alphabets directly on a $\log-\log$ scale, the predicted power-law relationship of (\ref{eq:pwralphabets}) is clearly visible.  R reports that the associated p-value matching the power-law tail linearity in the ccdf of Fig. \ref{fig:musicalphabets_pdf} is $< (2.2) \times e^{-16}$ over the range $10.0-500.0$, with an adjusted R-squared value of $0.9879$.  The slope is $1.181 \pm 0.002$, also consistent with (\ref{eq:pwralphabets}).  This too is an emphatic result.}
\end{minipage}}
\paragraph{}

We believe that this throws some light (but does not necessarily explain) why the predicted reciprocal relationship between the power-law slope of the unique alphabet distribution and that of the length distribution is not adhered to closely in our data.  There are a potentially infinite number of alphabets related themselves by power-laws, but only one length distribution.

\subsection*{Protein alphabets}
\label{app:proteinalphabets}
For proteins, with increasing sophistication we are able to recognise not just the 22 amino acids transcribed directly from DNA but also the increasingly large number of known post-translational modifications (PTM) which dramatically extend and continue to extend the size of the unique amino acid alphabet that allows us to categorise proteins.  Will this process of discovery stop ?  We argue that it cannot as it is intimately linked to the total number of proteins known, and this continues to grow apace.

We can gain insight into the growth of the unique protein alphabet by studying collections, such as the SwissProt database, over different revisions \cite{SwissProt2013,SwissProt2015,SwissProt2017} as it incorporates PTM information from the Selene project \cite{Selene2013}.

\begin{figure}[H]
\includegraphics[width=8cm,height=6cm]{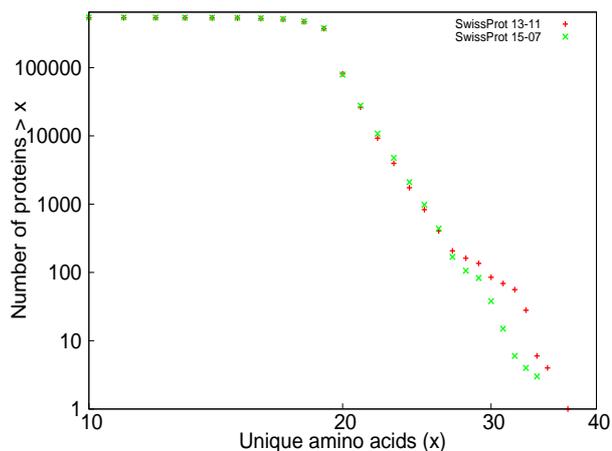}
\caption{The classic linear signature of a power-law distributions of unique amino acid alphabets for SwissProt 13-11 and the more up to date SwissProt 15-07 on a $\log-\log$ ccdf.}
\label{fig:selenealphabets}
\end{figure}

Fig. \ref{fig:selenealphabets} shows the frequencies of the unique amino acid alphabet recorded in the proteins of SwissProt release 13-11 and SwissProt release 15-07 as ccdfs in $\log-\log$ form.  Although the maximum unique alphabet for amino acids is 22 for those decoded directly from DNA, we should note that finding a protein with all 22 would be unlikely as both the $21^{st}$ and $22^{nd}$ amino acids, selenocysteine and pyrollysine, are rare in proteins.  Pyrrolysine is found in methanogenic archaea and bacteria and is encoded by a re-purposed stop codon (UAG),  requiring the action of  additional gene products to accomplish its incorporation \cite{Quitterer2012}. Thus it is not easy to annotate pyrrolysine from the gene sequence alone, and direct chemical analysis of proteins would be more informative.

Selenocysteine is found in all domains of life, but the selenoproteome is small \cite{ReevesHoffman2009} and an additional concern is misannotation in the databases, because a stop codon (UGA) is re-purposed from ``halt translation'' to  ``incorporate    seleocysteine'' by additional sequences downstream of the gene as well as other trans-acting factors \cite{Mehta2004}.

As a result, any unique amino acid count beyond 21 must contain post-translationally modified amino acids and we note the following:

\begin{itemize}
 \item Almost the whole of the tail of Fig. \ref{fig:selenealphabets} consists of proteins in which there must be post-translationally modified amino acids, effectively doubling the unique alphabet derived directly from DNA.
 \item The increase in numbers between 20 and 26 unique amino acids can be seen by the slightly displaced points upwards in the SwissProt 15-07 dataset compared with the SwissProt 13-11 dataset.
 \item The remainder of the tail significantly straightens with the more comprehensively annotated SwissProt 15-07 presumably due to reduction in noise and increasing numbers.
\end{itemize}

\paragraph{}
\fbox{%
\begin{minipage}{12cm}
\textbf{The linearity in each tail strongly supports equation (\ref{eq:pwrlaw}) even though the range is less than 1 decade because the slope is so steep arising from the current paucity of the unique alphabet.  For SwissProt 13-11 of  Fig. \ref{fig:selenealphabets}, R lm() reports that the associated p-value matching the power-law tail linearity is $8.124 \times e^{-13}$ over the range $19 - 35$, with an adjusted R-squared value of $0.9698$.  The slope is $-15.91 \pm 0.56$.  For SwissProt 15-07 of  Fig. \ref{fig:selenealphabets}, R lm() reports that the associated p-value matching the power-law tail linearity is $< 2.2 \times e^{-16}$ over the range $19-33$, with an adjusted R-squared value of $0.9968$.  The slope is $-18.56 \pm 0.19$.}
\end{minipage}}
\paragraph{}

We should also clear up a potential point of confusion here.  One reviewer stated that the distribution of number of proteins against unique amino acid count had a ``tiny power-law tail'' and that the distribution was uniform.  The reviewer reasoned that this explained why the average length of proteins was highly conserved in contrast to our explanation in the Appendix p. \pageref{app:averagecomplength}.

\begin{figure}[H]
    \captionsetup[subfigure]{labelformat=empty}
    \centering
    \begin{subfigure}[t]{0.5\textwidth}
        \centering
        \caption{A}
        \includegraphics[width=6cm]{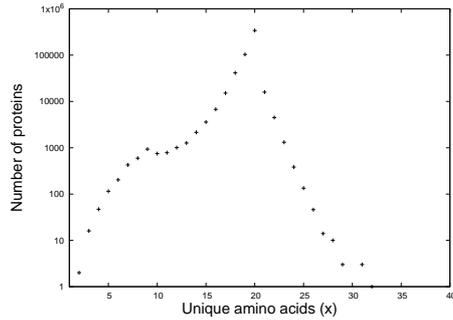}
        \label{fig:selene13-11_logpdf}
    \end{subfigure}%
    ~ 
    \begin{subfigure}[t]{0.5\textwidth}
        \centering
        \caption{B}
        \includegraphics[width=6cm]{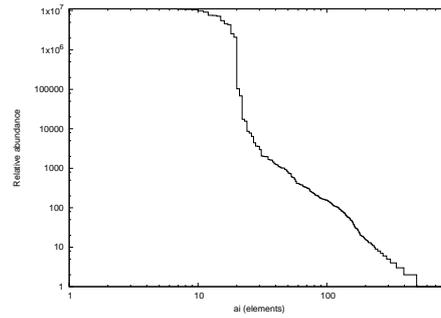}
        \label{fig:selene13-11_aa_ccdf}
    \end{subfigure}%

    \caption{The frequency of proteins plotted against the unique amino acid count for SwissProt 13-11 on a $\log-linear$ plot (A), and the frequency at which each amino acid occurs including PTM amino acids plotted in rank order on a $\log-\log$ ccdf (B).}
\end{figure}

It is indeed true at the present rate of knowledge that the distribution of unique amino acids has a tiny power-law tail but the distribution is anything but uniform as we can see in SwissProt 13-11 by considering two different kinds of plot.  Fig. \ref{fig:selene13-11_logpdf} plots the logarithmic frequency of proteins against their unique amino acid alphabet.  Whatever this distribution is, it is certainly not uniform, although we know the tail from $19-35$ amino acids is an accurate power-law from the analysis of the data shown in Fig. \ref{fig:selenealphabets}.

In contrast, Fig. \ref{fig:selene13-11_aa_ccdf} plots the occurrence rate of each unique amino acid including post-translational modification across the entire SwissProt 13-11 distribution, of which there are more than $800$ recorded by the Selene project.  In other words it shows in how many proteins each amino acid appears, organised in rank order.  This matches the homogeneous model discussed in Appendix A p. \pageref{app:homogeneous}, and a power-law in the tail is evident as expected.

We note in passing a possible intriguing relationship between the overhang in Fig. \ref{fig:selene13-11_aa_ccdf} between around 10 and 30 on the x-axis and the contemporary question of PTM undercounting \cite{ThaysenAndersen2014}.

\paragraph{}
\fbox{%
\begin{minipage}{12cm}
\textbf{R lm() reports that the associated p-value matching the power-law tail linearity in the ccdf of Fig. \ref{fig:selene13-11_aa_ccdf} is $< (2.2) \times e^{-16}$ over the range $22.0-800.0$, with an adjusted R-squared value of $0.9778$.  The slope is $-2.63 \pm 0.31$.  This too is an emphatic result.}
\end{minipage}}
\paragraph{}

\newpage
\section*{Appendix D: Power-laws, Statistical Rigour and Rules of Thumb}
\label{app:powerlaws}
\begin{figure}[H]
\centering
\includegraphics[width=8cm,height=6cm]{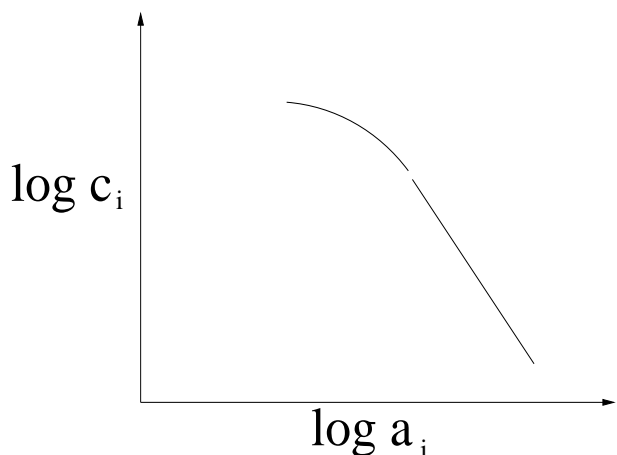}
\caption{The classic linear signature of a power-law in the tail of a $\log-\log$ ccdf.}
\label{fig:prediction}
\end{figure}

Power-laws are ubiquitous in nature and are generated by a number of mechanisms, \cite{Newman2006}.  In essence, power-law behaviour can be represented by the pdf (probability density function) p(s) of entities of size s appearing in some process, given by a relationship like

\begin{equation}
p(s) = \frac{k}{s^b}   \label{eq:power}
\end{equation}

where $k, b$ are constants.  On a $\log p - \log s$ scale the pdf is a straight line with negative slope $- b$.  It can easily be verified that the equivalent cdf (cumulative density function) $c'(s)$ derived by integrating (\ref{eq:power}) also obeys a power-law $\sim s^{-b+1}$, (for $b \neq 1$).  The classic linear signature of a power-law tail in a ccdf (complementary cumulative distribution function) is usually shown as in Fig. \ref{fig:prediction} which displays $c(s) = 1 - c'(s)$.

\begin{figure}[H]
    \captionsetup[subfigure]{labelformat=empty}
    \centering
    \begin{subfigure}[t]{0.5\textwidth}
        \centering
        \caption{A}
        \includegraphics[width=6cm]{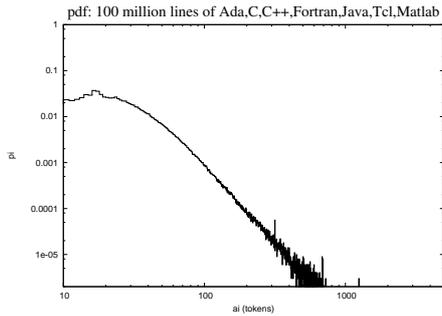}
        \label{fig:universe_pdf}
    \end{subfigure}%
    ~ 
    \begin{subfigure}[t]{0.5\textwidth}
        \centering
        \caption{B}
        \includegraphics[width=6cm]{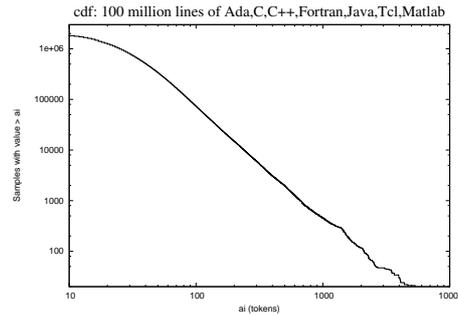}
        \label{fig:universe_cdf}
    \end{subfigure}%

    \caption{The pdf (A) and the ccdf (B) of the length distributions of the same large population of software.}
\end{figure}

For noisy data, the ccdf form is used most often because of its fundamental property of reducing noise present in the pdf, as noted by \cite{Newman2006}.  This effect is because the ccdf is obtained by integration.  This reduces noise inherent in the pdf preserving any power-law behaviour while allowing any linearity to be measured more accurately.  The benefit of this can be clearly seen in data extracted from software systems as shown in Figs. \ref{fig:universe_pdf} (the pdf) and \ref{fig:universe_cdf} (the corresponding ccdf).  The effect is even more pronounced in the inherently more noisy protein data.

Whilst on the subject of significance, a rule of thumb often used to determine the existence of a power-law is that it should appear over two or more decades in the x-axis of the ccdf.  This is useful only as a rule of thumb when the slope is not too steep.  Since the scale of the y-axis on the $\log-\log$ ccdf is effectively the scale of the x-axis times the slope of the power-law, then a steep slope would require a large scale of y-axis frequency measurements to provide the rule of thumb of 2-3 decades in the x-axis.  For example, a slope of around 3 would require y-axis frequency measurement only over some 6-9 decades to give the rule of thumb of 2-3 decades in the x-axis, which is reasonable.  On the other hand, if the slope is 10, y-axis frequency measurement over some unreasonably large 20-30 decades would be required to give the rule of thumb of 2-3 decades in the x-axis.

This is an important point for the protein studies considered here wherein we are investigating the predicted power-law in unique alphabet.  Here, the x-axis is the unique alphabet of amino acids.  The size of this alphabet is small \textit{at the current state of knowledge}, leading to a steep power-law slope.  In such situations, we fall back on normal procedures of statistical inference to replace subjective belief with objective perception and therefore all that is required is that there is statistically significant linearity in the tail of the distribution of the $\log-\log$ ccdf for the number of measurement points used.  A rule of thumb guides but does not replace normal statistical inference whereby a result is either significant at some level or it is not for a given model and data.

\begin{figure}[H]
\centering
\includegraphics[width=8cm,height=6cm]{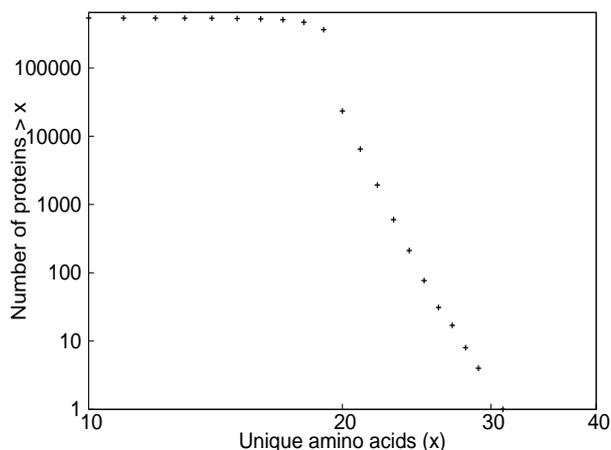}
\caption{The highly linear tail of the occurrence frequency of unique alphabet sizes in the SwissProt 13-11 protein distribution merged with the Selene 2013 post-translational modification annotation, extending the range of the natural unique alphabet of 22 amino acids directly coded from DNA to just over 30 in this dataset.}
\label{fig:swissprot_selene_13-11}
\end{figure}

This effect can be seen in Fig. \ref{fig:swissprot_selene_13-11},  a $\log-\log$ ccdf of the occurrence rate in the size of the unique alphabet in SwissProt version 13-11 \cite{SwissProt2013}, which is merged with the Selene post-translational modification data \cite{Selene2013,PTM2014}.  These represent amongst the best annotated protein data including the rapidly growing field of post-translational modification (PTM), a process whereby nature alters some of the amino acids by covalent processes such as glycosylation, phosphorylation, methylation, acylation, etc., thereby extending the unique alphabet beyond the 22 amino acids directly coded from DNA \cite{ApweilerHermjakobSharon1999,KhouryBalibanFloudas2011,ZafarNasirBokhari2011,PrabakarnLippensSteenGunawardena2012,CampbellEtAl2014}.

As can be seen, the tail of SwissProt 13-11 in Fig. \ref{fig:swissprot_selene_13-11}, covers only a range up to just over 30 even though there are thousands of PTM known or predicted by existing research.  As a result there are only nine data points for unique alphabets of size greater than 20, although each point is an aggregate of a large number of observations.

\paragraph{}
\fbox{%
\begin{minipage}{12cm}
\textbf{An R lm() analysis on this tail reports that the associated p-value matching the power-law tail linearity in the ccdf of Fig. \ref{fig:musicpowerlaws_pdf} is $6.576 \times e^{-12}$ over the range $21.0-30.0$, with an adjusted R-squared value of $0.9951$.  The slope is $-22.9 \pm 0.2$.}
\end{minipage}}
\paragraph{}

This is a statistically emphatic result for the existence of a power-law, even though the size of the slope is steep because the x-axis is restricted.  Later versions of SwissProt with Selene annotations increase the PTM alphabet, Appendix C p. \pageref{app:proteinalphabets}.
  
Power-law behaviour has been studied in a wide variety of environments starting with the pioneering work of \cite{Zipf35} (linguistics) and followed by \cite{Rawl04} (economic systems) and the excellent reviews by \cite{Mitzenmacher2003} and \cite{Newman2006}.  In software systems significant activity, much of it recent, \cite{ClarkGreen77}, \cite{Mitzenmacher2002}, \cite{Myers2003}, \cite{Challet2004}, \cite{Gorshenev2004}, \cite{Potanin05}, \cite{Baxter06},  \cite{Concas2007}, \cite{Louridas:2008:PLS:1391984.1391986} and \cite{HatTSE08} has addressed power-law behaviour in various contexts.

To give some idea of the scope of these, Mitzenmacher \cite{Mitzenmacher2002} considers the distributions of file sizes in general filing systems and observed that such file sizes were typically distributed with a lognormal body and a Pareto (i.e. power-law) tail. Gorshenev and Pis'mak \cite{Gorshenev2004} studied the version control records of a number of open source systems with particular reference to the number of lines added and deleted at each revision cycle. Louridas et. al. \cite{Louridas:2008:PLS:1391984.1391986} show evidence that power laws appear in software at the class and function level and that distributions with long, fat tails in software are much more pervasive than previously established.

\newpage
\section*{Appendix E: Hartley-Shannon Information, parsimony and token-agnosticism}
\label{app:hsinfo}
\begin{figure}
\includegraphics[width=8cm]{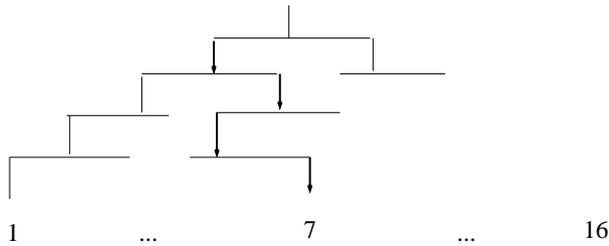}
\caption{A binary tree.  Each level proceeding down can either go left or right.  There are four levels leading down to one of $2^{4}$ = 16 possibilities.  Only four choices are needed to reach any of the possibilities.  We note that $\log_{2}(16) = 4$.  Here the number 7 has been singled out by the choices left, right, left, right as the tree is descended.}
\label{fig:choices}
\end{figure}

Information theory has its roots in the work of Hartley \cite{Hartley1928} who showed that a message of N signs (i.e. tokens) chosen from an \textit{alphabet} or code book of S signs has $S^{N}$ possibilities and that the \textit{quantity of information} is most reasonably defined as the \textit{logarithm} of the number of possibilities or choices $\log S^{N} = N \log S$.  To gain insight into the reason why the logarithm makes sense, consider Fig. \ref{fig:choices}.  The number of choices necessary to reach any of the 16 possible targets is the number of levels which is $\log_{2}$(number of possibilities).  The base of the logarithm is not important here.

Information theory was developed substantially by the pioneering work of Shannon \cite{Shannon1948}, \cite{Shannon1949} and many researchers since but we have remained with Hartley's original clear vision and most importantly its token-agnosticism.  We re-iterate that is important not to conflate information content with functionality or meaning and Cherry \cite{Cherry1963} specifically cautions against this noting that the concept of information based on alphabets as extended by Shannon and Wiener amongst others, \textit{relates only to the symbols themselves} and not their \textit{meaning}.  Indeed, Hartley in his original work, defined \textit{information} as the successive selection of signs, rejecting all meaning as a mere subjective factor. In the sense used here therefore, Conservation of H-S Information will be synonymous with Conservation of Choice, not meaning.  This turns out to be enough to predict the important system properties detailed in this paper.  In other words, those properties depend only on the alphabet and not on what combining tokens of the alphabet might mean in any human sense.

We believe CoHSI therefore represents the most parsimonious theory capable of explaining all the observed features of the numerous disparate datasets analysed in this paper.

\nolinenumbers

%
%
%
%
%
%
%
%
%
%
%
%
%


\end{document}